\documentclass[12pt]{amsart}
\usepackage{amsmath,amssymb,cite,enumitem,graphicx,hyperref}

\hypersetup{pdftex,colorlinks=true,allcolors=blue}
\usepackage[all]{hypcap}

\usepackage[foot]{amsaddr}
\usepackage[sc]{mathpazo}

\newcommand\ack{\subsection*{Acknowledgment}}

\makeatletter
\def\paragraph{\@startsection{paragraph}{4}%
  \z@\z@{-\fontdimen2\font}%
  {\normalfont\bfseries}}
\makeatother

\DeclareMathAlphabet\mathsfbi{T1}{phv}{b}{it}

\numberwithin{equation}{section}

\newcommand\BV{\boldsymbol} 
\newcommand\BM{\mathsfbi} 

\newcommand\parderiv[2]{\frac{\partial #1}{\partial #2}}

\def\HS{{HS}}

\begin{document}

\author[Rafail V. Abramov]{Rafail V. Abramov}

\address{Department of Mathematics, Statistics and Computer Science,
University of Illinois at Chicago, 851 S. Morgan st., Chicago, IL 60607}

\email{abramov@uic.edu}

\title[Power spectra in the two-dimensional Poiseuille and Couette
  flow]{Power spectra via the van der Waals effect in the
  two-dimensional Poiseuille and Couette flow}

\begin{abstract}
We numerically simulate the two-dimensional inertial flow with the van
der Waals effect in a straight periodic channel around the Poiseuille
and Couette stationary states. Even though the flow remains laminar
macroscopically, we observe complex dynamics and power decay of the
Fourier spectra of small fluctuations of the density, velocity
divergence, vorticity and kinetic energy of the flow near their
respective stationary background states. Remarkably, pinning the
vorticity to its background state, and leaving only the density and
velocity divergence as the variables, results in the dynamics and
power decay of the Fourier spectra qualitatively similar to those of
the full system. This strongly indicates that the underlying physics
of the power spectra reside primarily in the density and velocity
divergence variables, and are not directly related to the vorticity of
the flow.
\end{abstract}

\maketitle

\section{Introduction}

The power decay of the Fourier spectra of various quantities in gas
flows have been confirmed by numerous atmospheric observations on
Earth \cite{FicMcV,NasGag,Zou2021} (also see the reference section of
\cite{Zou2021} for a more comprehensive list) and Jupiter
\cite{ChoSho,MorMigAlt}, as well as through laboratory experiments
\cite{BucVel}. However, only empirical, {\em ad hoc} dimensional
hypotheses have been suggested as an explanation for this phenomenon
thus far
\cite{Kol41a,Kol41b,Kol41c,Obu41,Obu42,Obu49,Cha49,Cor51,Kol62}.

In our recent works \cite{Abr22,Abr23,Abr24,Abr26,Abr27,Abr25} we
proposed a new model of a compressible gas flow, which relies upon the
empirically observed equilibration of pressure at low Mach numbers. The
pressure equilibration has been previously used to explain large
density and temperature variations in a low Mach number flow in the
presence of gravity \cite{RehBau,Hor,MlaTsuNag,MatOhbMun}, where the
traditional Boussinesq approximation \cite{Bou2} fails. However, the
novel observation in our works
\cite{Abr22,Abr23,Abr24,Abr26,Abr27,Abr25} is that, in the absence of
gravity, the pressure equilibration leads to the constant pressure
(inertial flow), which causes the van der Waals effect (a.k.a.~the
second coefficient of the virial expansion of the equation of state,
or the mean field effect of the intermolecular potential \cite{Abr22})
to become the leading order term under the pressure gradient in the
momentum transport equation. We subsequently found that the inclusion
of the van der Waals effect creates an instability in the numerically
simulated inertial flow, and leads to a spontaneous development of
turbulent dynamics. We also observed the power decay of the Fourier
spectra in the resulting turbulent flow.

Remarkably, while it is impossible to have a two-dimensional (2D) flow
in the real world, both our simulations \cite{Abr23} and theory
\cite{Abr27} suggest that the underlying physics of turbulence are
two-dimensional.  Namely, in \cite{Abr23} we conducted a numerical
simulation of a 2D inertial jet in a channel, and observed that it
exhibits major features of a typical turbulent flow -- namely, in the
same fashion as a 3D flow does, the simulated 2D laminar jet
spontaneously breaks down into chaotic dynamics whose time averages of
the Fourier spectra of the kinetic energy exhibit the power decay.
Then, we found in \cite{Abr27} that the turbulent instability, created
by the van der Waals effect in the velocity divergence by coupling it
to the large scale background vorticity, manifests in the 2D space as
well.

Even in a linearized two-dimensional setting, the analysis of
turbulent dynamics remains a challenging task \cite{Abr27}. The main
reason behind it is that, even with a simple Couette flow serving as a
stationary background state, linearized dynamics of small
perturbations along a characteristic in the Fourier space are
represented by a 3$\times$3 system of linear non-autonomous ODE, which
we were unable to solve explicitly.  Instead, we only managed to
examine its solutions for relatively short times, and in the
asymptotic limit. The most interesting, ``inertial'' regime, which
lies between the initial unstable stage, and the asymptotic decay,
remained inaccessible thus far.

In the current work, we examine the dynamics of small perturbations
around the Poiseuille and Couette stationary background states via
numerical simulations. For that, we re-cast the transport equations
into the divergence--vorticity variables much like we did in
\cite{Abr27}, which allows to separate the effects of compression and
rotation of the flow into two distinct variables. In the initial
conditions, the divergence and vorticity variables are set to their
respective background states, while the density variable is ``nudged''
from its background state by a small perturbation.

Remarkably, the flow in our simulations remains macroscopically
laminar -- that is, while the small fluctuations around the studied
stationary background states indeed exhibit a complex behavior, the
total flow never breaks down on a macroscopic scale, which is
confirmed by the streaks of passive tracers seeded in the initial
condition. Yet, the power decay is observed in the time averages of
the Fourier transforms of the system variables, much like in our
preceding works \cite{Abr22,Abr23,Abr24,Abr25,Abr26}.

Even more remarkably, we find that pinning the vorticity to its
constant background state (Poiseuille or Couette), and leaving only
the density and velocity divergence as the variables, results in a
qualitatively similar chaotic dynamics and power decay of the Fourier
spectra, indicating that the vorticity fluctuations are not a key part
of the physics underpinning the power spectra. Therefore, should we
carry out the same linearization and analysis as in our work
\cite{Abr27}, the resulting system of ODE along a characteristic in
the Fourier space will be $2\times 2$, rather than $3\times 3$. Such a
reduction in size likely makes the system more receptive to further
analysis, which may improve our understanding of the physics behind
the power decay of the Fourier spectra in real-world gas flows.

\section{The divergence--vorticity formulation in two dimensions}

The inertial gas flow equations \cite{Abr22,Abr23,Abr24,Abr26,Abr27} are
\begin{equation}
\parderiv\rho t+\nabla\cdot(\rho\BV u)=0,\qquad\parderiv{(\rho\BV
  u)}t+\nabla\cdot(\rho\BV u^2)+\frac{4p_0}{\rho_\HS}\nabla\rho
=\nabla\cdot\left[\mu\left(\nabla\BV u+\nabla\BV u^T-\frac
  23(\nabla\cdot\BV u)\BM I\right)\right].
\end{equation}
Above, $\rho$ and $\BV u$ are the density and velocity variables,
respectively, $\mu$ is the dynamic viscosity, $p_0$ is the constant
pressure parameter, and $\rho_\HS$ is the density of a hard sphere.
In what follows, we assume that the dynamic viscosity $\mu$ is
constant, for simplicity. First, in the momentum equation, we factor
$\rho$ out of the advection terms with the help of the density
equation:
\begin{equation}
\rho\left(\parderiv{\BV u}t+\BV u\cdot\nabla\BV u\right)+\frac{4p_0}{
  \rho_\HS}\nabla\rho=\mu\left(\Delta\BV u+\frac 13\nabla(\nabla\cdot
\BV u)\right).
\end{equation}
Next, we divide both sides by $\rho$, and assume that the kinematic
viscosity $\nu=\mu/\rho$ is also constant (that is, we neglect the
variations of $\rho$ in the viscous term). The result is
\begin{equation}
\parderiv{\BV u}t+\BV u\cdot\nabla\BV u+\frac{4p_0}{\rho_\HS}\frac{
  \nabla\rho}\rho=\nu\left(\Delta\BV u+\frac 13\nabla(\nabla\cdot \BV
u)\right).
\end{equation}
At this point, we introduce the perpendicular gradient $\nabla^\perp$
via
\begin{equation}
\nabla^\perp=\begin{pmatrix}-\parderiv{}y \\ \parderiv{}x\end{pmatrix},
\end{equation}
and define the divergence $\chi$, vorticity $\omega$, potential
$\varphi$ and stream function $\psi$ of the flow, respectively, via
\begin{equation}
\chi=\nabla\cdot\BV u,\qquad\omega=\nabla^\perp\cdot\BV u,\qquad
\Delta\varphi=\chi,\qquad\Delta\psi=\omega.
\end{equation}
With the help of the Helmholtz decomposition, we have
\begin{equation}
\label{eq:Helmholtz}
\BV u=\BV u^\varphi+\BV u^\psi,\qquad\BV u^\varphi=\nabla\varphi,
\qquad\BV u^\psi=\nabla^\perp\psi,
\end{equation}
so that the momentum equation above can be expressed entirely in terms
of $\chi$, $\omega$, $\varphi$ and $\psi$. For that, we compute
$\nabla\cdot$ and $\nabla^\perp\cdot$ of the momentum equation, which,
after some manipulations, leads to the complete system
\begin{subequations}
\label{eq:div_vort}
\begin{equation}
\parderiv\rho t+\nabla\cdot(\rho\BV u)=0,\qquad \parderiv{\chi}t+
\nabla\cdot(\chi\BV u)-2\det(\nabla\BV u)+\frac{4p_0}{\rho_\HS}\nabla
\cdot\left(\frac{\nabla\rho}\rho\right)=\frac 43\nu\Delta\chi,
\end{equation}
\begin{equation}
\parderiv{\omega}t+\nabla\cdot(\omega\BV u)=\nu\Delta\omega,\qquad
\Delta\varphi=\chi,\qquad\Delta\psi=\omega,\qquad\BV u=\nabla
\varphi+\nabla^\perp\psi.
\end{equation}
\end{subequations}
The turbulent instability, which we uncovered in our work
\cite{Abr27}, is comprised of the term $-2\det(\nabla\BV u)$ and the
van der Waals effect $4p_0\nabla\cdot(\rho^{-1}\nabla\rho)/\rho_\HS$,
both in the $\chi$-equation. The former creates linearly unstable
fluctuations of $\chi$ at the inertial range wavenumbers, while the
latter couples the $\rho$- and $\chi$-equations into a wave-like
equation.

\subsection{Computational set-up}

We compute the numerical solutions of \eqref{eq:div_vort} using
OpenFOAM \cite{WelTabJasFur}.  The equations in \eqref{eq:div_vort}
are time-discretized using the implicit Euler scheme, where the
variables are updated from the time level $n$ to $n+1$ in the order
listed:
\begin{subequations}
\label{eq:div_vort_num}
\begin{equation}
\frac{\rho_{n+1}-\rho_n}{\Delta t}+\nabla\cdot(\rho_{n+1}\BV u_n)=0,
\end{equation}
\begin{equation}
\frac{\chi_{n+1}-\chi_n}{\Delta t}+\nabla\cdot(\chi_{n+1}\BV u_n)-2
\det(\nabla\BV u_n)+\frac{4p_0}{\rho_\HS}\nabla\cdot\left(\frac{
  \nabla\rho_{n+1}}{\rho_{n+1}}\right)=\frac 43\nu\Delta\chi_{n+1},
\end{equation}
\begin{equation}
\frac{\omega_{n+1}-\omega_n}{\Delta t}+\nabla\cdot(\omega_{n+1}\BV
u_n)=\nu\Delta\omega_{n+1},
\end{equation}
\begin{equation}
\Delta\varphi_{n+1}=\chi_{n+1},\qquad
\Delta\psi_{n+1}=\omega_{n+1},\qquad\BV
u_{n+1}=\nabla\varphi_{n+1}+\nabla^\perp\psi_{n+1}.
\end{equation}
\end{subequations}
The time step $\Delta t$ is chosen adaptively at each time step to
correspond to 20\% of the Courant number. The spatial discretization
of \eqref{eq:div_vort_num} is implemented via the standard
second-order finite volume scheme using the van Leer flux limiter
\cite{vanLee}.

Our two-dimensional computational domain is a rectangular channel of
length $L=40$ cm, and width $W=25$ cm. The spatial discretization is
uniform with the cell size of $0.25\times 0.25$ mm, so that the number
of spatial steps is 1600 along the channel, and 1000 across. The
constant parameters of the simulation are
\begin{itemize}
\item Constant pressure $p_0=1.013\cdot 10^5$ Pa, which corresponds to
  sea level;
\item The density of a hard sphere $\rho_\HS=1850$ kg/m$^3$ (for
  details, see \cite{Abr23});
\item Kinematic viscosity $\nu=1.525\cdot 10^{-5}$ m$^2$/s, which
  corresponds to air at sea level.
\end{itemize}
The boundary conditions are periodic at the inlet and outlet of the
channel (so that, effectively, the channel is a cylinder). At the
channel walls, the boundary conditions are:
\begin{itemize}
\item The density $\rho$ is set to $\rho_0=1.204$ kg/m$^3$, which
  corresponds to air at sea level;
\item The potential $\varphi$ is set to zero;
\item The velocity divergence $\chi$ is set according to Thom's
  formula \cite{Thom1933};
\item The stream function $\psi$ and vorticity $\omega$ are set to the
  values of the stationary background profile, depending on the type
  of the flow (Poiseuille or Couette).
\end{itemize}

\section{Numerical simulation of the Poiseuille flow}
\label{sec:Poiseuille}

The Poiseuille flow is a stationary state of \eqref{eq:div_vort},
represented by a parabolic velocity profile in a straight channel:
\begin{equation}
\label{eq:u_Poiseuille}
\BV u_0=U_0\left[1-\left(\frac{2y}W\right)^2\right]\begin{pmatrix} 1
  \\ 0\end{pmatrix},
\end{equation}
where $W=25$ cm is the width of the channel, and we choose $U_0=30$
m/s. As we can see, the velocity is zero at the walls of the channel,
and reaches 30 m/s in the center of the channel. In the variables of
\eqref{eq:div_vort}, the Poiseuille state becomes
\begin{equation}
\label{eq:Poiseuille_rho_chi_omega}
\rho=\rho_0,\qquad\chi=0,\qquad\omega=\frac{8U_0}{W^2}y,
\end{equation}
that is, $\rho$ is set to its background value of $1.204$ kg/m$^3$,
and $\omega$ varies linearly between $-480$ and $480$ m/s$^2$ across
the width of the channel.

\begin{figure}[t]%
\includegraphics[width=0.49\textwidth]{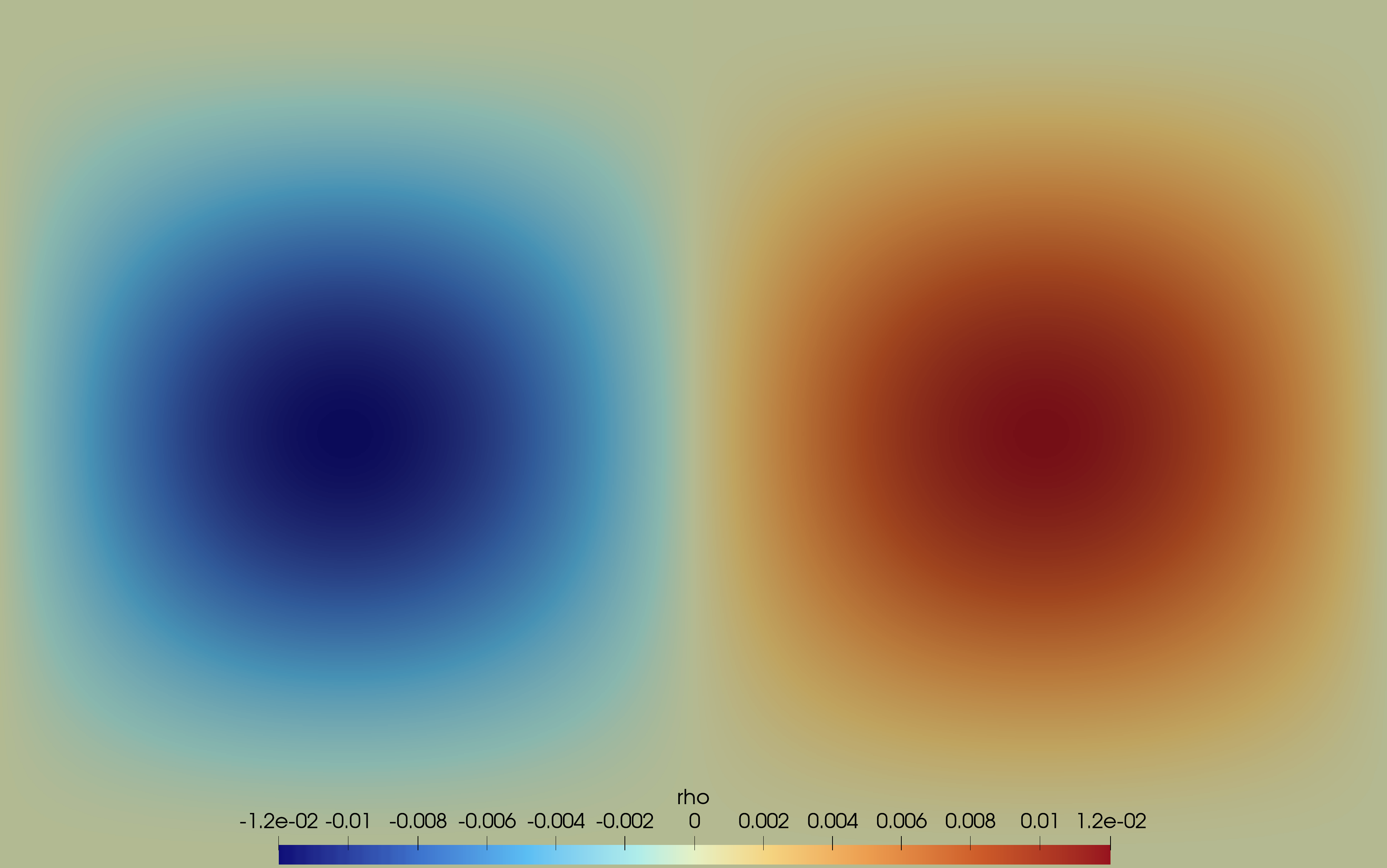}\,%
\includegraphics[width=0.49\textwidth]{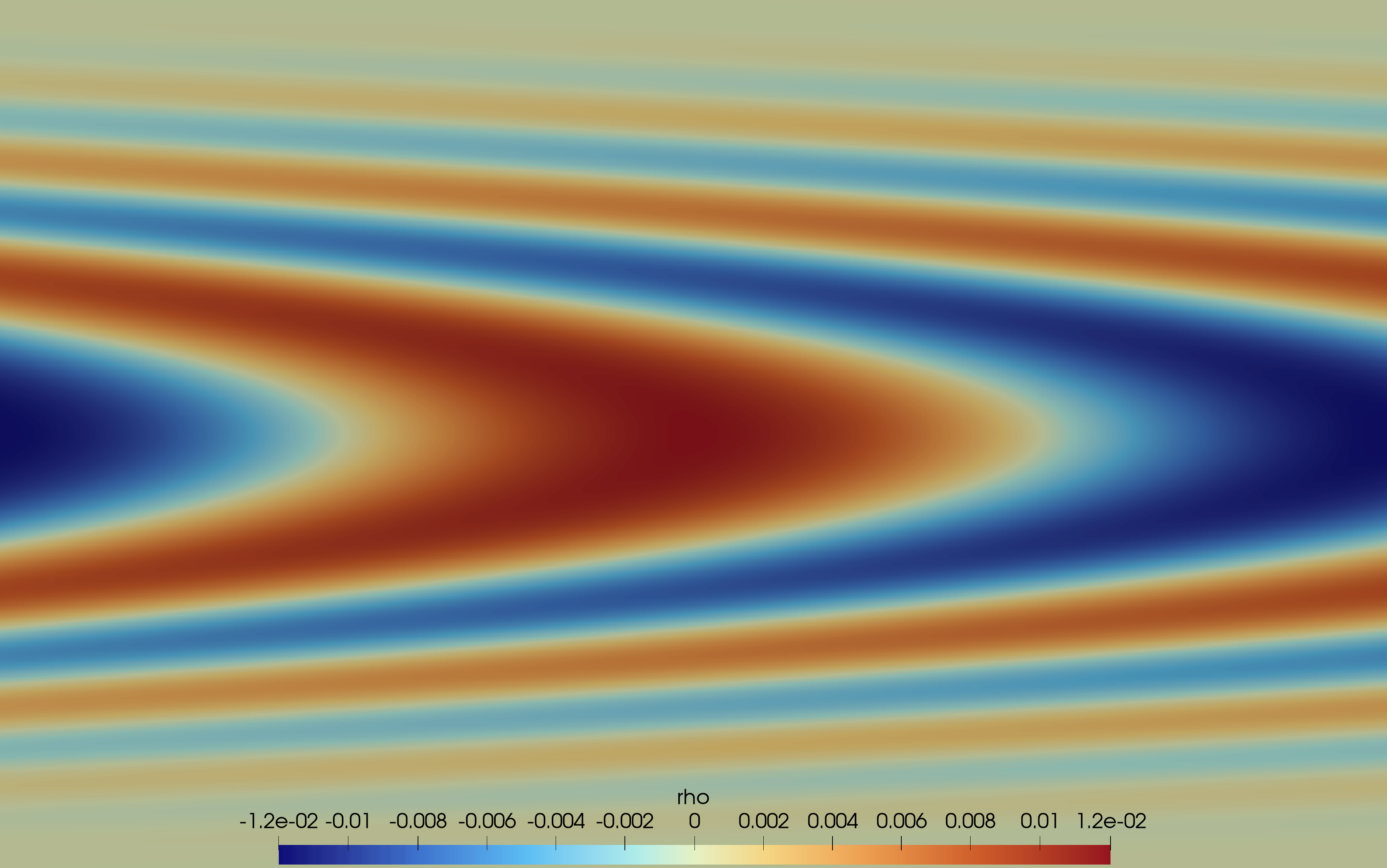}%
\caption{Density fluctuations in the Poiseuille flow in the absence of
  the van der Waals effect. Left -- starting time, right -- at 0.05
  seconds.}
\label{fig:density_Poiseuille_novdW}
\end{figure}
\begin{figure}[t]%
\includegraphics[width=0.49\textwidth]{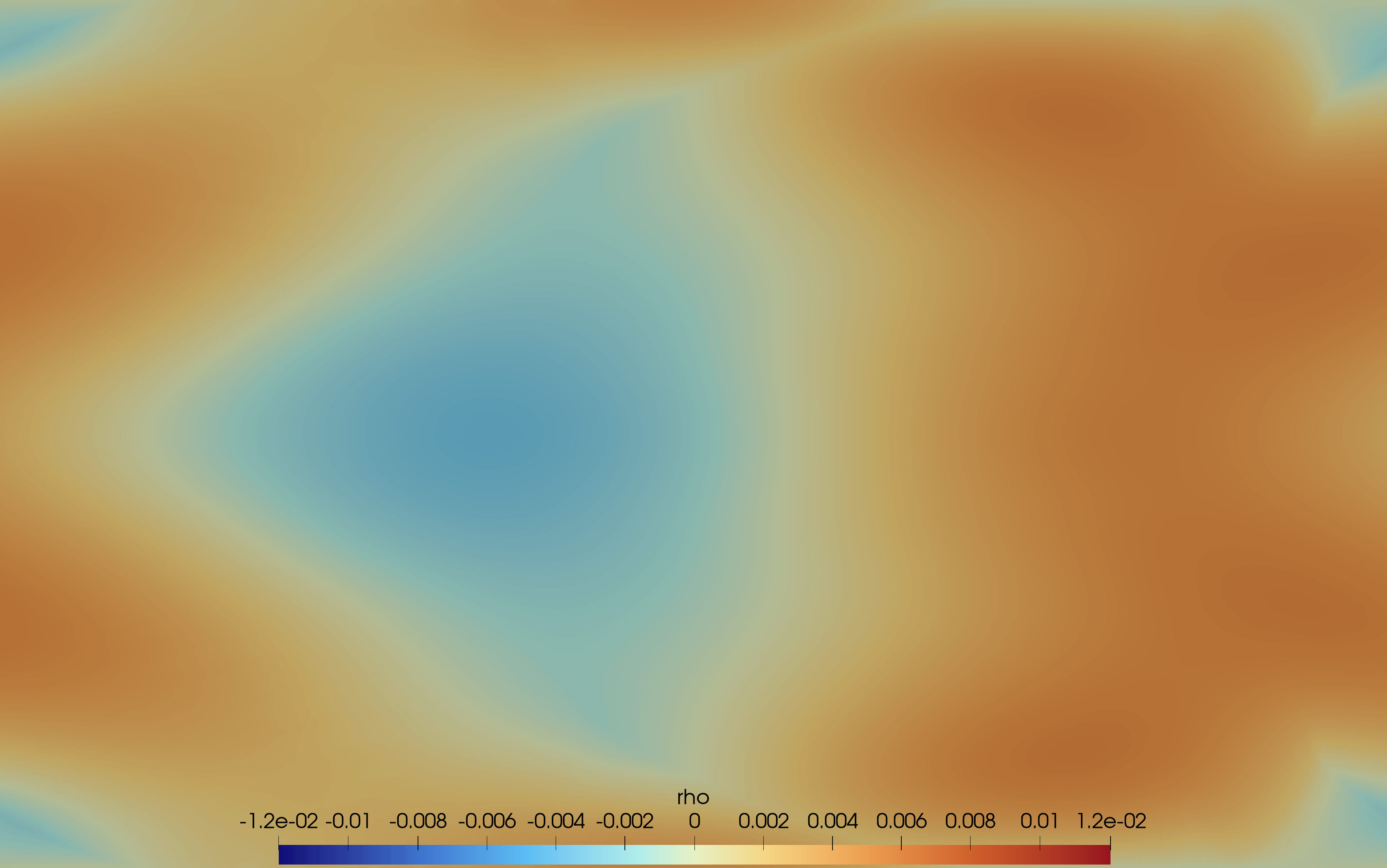}\,%
\includegraphics[width=0.49\textwidth]{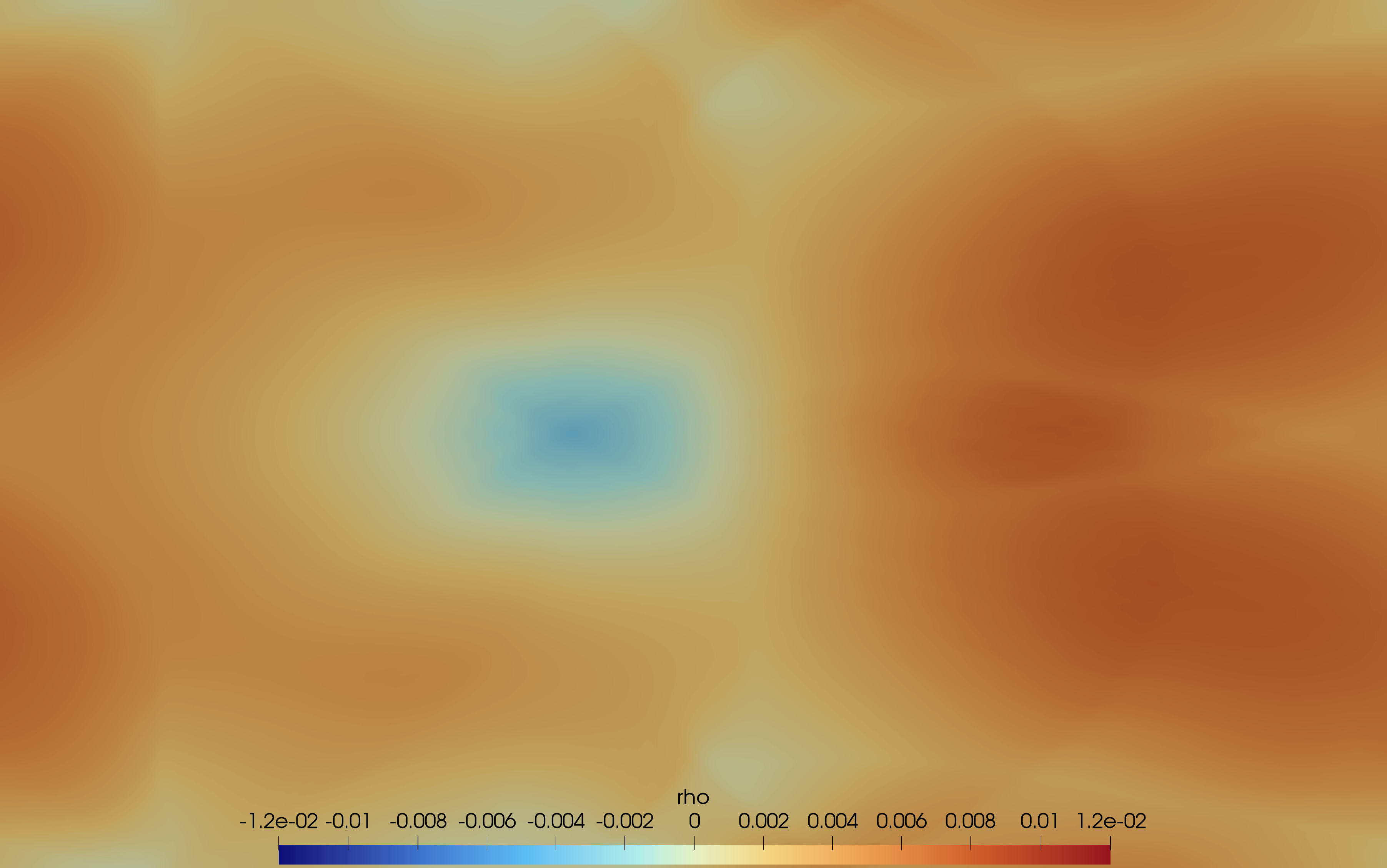}\\%
\vspace{1.5pt}%
\includegraphics[width=0.49\textwidth]{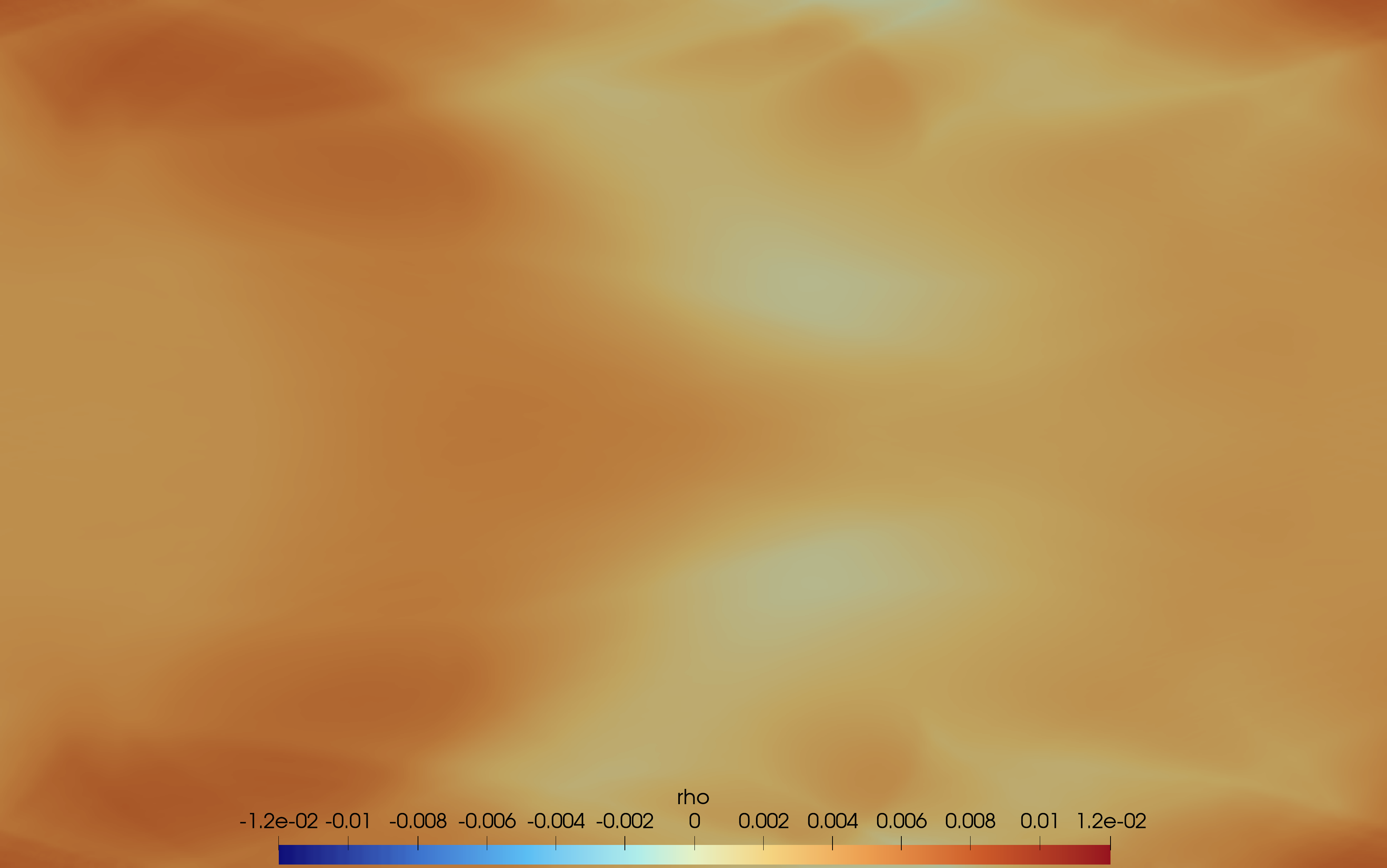}\,%
\includegraphics[width=0.49\textwidth]{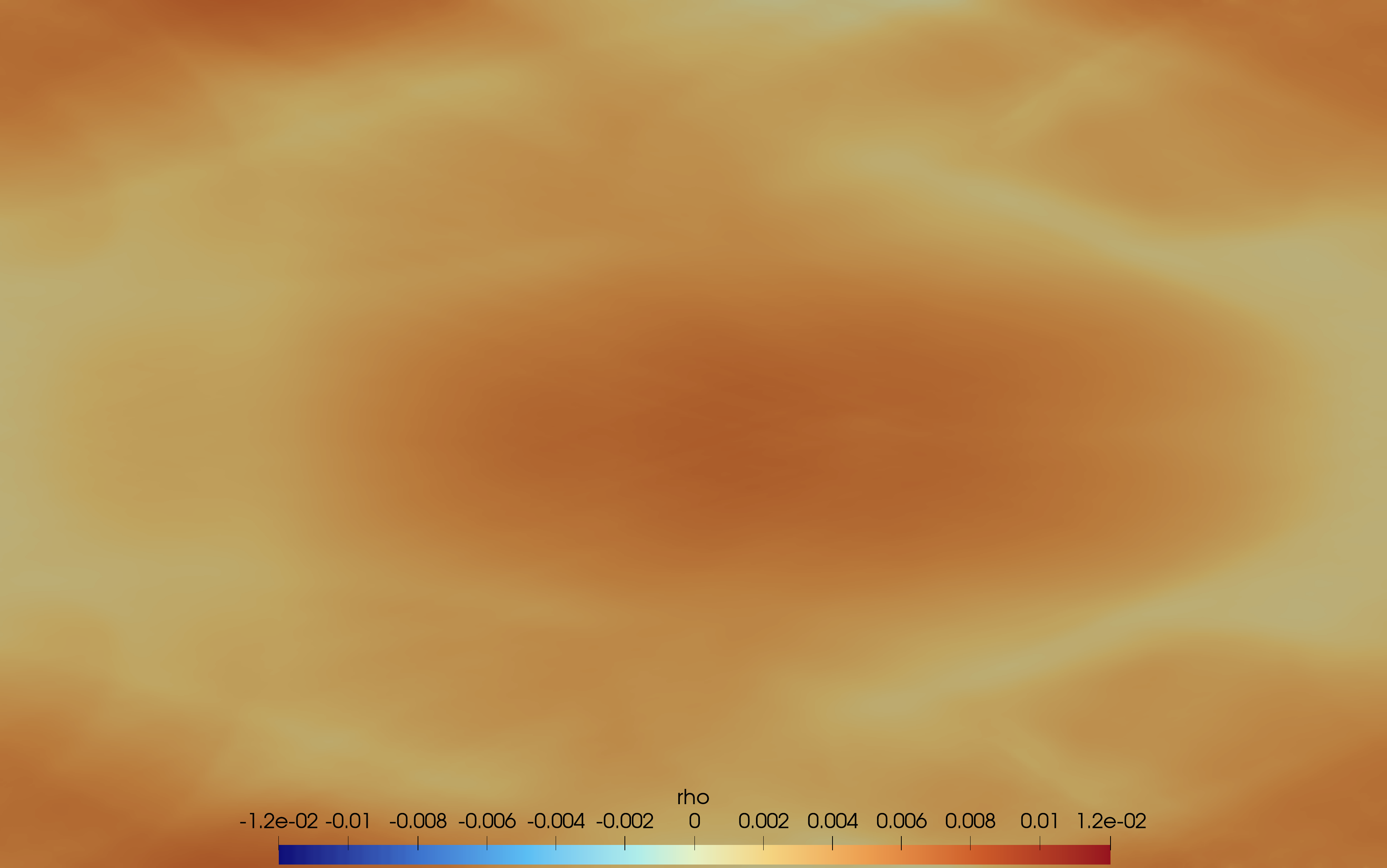}%
\caption{Density fluctuations in the Poiseuille flow in the presence
  of the van der Waals effect. Upper-left -- at 0.01 seconds,
  upper-right -- at 0.02 seconds, lower-left -- at 0.03 seconds,
  lower-right -- at 0.05 seconds.}
\label{fig:density_Poiseuille}
\end{figure}

In the initial condition of our simulation, the Poiseuille state is
perturbed by introducing a small (1\% of the background magnitude at
most), but large-scale density deviation into $\rho$ from the
background state $\rho_0$, given by
\begin{equation}
\label{eq:density_deviation}
\rho=\rho_0\left\{1+0.005\sin\left(\frac{2\pi x}L\right)\left[1+\cos\left(
  \frac{2\pi y}W\right)\right]\right\}.
\end{equation}
Graphically, the deviation of $\rho$ from its background state
$\rho_0$ is shown on the left-hand pane of Figure
\ref{fig:density_Poiseuille_novdW}.  The initial values of $\chi$ and
$\omega$ are set to their respective Poiseuille background values in
\eqref{eq:Poiseuille_rho_chi_omega}.

\subsection{The flow in the absence of the van der Waals effect}

First, we conduct a numerical simulation in the absence of the van der
Waals effect, which amounts to setting $p_0=0$ in the divergence
equation of \eqref{eq:div_vort}. This change effectively decouples the
equations for $\chi$ and $\omega$ from $\rho$. Since the only
difference between the initial condition and the Poiseuille state is
the fluctuation in $\rho$, which is now decoupled from $\chi$ and
$\omega$, the latter two remain at their Poiseuille stationary
states. As a result, only the nonuniformities of $\rho$ are
transported by the otherwise stationary Poiseuille flow along the
channel. The reason we conduct this trivial numerical simulation is to
ensure that the computational code works correctly. We find that,
within the machine round-off error, $\chi$ and $\omega$ indeed remain
at their Poiseuille steady states for the duration of the simulation,
and we show the density fluctuations at the terminal time $t=0.05$
seconds in the right-hand pane of Figure
\ref{fig:density_Poiseuille_novdW}.  It is not difficult to see that
the final state of the density variable indeed corresponds to the
Poiseuille flow transport.

\begin{figure}[t]%
\includegraphics[width=0.49\textwidth]{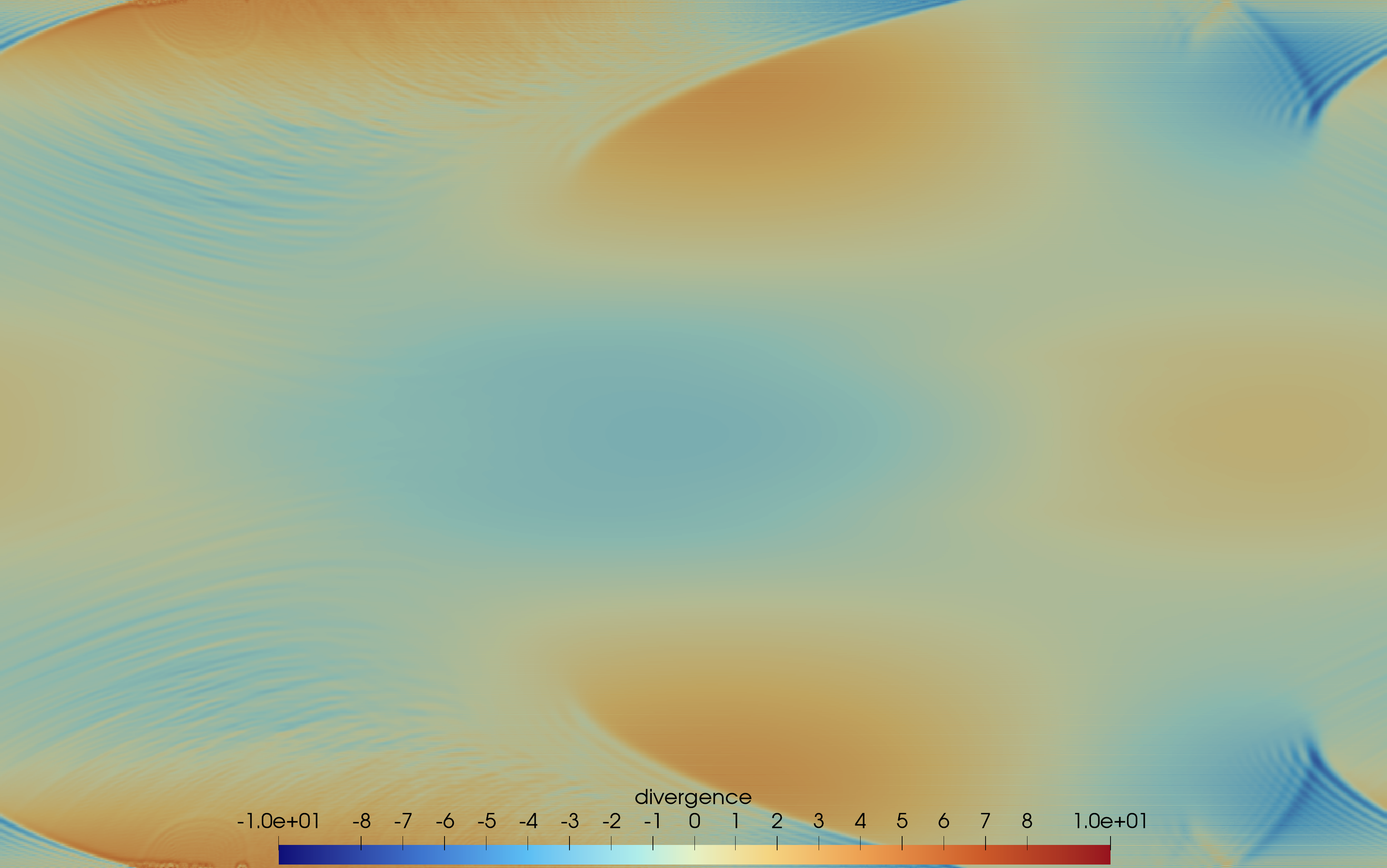}\,%
\includegraphics[width=0.49\textwidth]{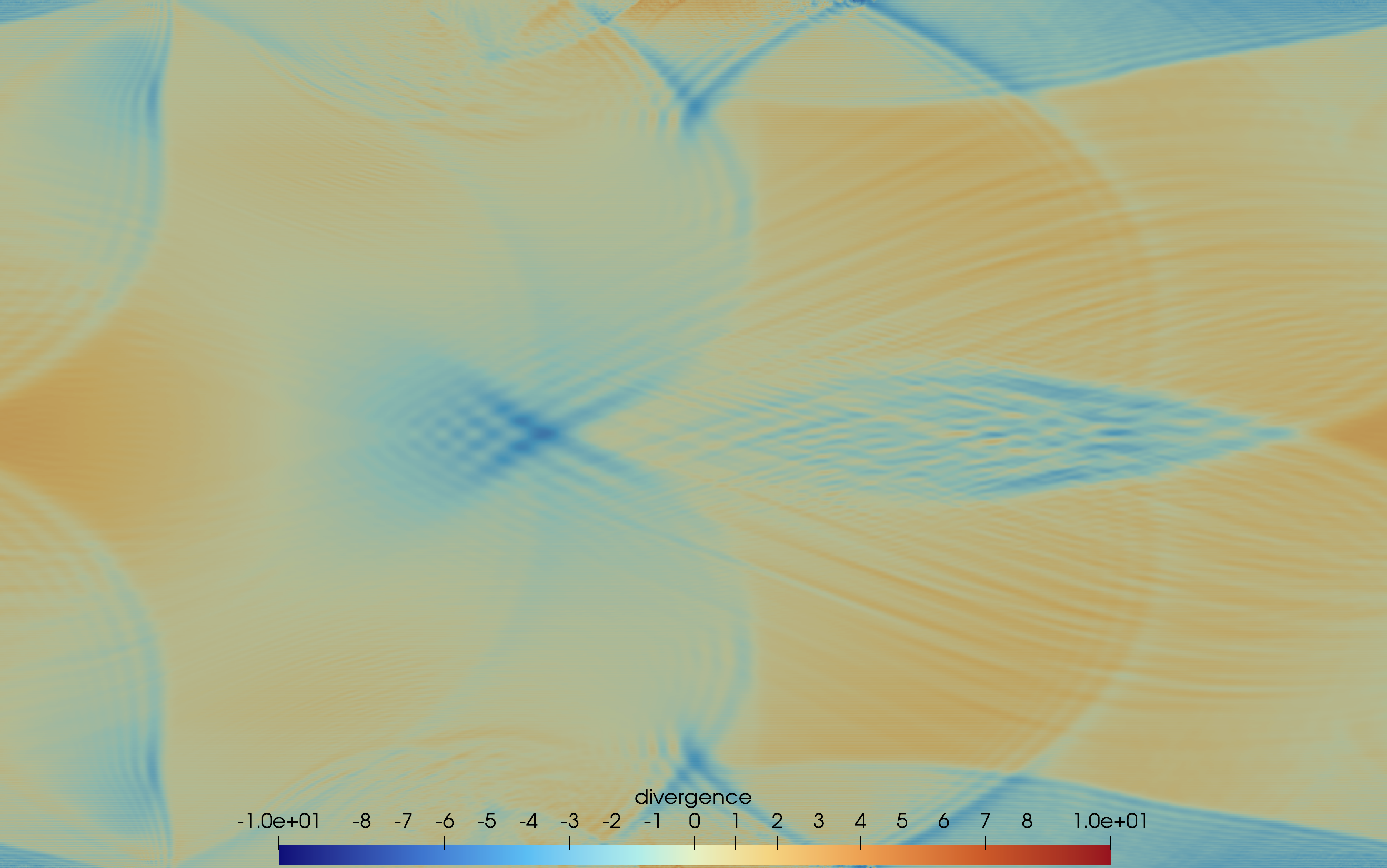}\\%
\vspace{1.5pt}%
\includegraphics[width=0.49\textwidth]{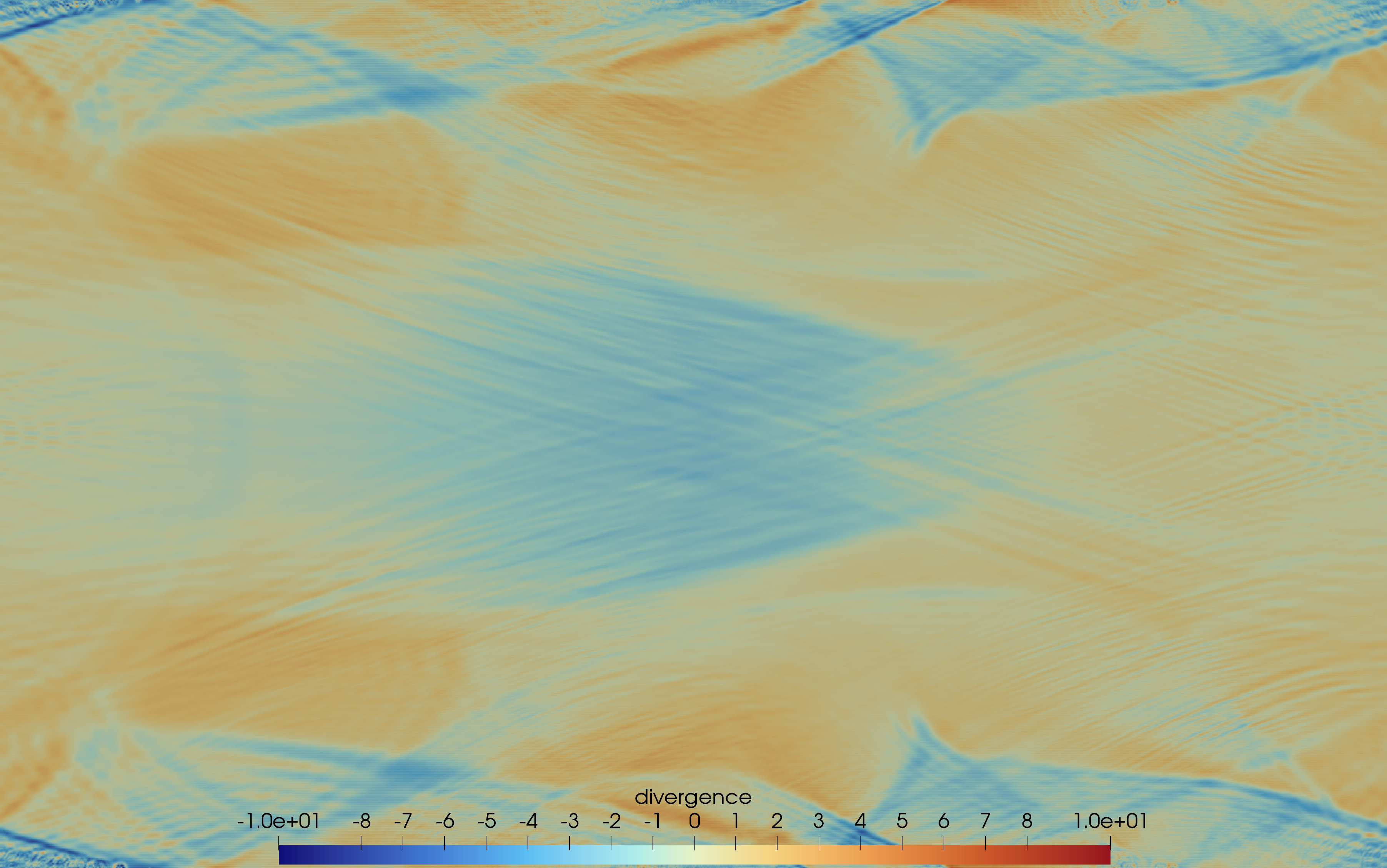}\,%
\includegraphics[width=0.49\textwidth]{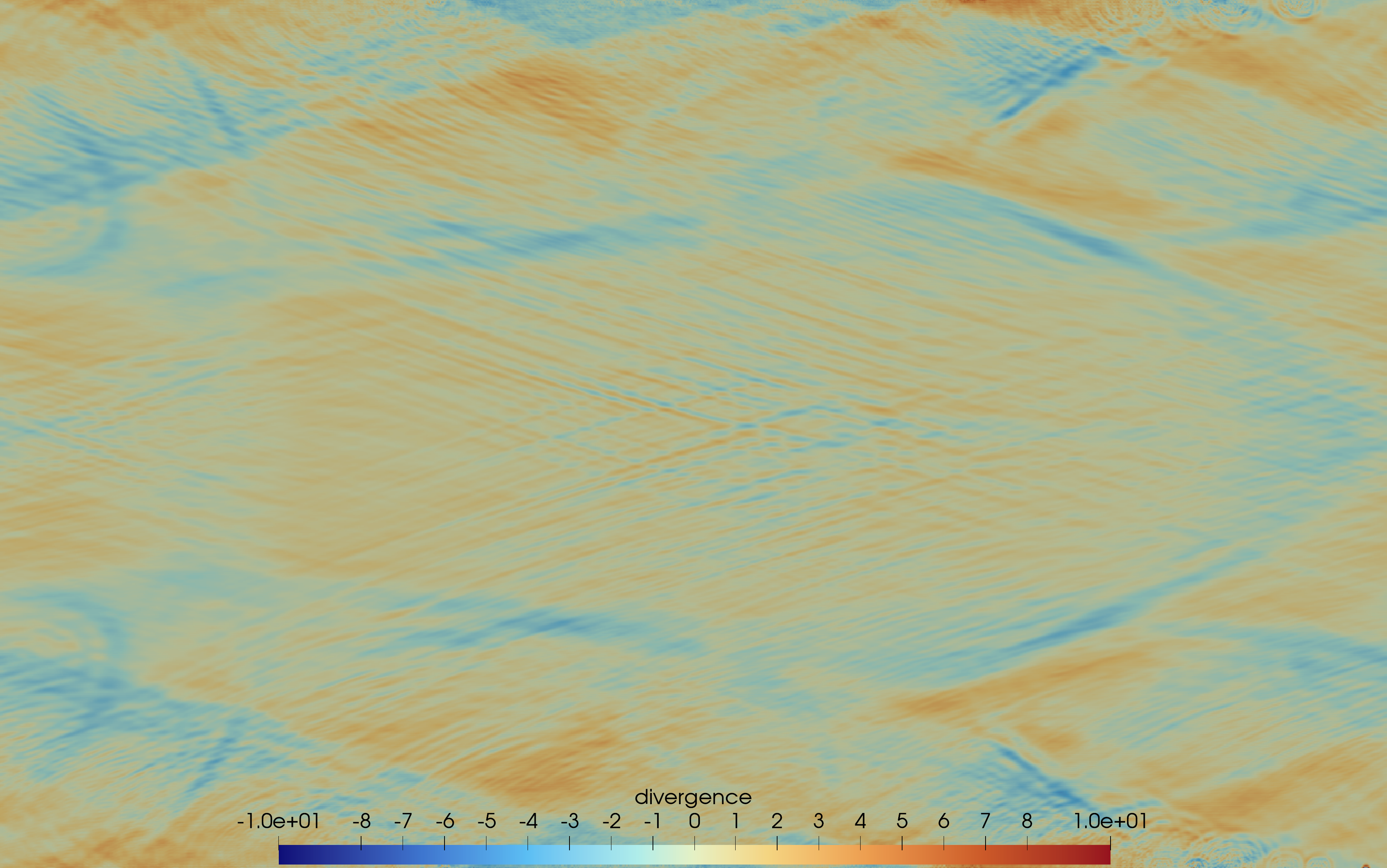}%
\caption{Divergence fluctuations in the Poiseuille flow in the
  presence of the van der Waals effect. Upper-left -- at 0.01 seconds,
  upper-right -- at 0.02 seconds, lower-left -- at 0.03 seconds,
  lower-right -- at 0.05 seconds.}
\label{fig:divergence_Poiseuille}
\end{figure}

\subsection{The flow in the presence of the van der Waals effect}

\begin{figure}[t]%
\includegraphics[width=0.49\textwidth]{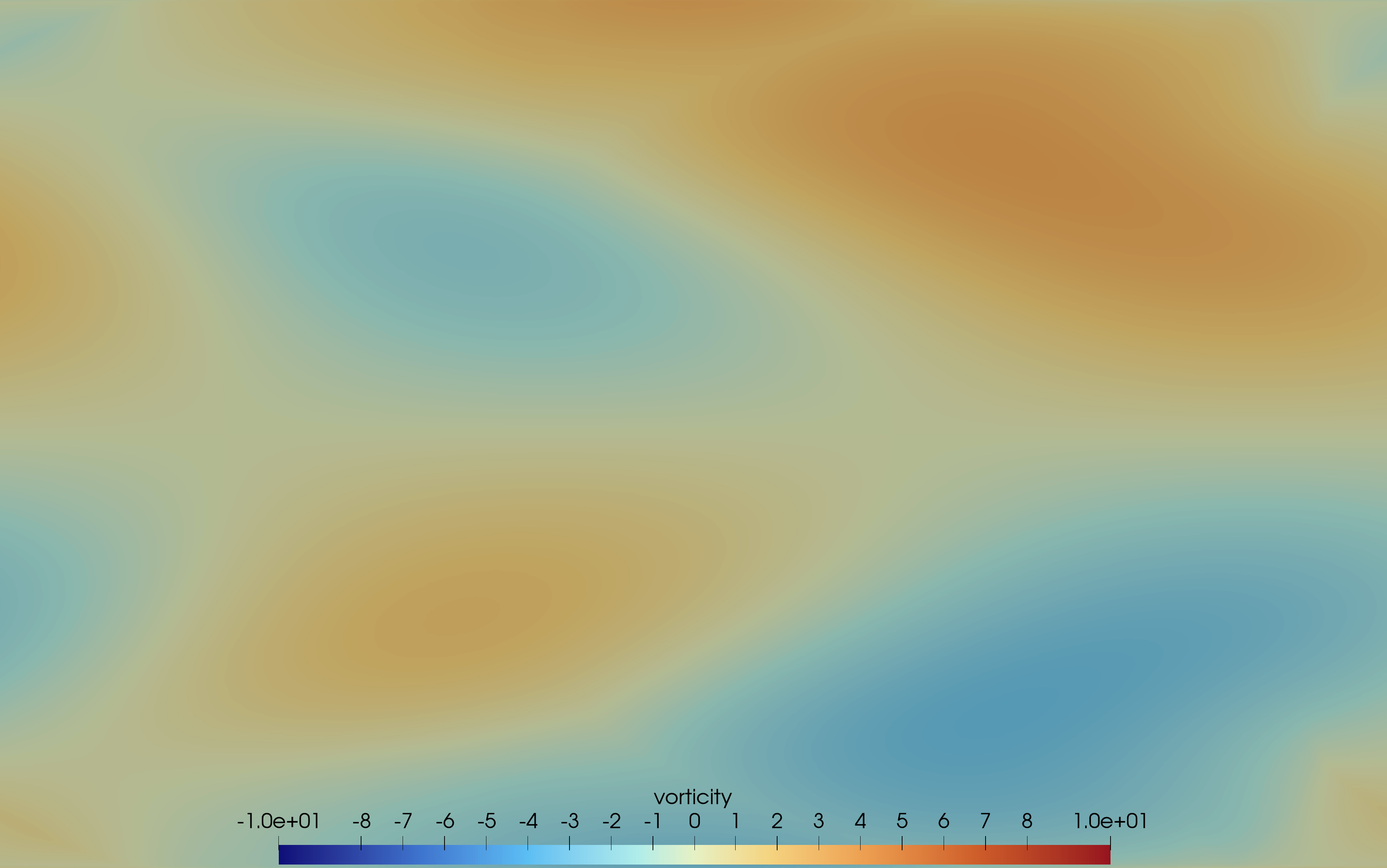}\,%
\includegraphics[width=0.49\textwidth]{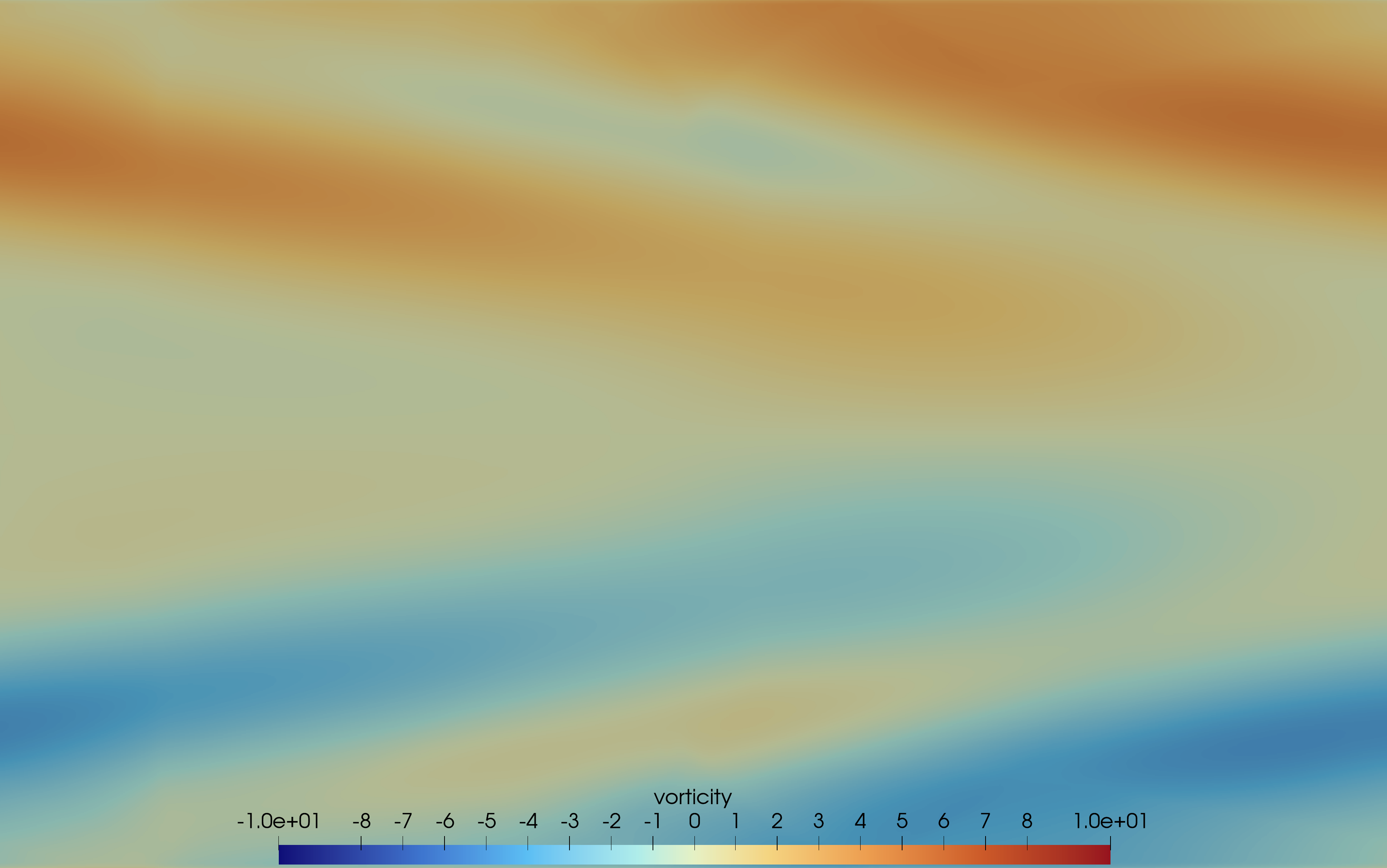}\\%
\vspace{1.5pt}%
\includegraphics[width=0.49\textwidth]{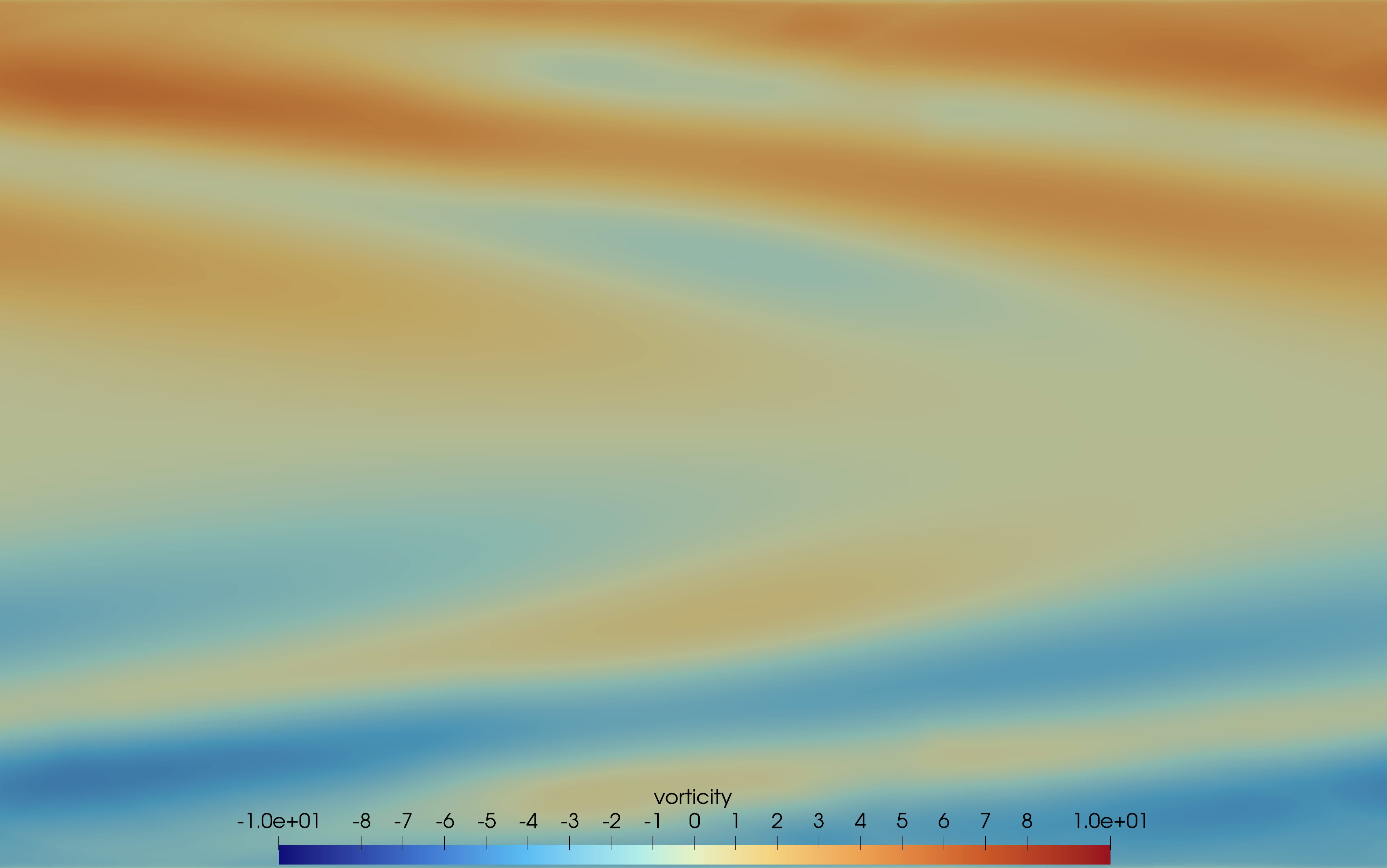}\,%
\includegraphics[width=0.49\textwidth]{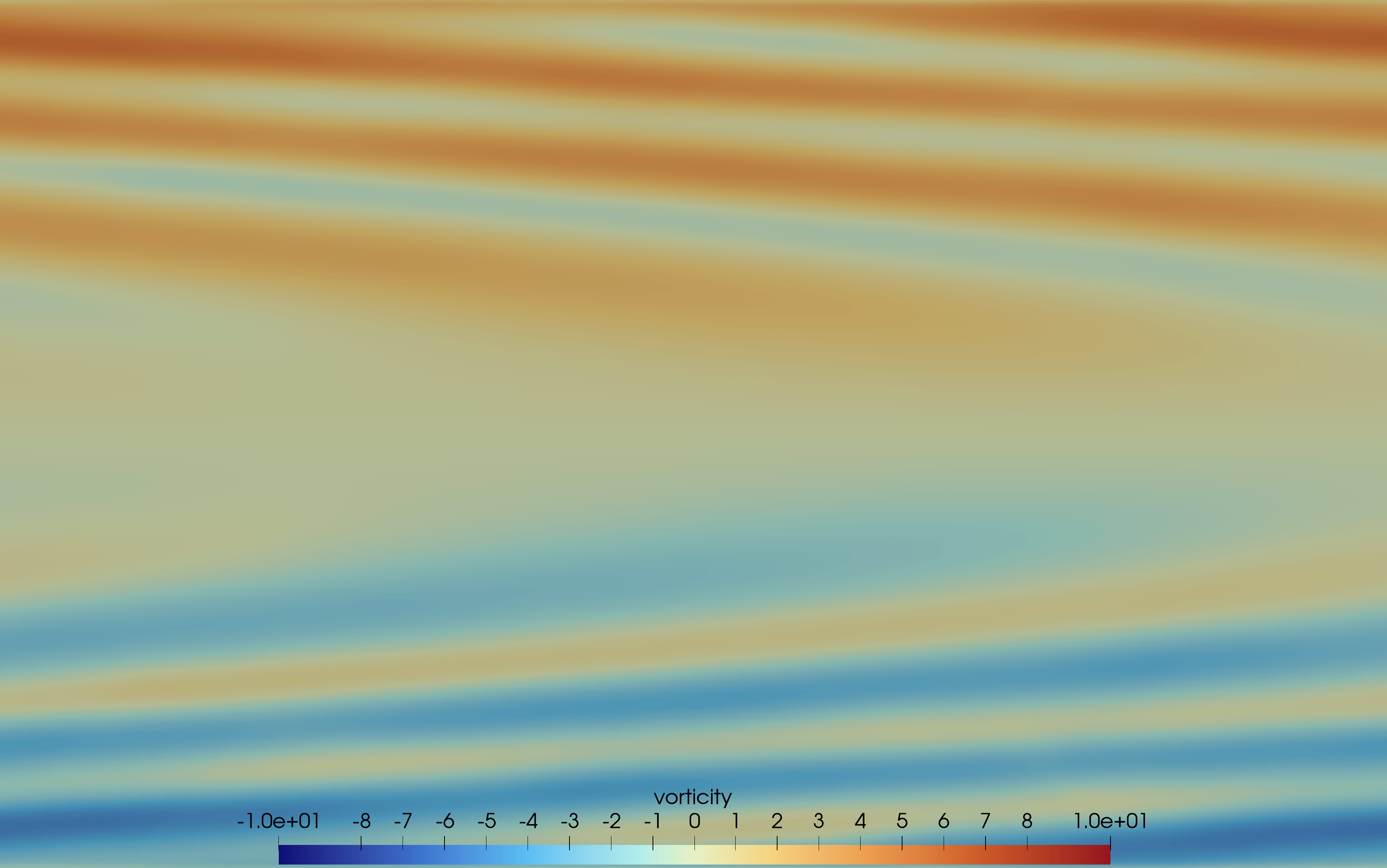}%
\caption{Vorticity fluctuations in the Poiseuille flow in the presence
  of the van der Waals effect. Upper-left -- at 0.01 seconds,
  upper-right -- at 0.02 seconds, lower-left -- at 0.03 seconds,
  lower-right -- at 0.05 seconds.}
\label{fig:vorticity_Poiseuille}
\end{figure}

Next, we simulate the same initial condition as above, however this
time the van der Waals effect is included in \eqref{eq:div_vort}. The
resulting snapshots of the deviations in $\rho$, $\chi$ and $\omega$
from their background states are shown in
Figures~\ref{fig:density_Poiseuille}, \ref{fig:divergence_Poiseuille}
and~\ref{fig:vorticity_Poiseuille}, respectively, for the elapsed
times $t=0.01$, $0.02$, $0.03$ and $0.05$ seconds. Our theory in
\cite{Abr27} predicts the existence of a direct cascade, where
large-scale fluctuations transition into small-scale fluctuations with
time, which appears to be confirmed by
Figures~\ref{fig:density_Poiseuille}--\ref{fig:vorticity_Poiseuille}.
In particular, the snapshots of the velocity divergence $\chi$ contain
wave-like patterns, which transition from large scales to small
scales with time.

\paragraph{Macroscopic laminarity of the flow}

\begin{figure}[t]%
\includegraphics[width=0.49\textwidth]{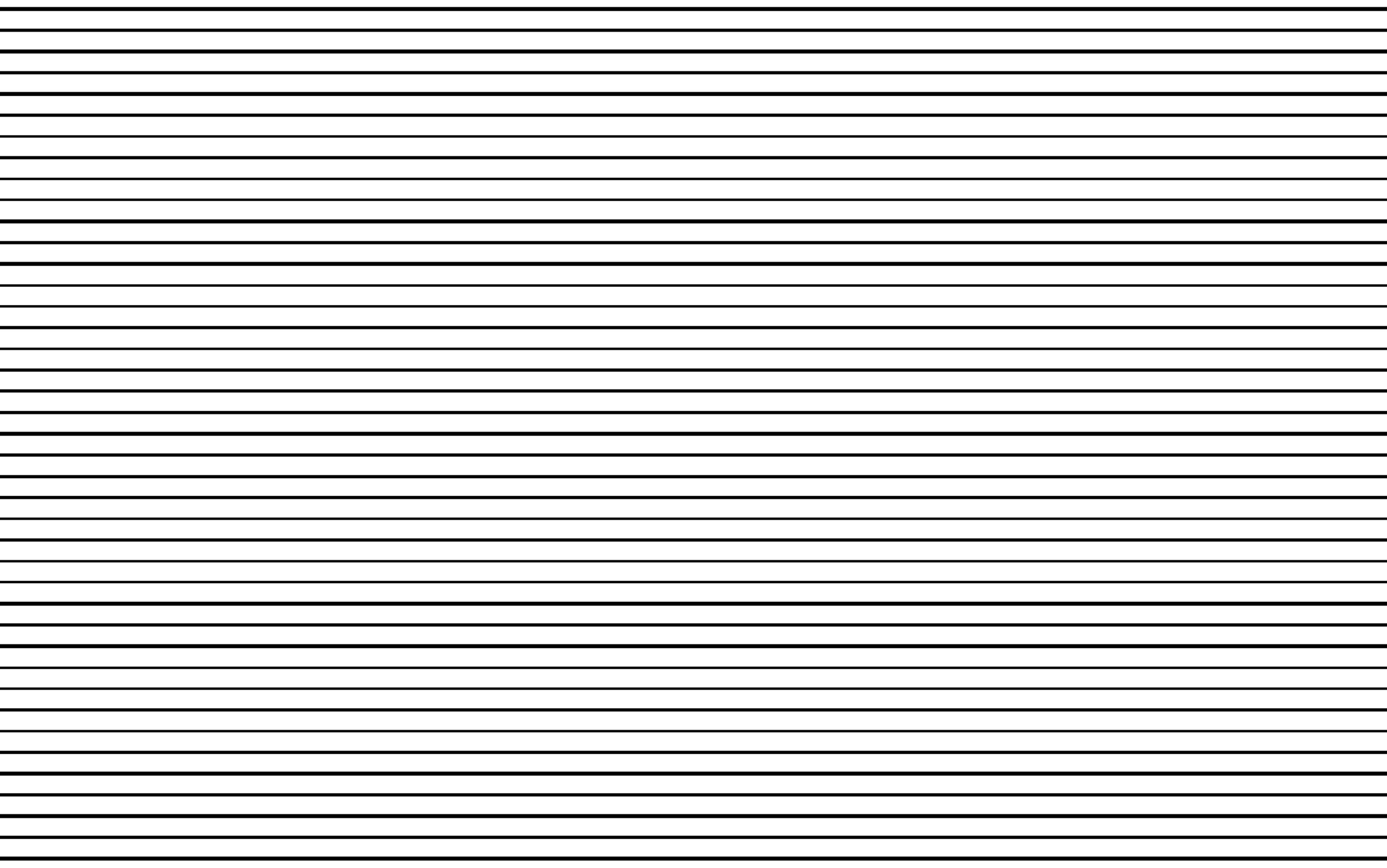}\,%
\includegraphics[width=0.49\textwidth]{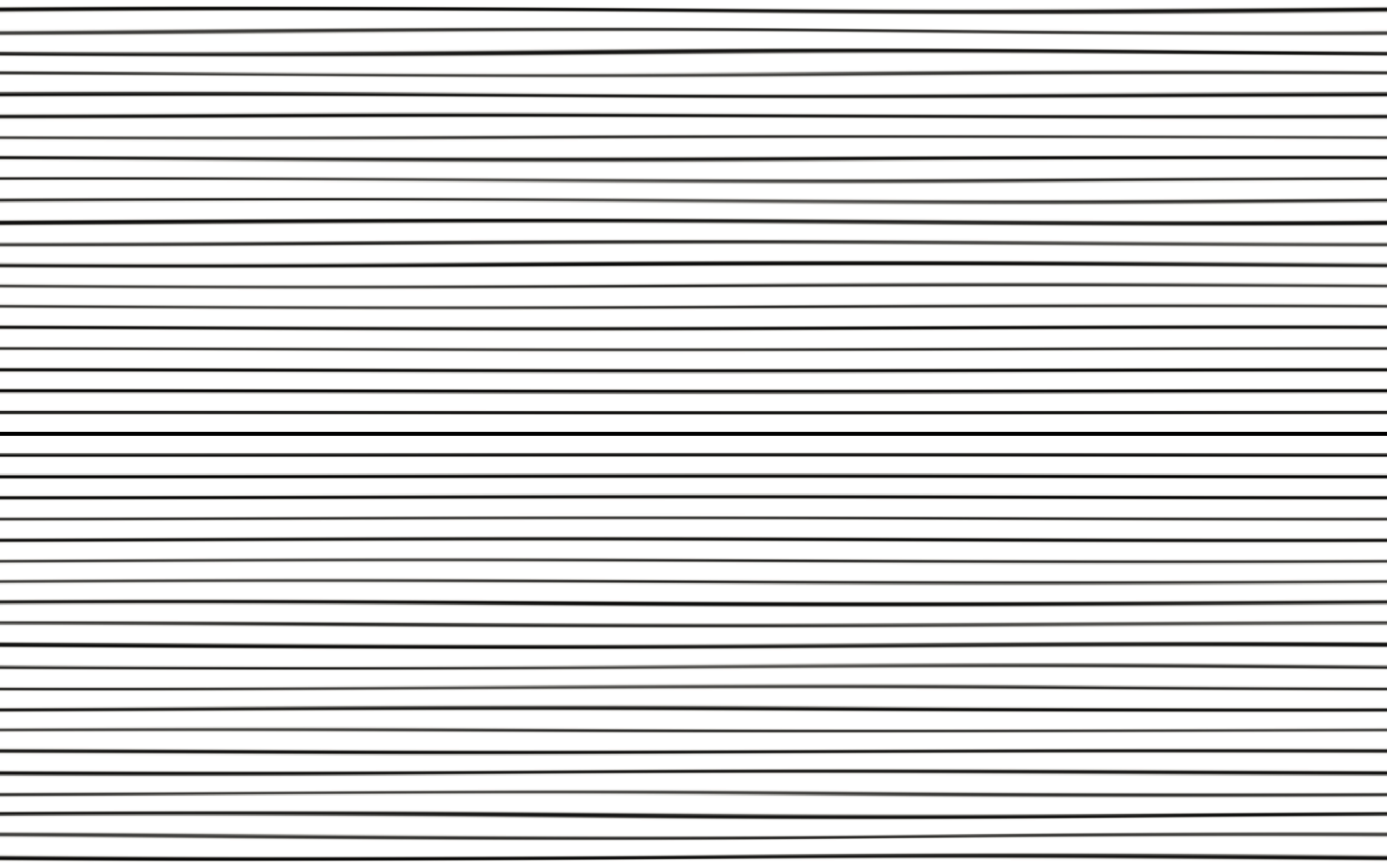}%
\caption{Tracer streaks in the Poiseuille flow in the presence of the
  van der Waals effect. Left -- starting time, right -- at 0.05
  seconds.}
\label{fig:tracer_Poiseuille}
\end{figure}

Observe, that, while the fluctuations in $\rho$ and $\omega$ remain
small relative to their background states ($\sim$ 1\% of the total
magnitude), the developed patterns look rather chaotic, especially in
the velocity divergence. Remarkably, at the same time the macroscopic
flow technically remains laminar. In the beginning of the numerical
simulation, we placed 41 equally spaced tracer streaks along the flow
(left-hand pane of Figure~\ref{fig:tracer_Poiseuille}). The thickness
of each streak is 1 mm, while the distance between the adjacent
streaks is 6 mm. At the end of the computation, that is, $t=0.05$
seconds, the streaks are essentially unchanged (right-hand pane of
Figure~\ref{fig:tracer_Poiseuille}), although minor distortion and
smudging can be discerned upon closer examination.


\subsection{Power decay of the Fourier spectra}

\begin{figure}[t]
\includegraphics[width=0.7\textwidth]{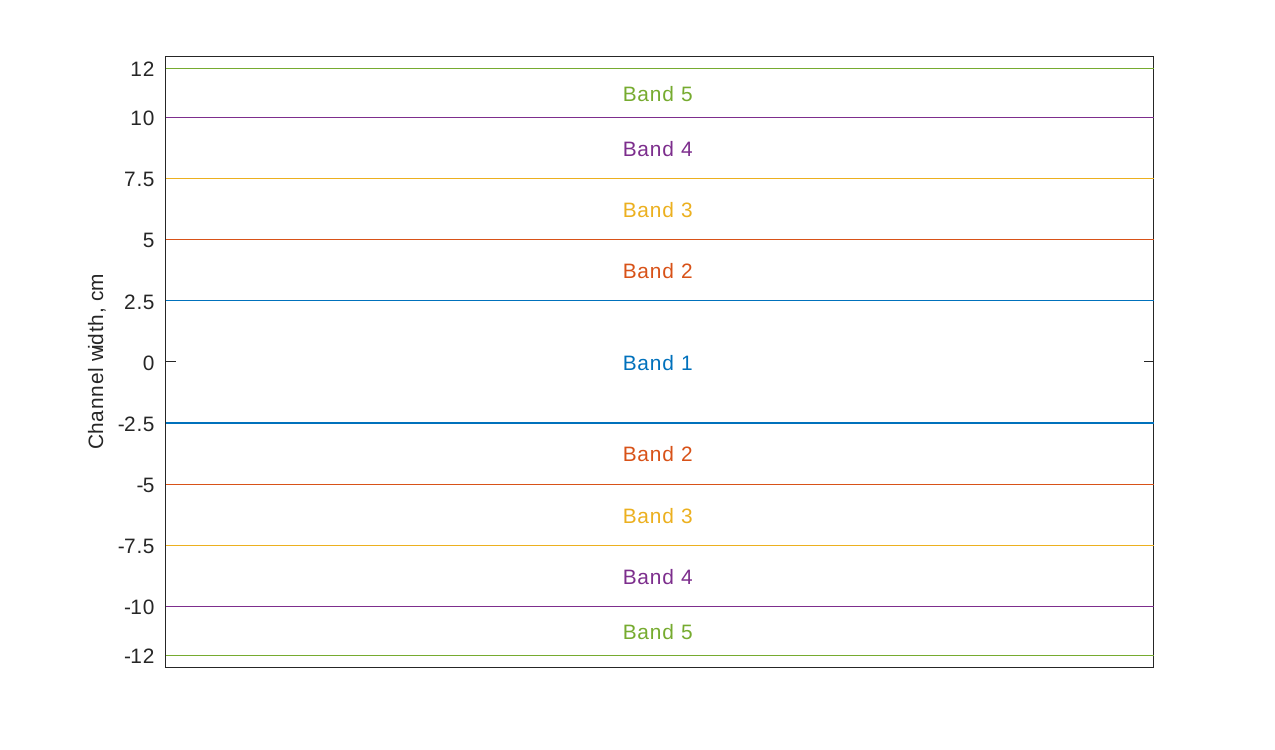}
\vspace{-2EM}
\caption{Channel bands.}
\label{fig:channel_bands}
\end{figure}
\begin{figure}[t]%
\includegraphics[width=0.5\textwidth]{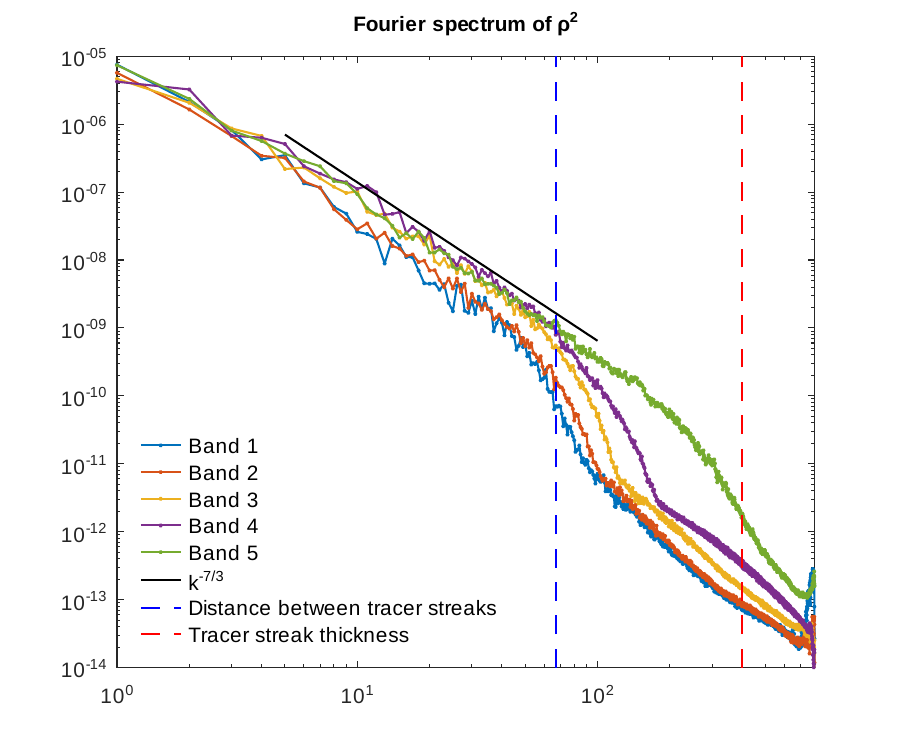}%
\includegraphics[width=0.5\textwidth]{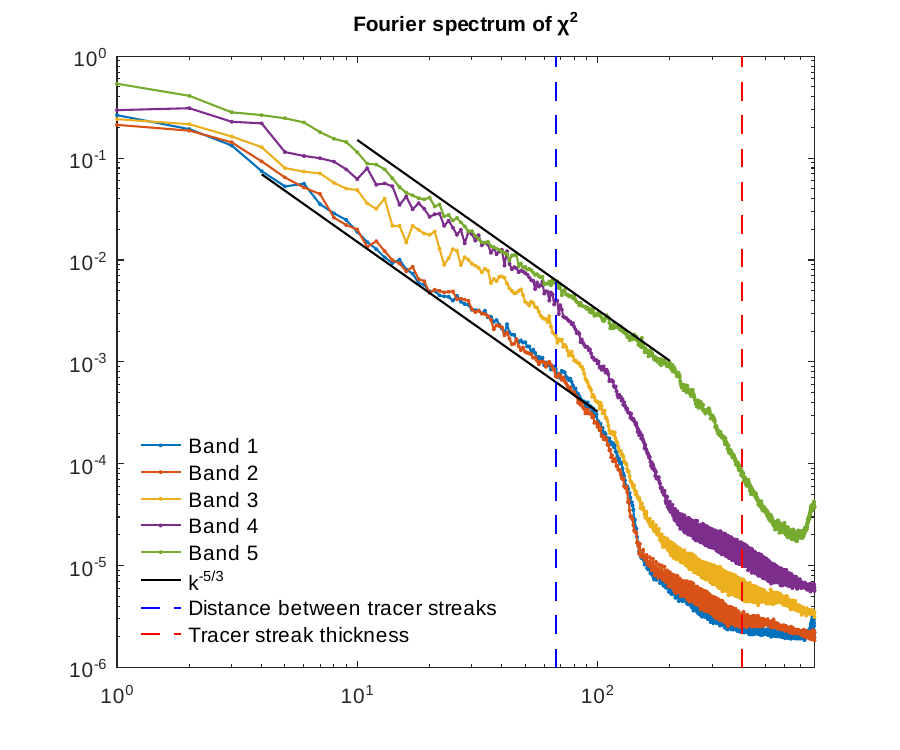}\\%
\includegraphics[width=0.5\textwidth]{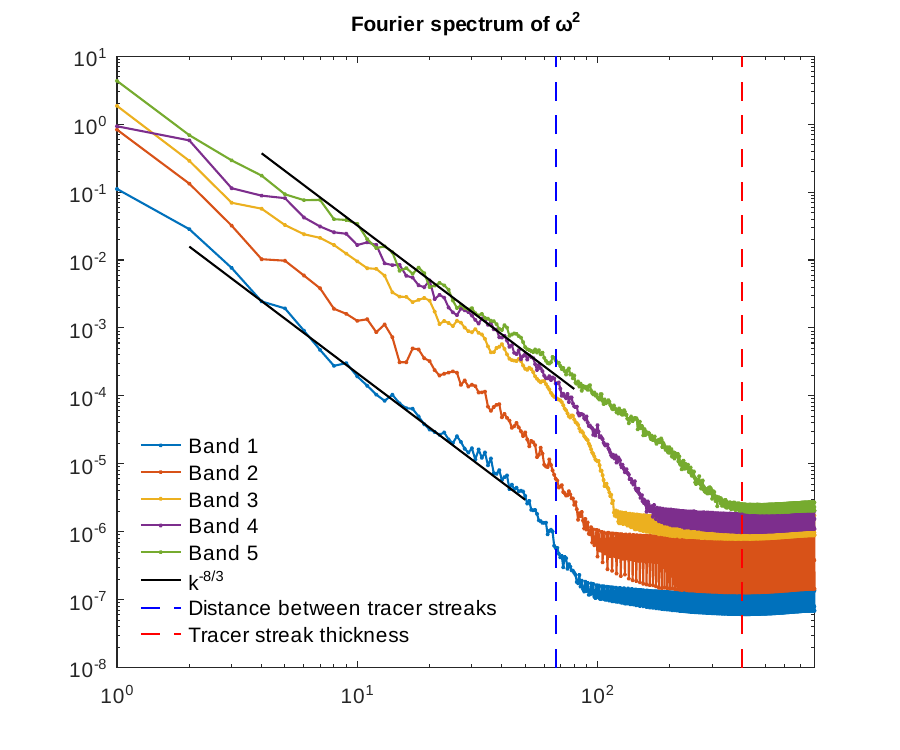}%
\vspace{-1EM}
\caption{The Fourier spectrum of $\rho^2$ (upper-left), $\chi^2$
  (upper-right), and $\omega^2$ (lower-center), Poiseuille flow.}
\label{fig:rho_div_omega_Poiseuille_spectrum}
\end{figure}

Here, we compute the time averages of the Fourier spectra of various
quantities in the same manner as we did in our preceding works
\cite{Abr22,Abr23,Abr24,Abr26,Abr25}. First, we divide the space in
the channel into the five streamwise bands, shown in
Figure~\ref{fig:channel_bands}.  Bands 1--4 are each 5 cm wide in
total, while Band 5, adjacent to the walls, is 4 cm wide in total. A
small space is left between each half of Band 5 and the adjacent wall
to avoid possible boundary effects ``polluting'' the free flow.

The time averages of the Fourier spectra are then computed as follows.
First, the quantity, of which the spectrum is computed, is spatially
averaged in the direction transversal to the flow (that is, across
each band), thus becoming the function of the streamwise coordinate
only. Then, the one-dimensional discrete Fourier transformation is
applied to the resulting spatial average. Finally, the complex phase
is discarded from the Fourier transform by computing its modulus. The
resulting modulus of the Fourier transform is time-averaged in the
window $0.03\leq t\leq 0.05$ seconds.

Thusly computed time-averages are shown in
Figure~\ref{fig:rho_div_omega_Poiseuille_spectrum} for the squares of
the fluctuations of $\rho$, $\chi$ and $\omega$. The density spectrum,
shown in the upper-left pane of
Figure~\ref{fig:rho_div_omega_Poiseuille_spectrum}, shows the $\sim
k^{-7/3}$ power decay for all five channel bands, up until the
wavenumber $k\sim 50$, which corresponds to the physical distance of
$~\sim 8$ mm. The spectrum of the velocity divergence, shown in the
upper-right pane of
Figure~\ref{fig:rho_div_omega_Poiseuille_spectrum}, also shows power
decay, however the rate of decay is different at $\sim k^{-5/3}$. The
vorticity spectrum has the most rapid decay (out of the three
aforementioned quantities) at $\sim k^{-8/3}$ rate, shown in the lower
pane of Figure~\ref{fig:rho_div_omega_Poiseuille_spectrum}.

Additionally, we compute the Fourier wavenumbers associated with the
spatial scales of the distance between the tracer streaks and the
thickness of a tracer streak, respectively, via dividing the length of
the channel (that is, 40 cm) by either the distance between the
streaks (6 mm) or the thickness of a streak (1 mm). The corresponding
wavenumbers are $k\approx 67$ and $k=400$, respectively, which are
shown as dashed vertical lines in all panes of
Figure~\ref{fig:rho_div_omega_Poiseuille_spectrum}, as well as all
subsequent figures with the Fourier spectra. As we can see, the power
decay of all Fourier spectra in
Figure~\ref{fig:rho_div_omega_Poiseuille_spectrum} occurs on larger
spatial scales than the distance between the tracer streaks. The
latter means that the Fourier spectrum with a power decay may exist in
a flow which is technically laminar at the corresponding spatial
scales.

\begin{figure}%
\includegraphics[width=0.5\textwidth]{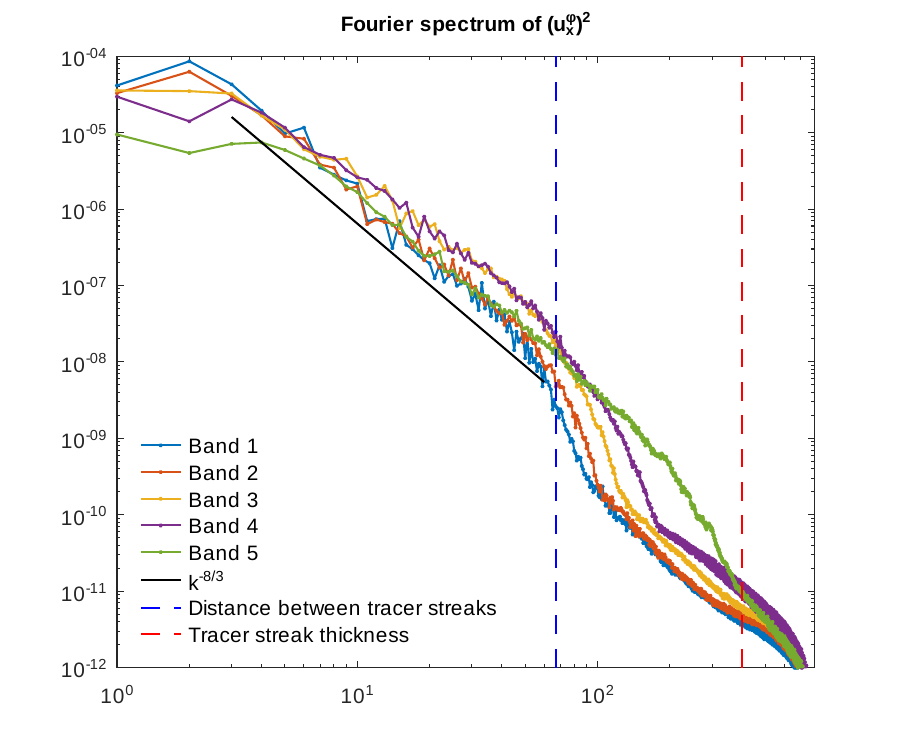}%
\includegraphics[width=0.5\textwidth]{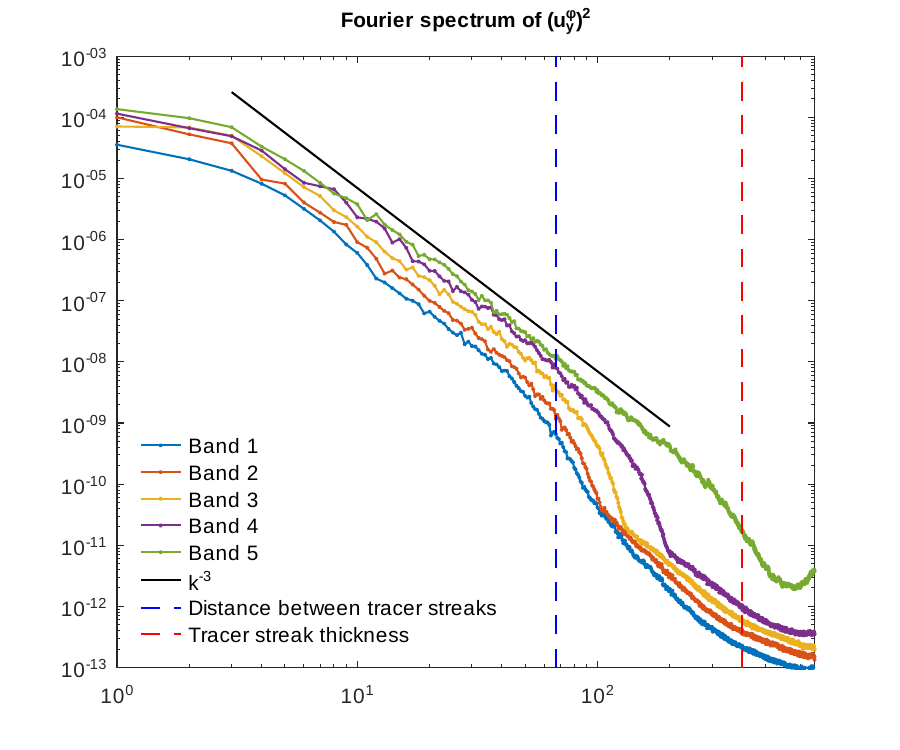}\\%
\includegraphics[width=0.5\textwidth]{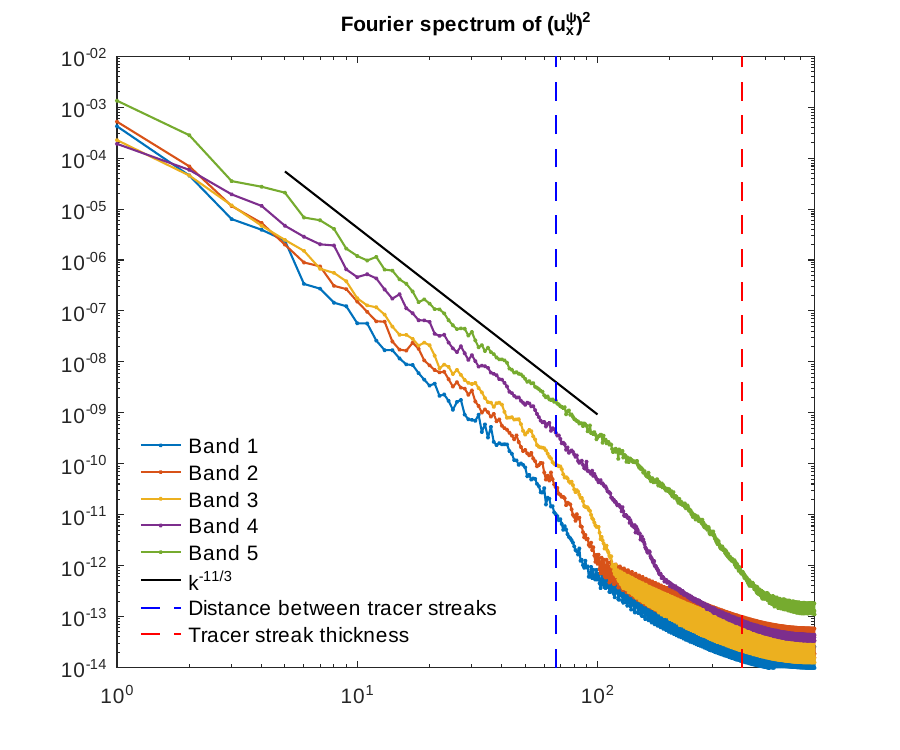}%
\includegraphics[width=0.5\textwidth]{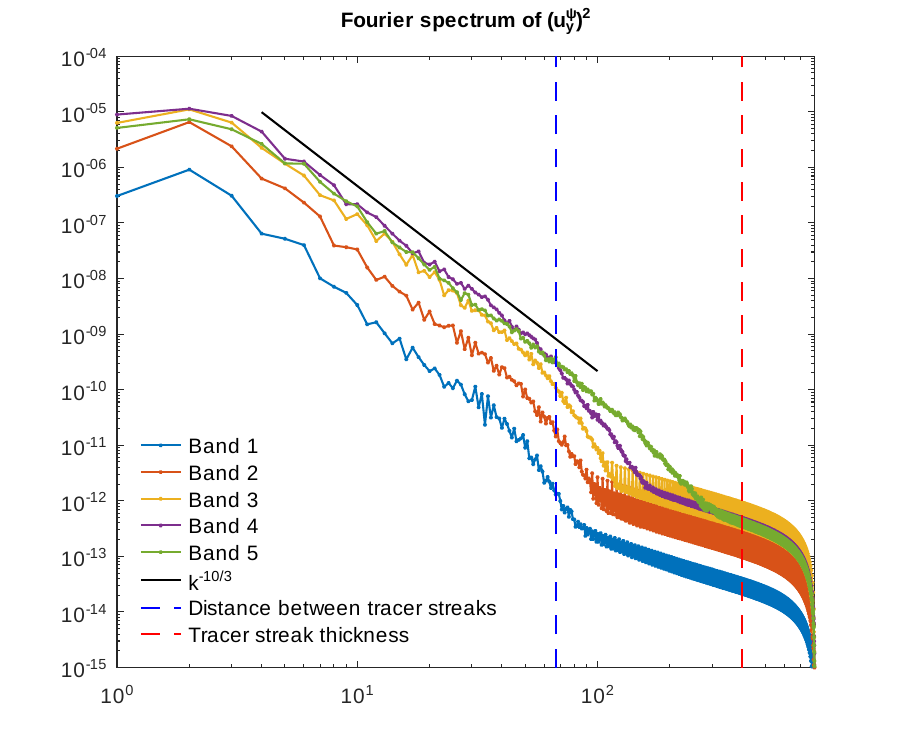}\\%
\vspace{-1EM}
\caption{The Fourier spectrum of $(u^\varphi_x)^2$ (upper-left),
  $(u^\varphi_y)^2$ (upper-right), $(u^\psi_x)^2$ (lower-left), and
  $(u^\psi_y)^2$ (lower-right), Poiseuille flow.}
\label{fig:U_Poiseuille_spectrum}
\end{figure}

\begin{figure}[t]%
\includegraphics[width=0.49\textwidth]{density_t0}\,%
\includegraphics[width=0.49\textwidth]{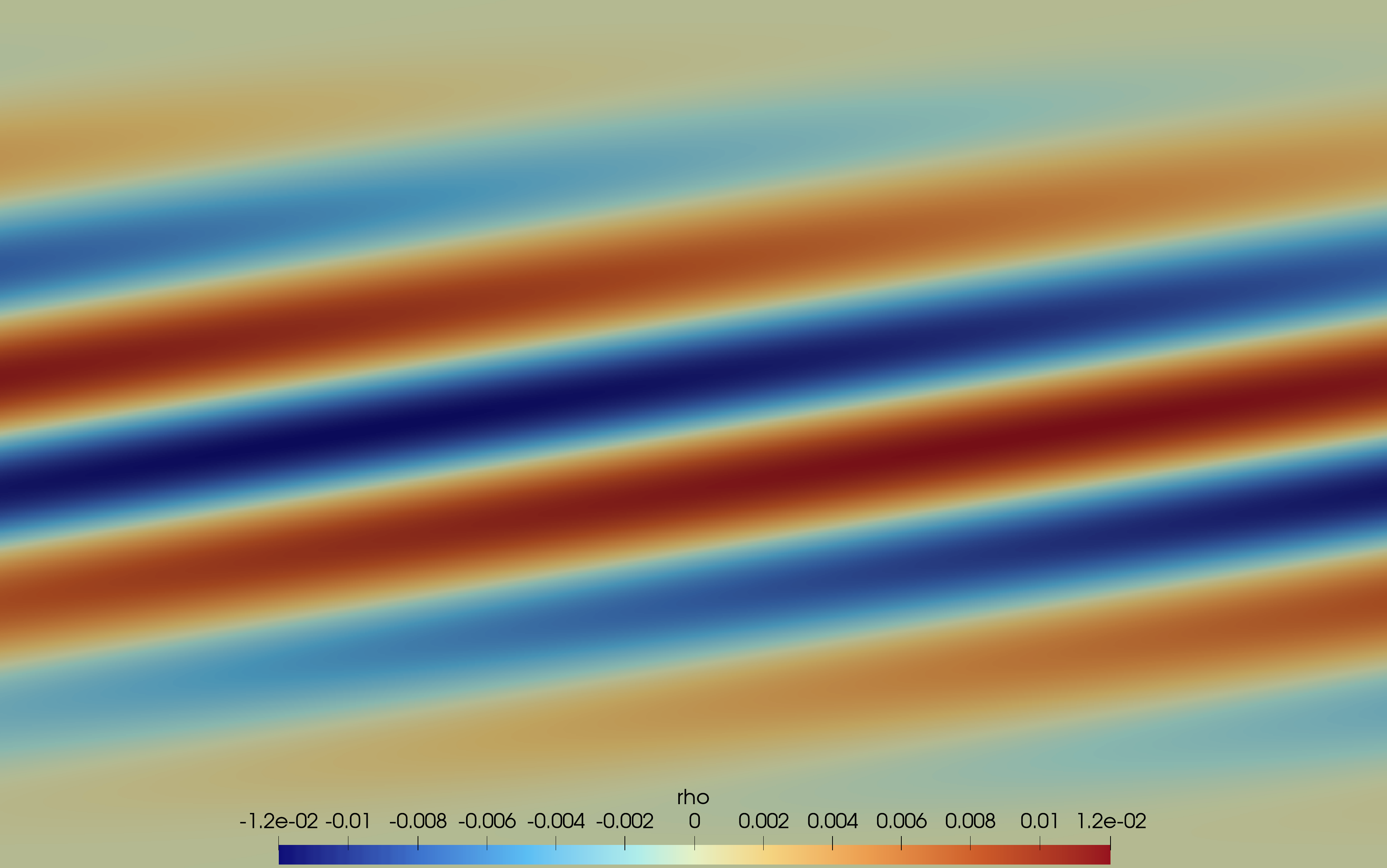}%
\caption{Density fluctuations of the Couette flow in the absence of
  the van der Waals effect. Left -- starting time, right -- at 0.05
  seconds.}
\label{fig:density_Couette_novdW}
\end{figure}
\begin{figure}[t]%
\includegraphics[width=0.49\textwidth]{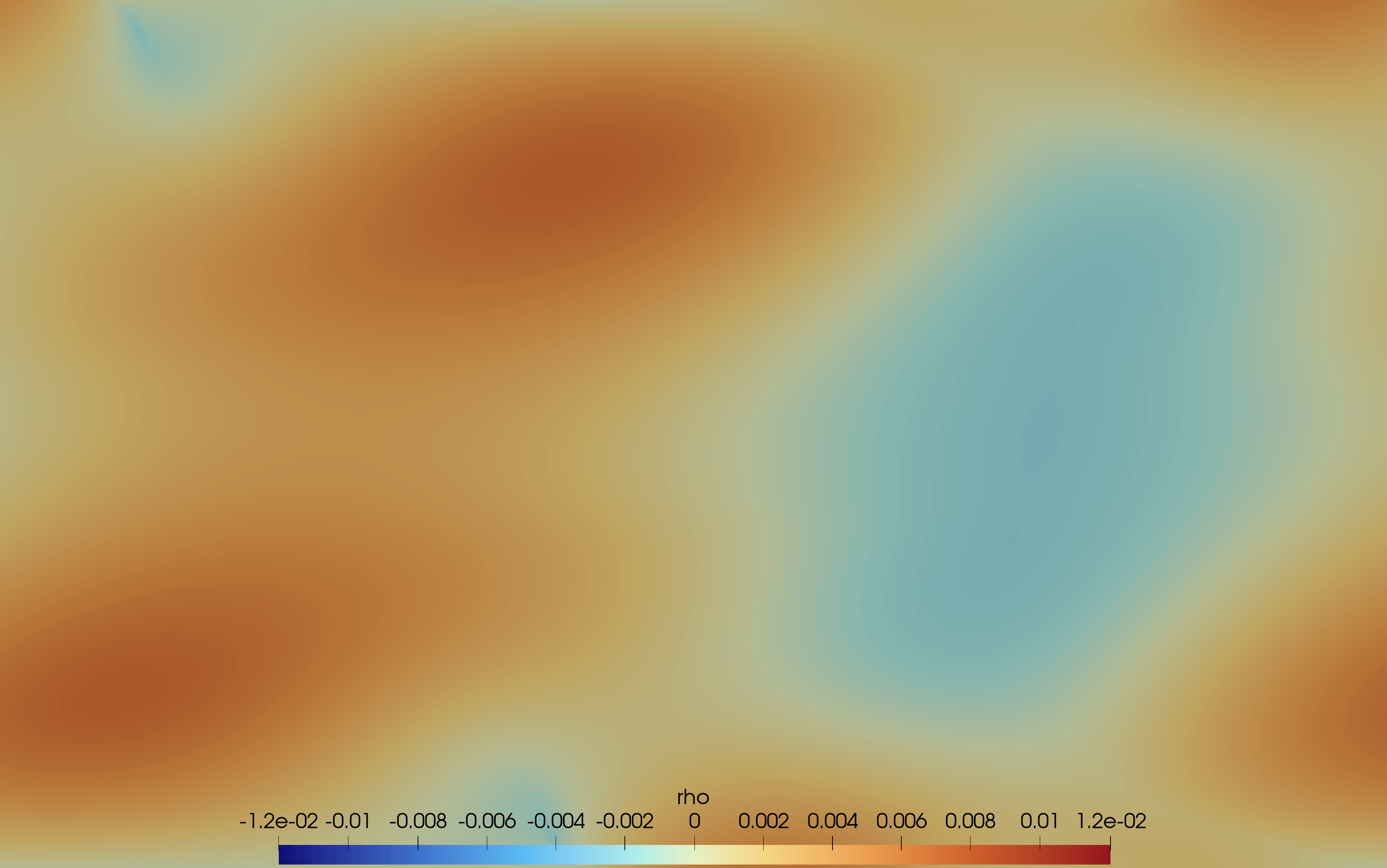}\,%
\includegraphics[width=0.49\textwidth]{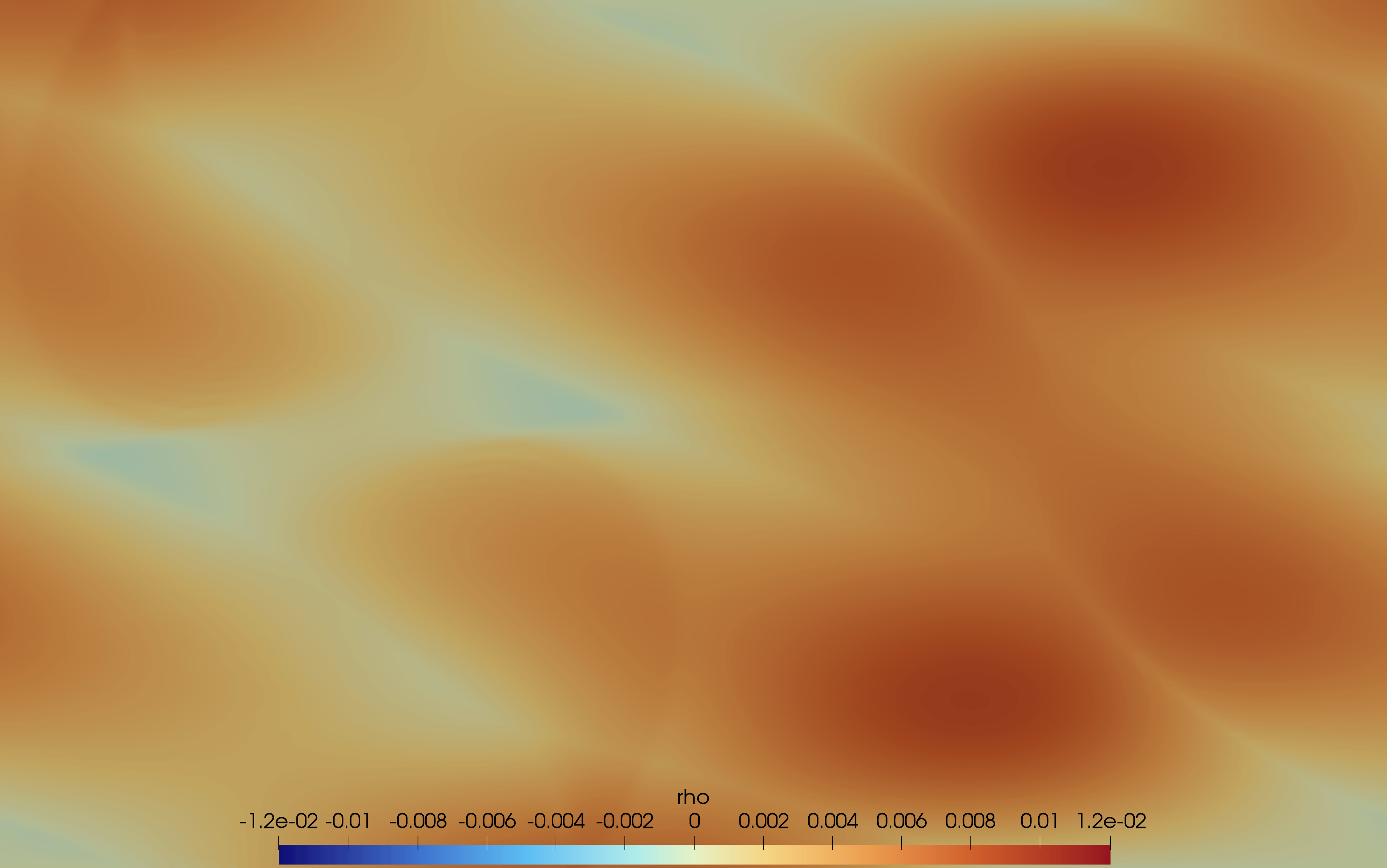}\\%
\vspace{1.5pt}%
\includegraphics[width=0.49\textwidth]{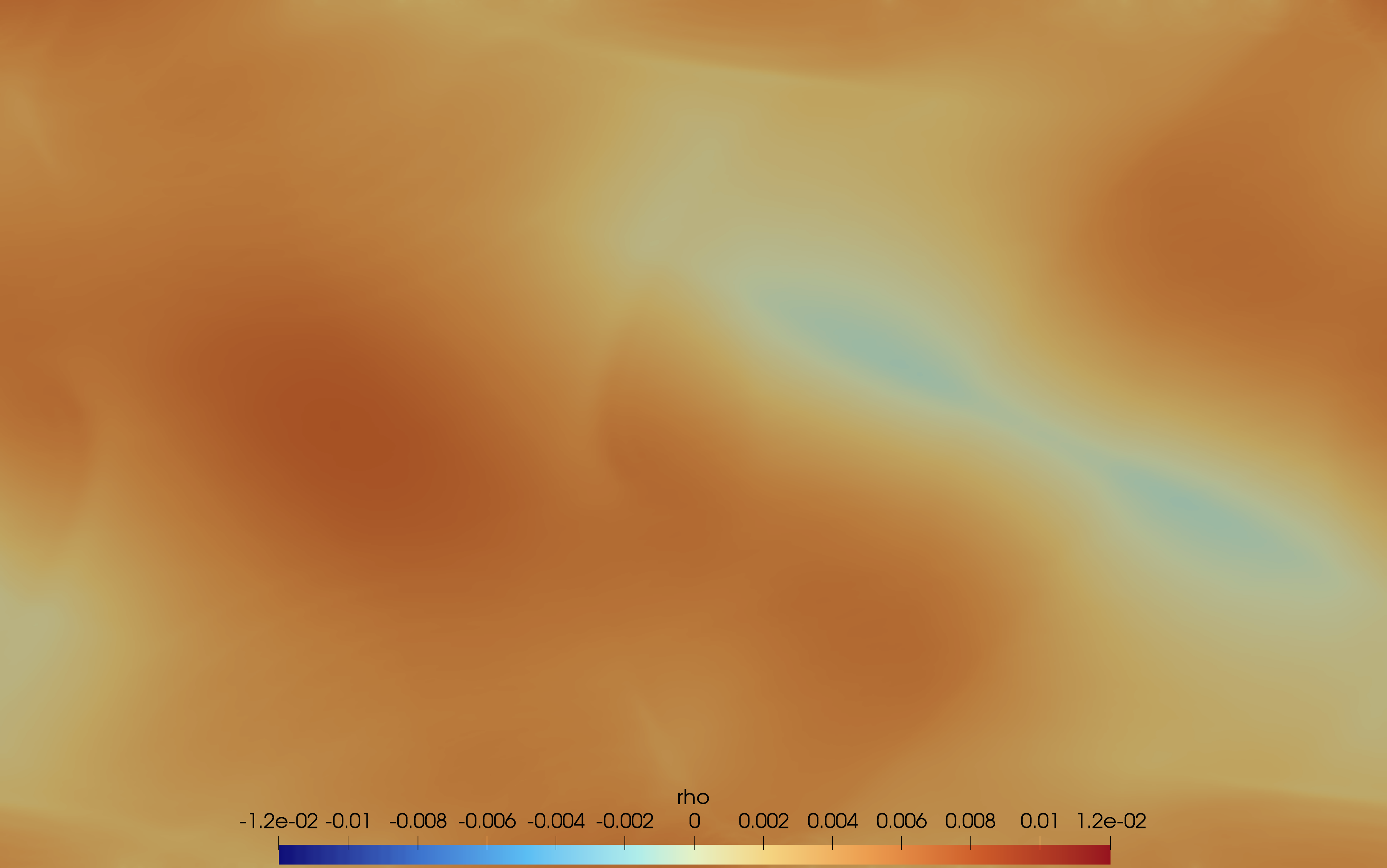}\,%
\includegraphics[width=0.49\textwidth]{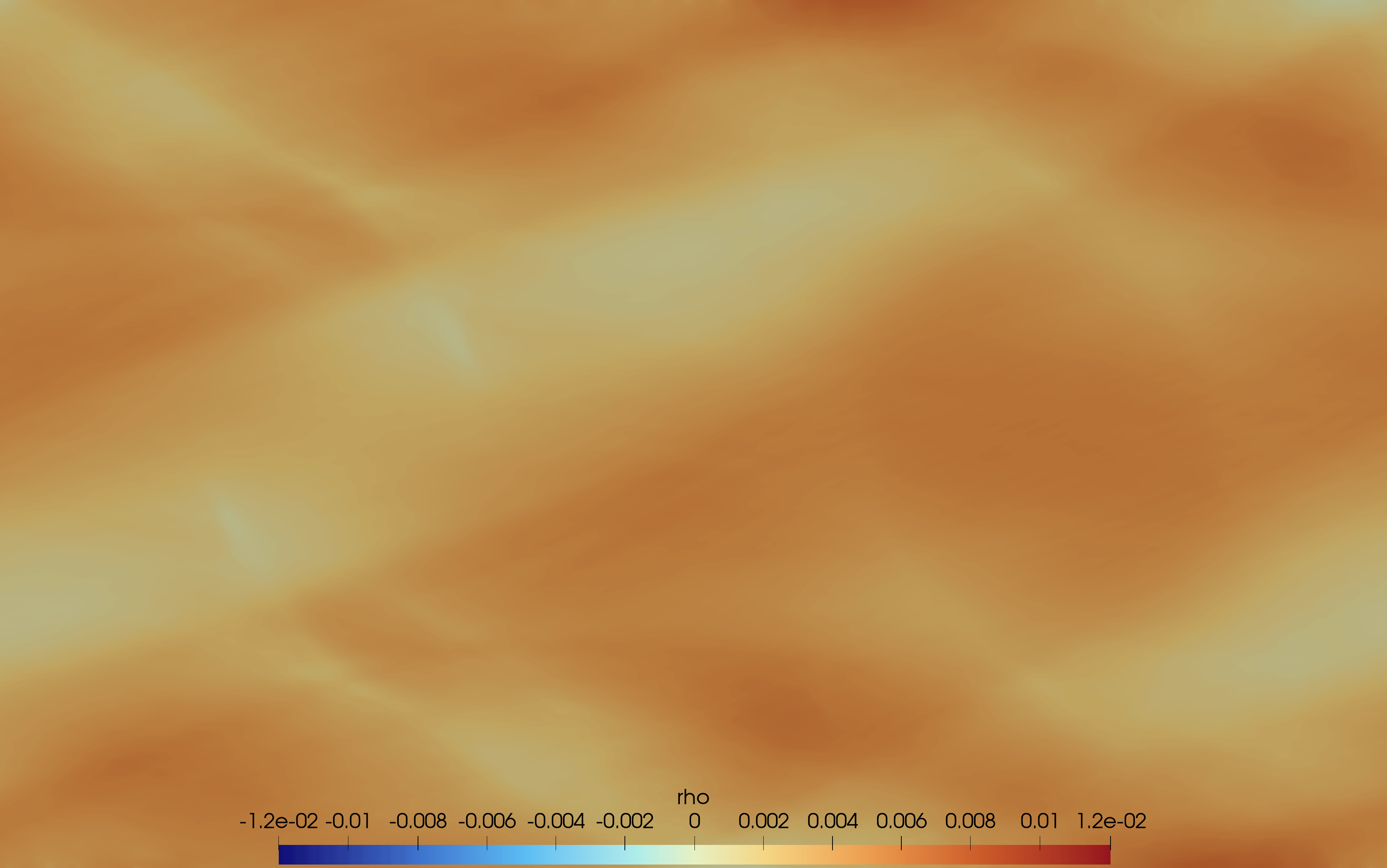}%
\caption{Density fluctuations in the Couette flow in the presence of
  the van der Waals effect. Upper-left -- at 0.01 seconds, upper-right
  -- at 0.02 seconds, lower-left -- at 0.03 seconds, lower-right -- at
  0.05 seconds.}
\label{fig:density_Couette}
\end{figure}

In Figure~\ref{fig:U_Poiseuille_spectrum} we show the time-averages of
the Fourier spectra for the squares of the fluctuations of the
streamwise and transversal components of velocities $\BV u^\varphi$
and $\BV u^\psi$, defined in \eqref{eq:Helmholtz}, which represent
different parts of the total kinetic energy of the flow. Just as in
Figure~\ref{fig:rho_div_omega_Poiseuille_spectrum}, here we can see
that the power decay of all displayed Fourier spectra occurs on the
spatial scales larger than the distance between the tracer streaks,
that is, where the flow is laminar. The spectrum of the streamwise
component $u^\varphi_x$, which corresponds to the potential part of
the streamwise kinetic energy of the flow, is shown in the upper-left
pane of Figure~\ref{fig:U_Poiseuille_spectrum}, and decays at the rate
of $\sim k^{-8/3}$ in all channel bands. The spectrum of the
transversal component $u^\varphi_y$, which corresponds to the
potential part of the transversal kinetic energy of the flow, is shown
in the upper-right pane of Figure~\ref{fig:U_Poiseuille_spectrum}, and
decays at the rate of $\sim k^{-3}$ in all channel bands. The spectrum
of the streamwise component $u^\psi_x$, which corresponds to the
stream function part of the streamwise kinetic energy of the flow, is
shown in the lower-left pane of
Figure~\ref{fig:U_Poiseuille_spectrum}, and decays at the rate of
$\sim k^{-11/3}$ in all channel bands. Finally, the spectrum of the
transversal component $u^\psi_y$, which corresponds to the stream
function part of the transversal kinetic energy of the flow, is shown
in the lower-right pane of Figure~\ref{fig:U_Poiseuille_spectrum}, and
decays at the rate of $\sim k^{-10/3}$ in all channel bands. It is
remarkable that the rates of decay of different parts of the kinetic
energy of the flow in Figure~\ref{fig:U_Poiseuille_spectrum} are all
distinctly different.

\section{Numerical simulation of the Couette flow}
\label{sec:Couette}

\begin{figure}[t]%
\includegraphics[width=0.49\textwidth]{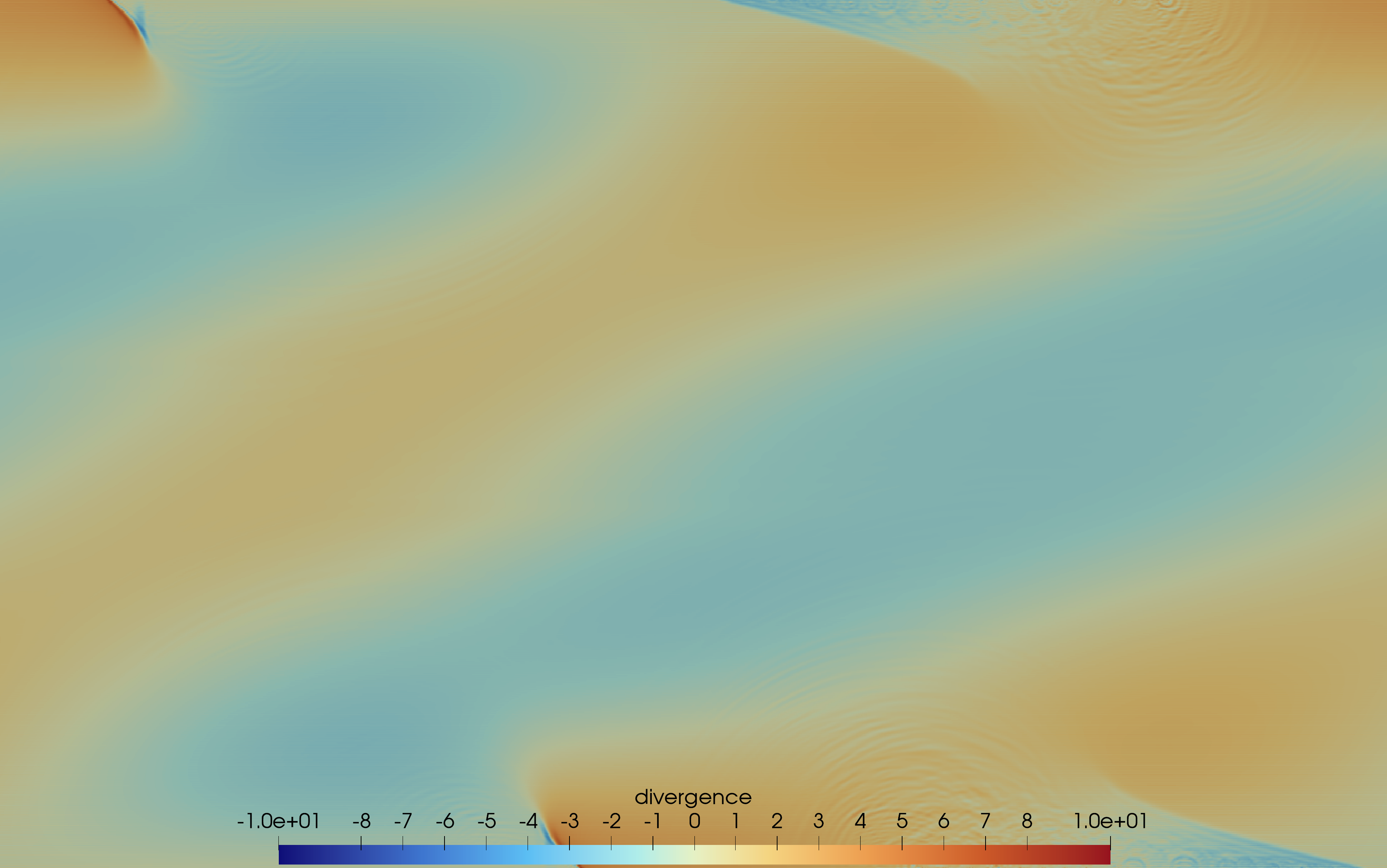}\,%
\includegraphics[width=0.49\textwidth]{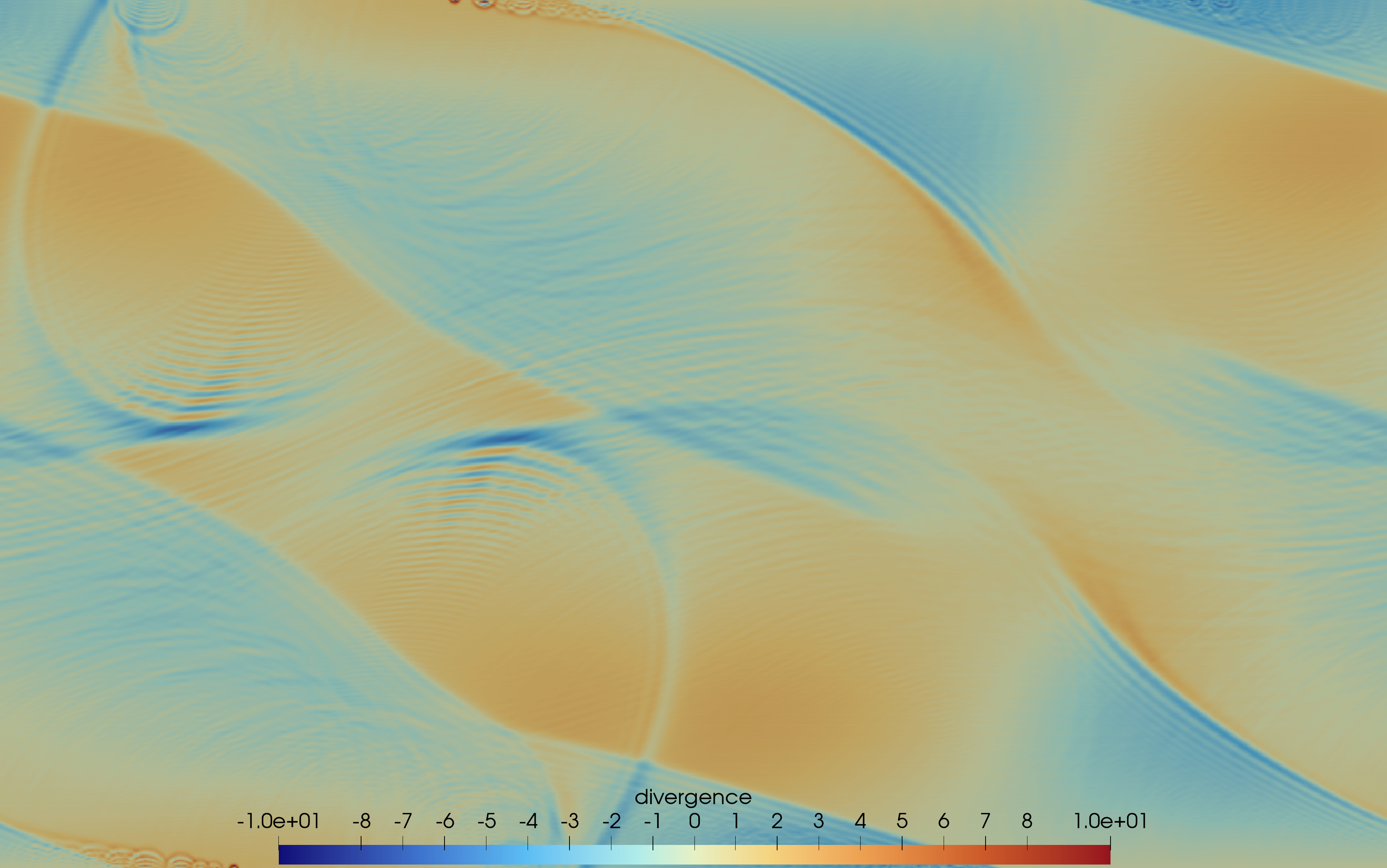}\\%
\vspace{1.5pt}%
\includegraphics[width=0.49\textwidth]{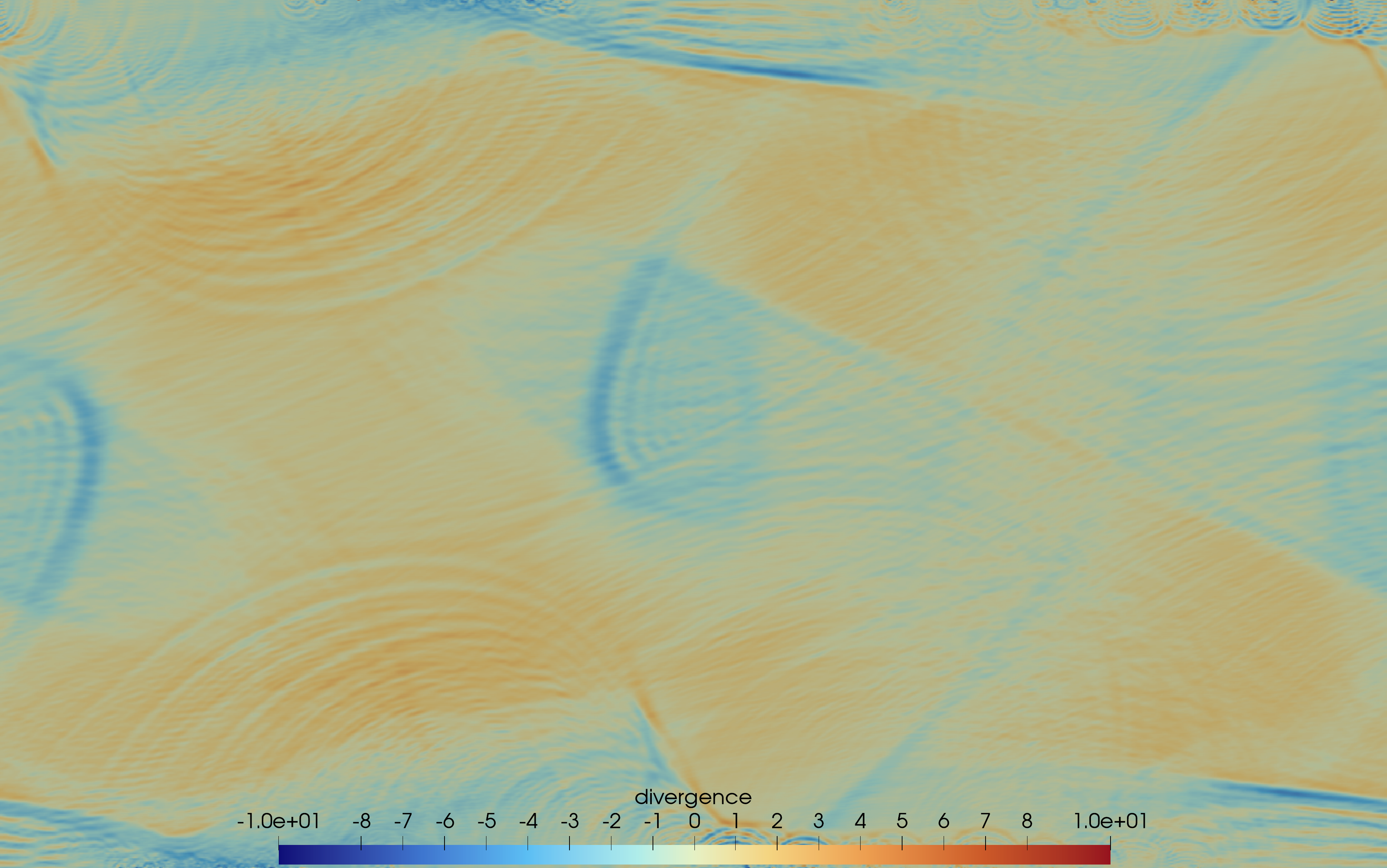}\,%
\includegraphics[width=0.49\textwidth]{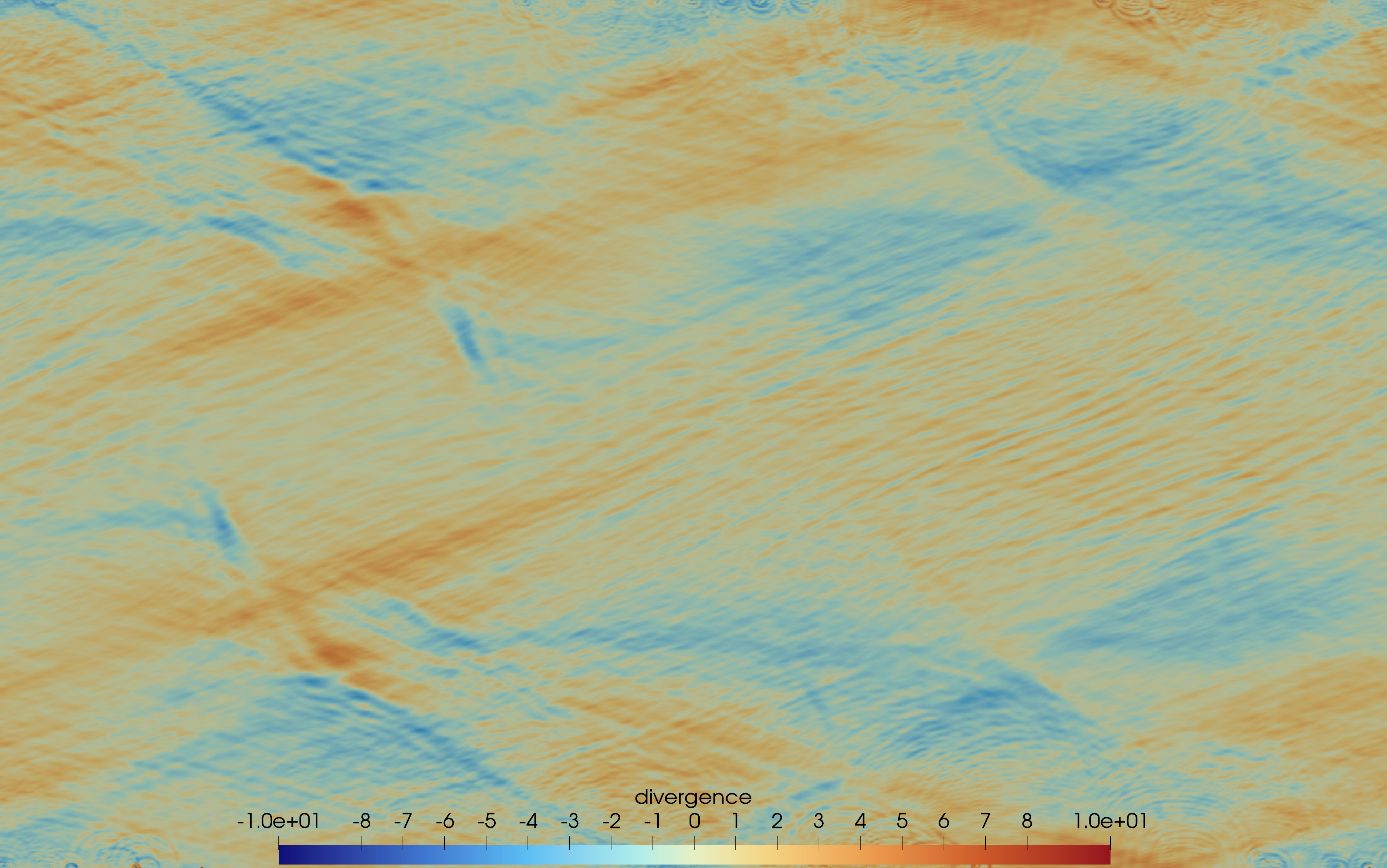}%
\caption{Divergence fluctuations in the Couette flow in the presence
  of the van der Waals effect. Upper-left -- at 0.01 seconds,
  upper-right -- at 0.02 seconds, lower-left -- at 0.03 seconds,
  lower-right -- at 0.05 seconds.}
\label{fig:divergence_Couette}
\end{figure}

The Couette flow is another stationary state of \eqref{eq:div_vort},
represented by a linear velocity profile in a straight channel:
\begin{equation}
\label{eq:u_Couette}
\BV u_0=\frac{U_0y}W\begin{pmatrix} 1 \\ 0\end{pmatrix},
\end{equation}
where $W=25$ cm is the width of the channel, and we choose $U_0=30$
m/s. As we can see, the velocity is zero in the center of the channel,
and reaches $\pm 15$ m/s at the walls of the channel, so that the
maximum velocity difference is 30 m/s just as it was for the
Poiseuille flow above in Section~\ref{sec:Poiseuille}. In the
variables of \eqref{eq:div_vort}, the Couette state becomes
\begin{equation}
\label{eq:Couette_rho_chi_omega}
\rho=\rho_0,\qquad\chi=0,\qquad\omega=-\frac{U_0}W,
\end{equation}
that is, $\rho$ is set to its background value of $1.204$ kg/m$^3$ as
it was for the Poiseuille flow, while $\omega=-120$ m/s$^2$ is set to
a constant throughout the channel.

In the initial condition of our simulation, the Couette state is
perturbed by the same density deviation \eqref{eq:density_deviation}
as was the Poiseuille flow above in Section~\ref{sec:Poiseuille}. The
initial density deviation is shown in the left-hand pane of
Figure~\ref{fig:density_Couette_novdW}. The initial values of $\chi$
and $\omega$ are set to their respective Couette background values in
\eqref{eq:Couette_rho_chi_omega}.

\subsection{The flow in the absence of the van der Waals effect}

As with the Poiseuille flow above in Section~\ref{sec:Poiseuille},
here we first conduct the numerical simulation in the absence of the
van der Waals effect, which amounts to setting $p_0=0$ in the
divergence equation of~\eqref{eq:div_vort}. As we explained before,
this change effectively decouples the equations for $\chi$ and
$\omega$ from that for $\rho$, which leads to $\chi$ and $\omega$
remaining at their stationary values, while $\rho$ being passively
transported by the Couette profile. The resulting density state at the
final time $t=0.05$ seconds is shown in the right-hand pane of
Figure~\ref{fig:density_Couette_novdW}, and indeed corresponds to the
passive Couette flow transport.

\subsection{The flow in the presence of the van der Waals effect}

\begin{figure}[t]%
\includegraphics[width=0.49\textwidth]{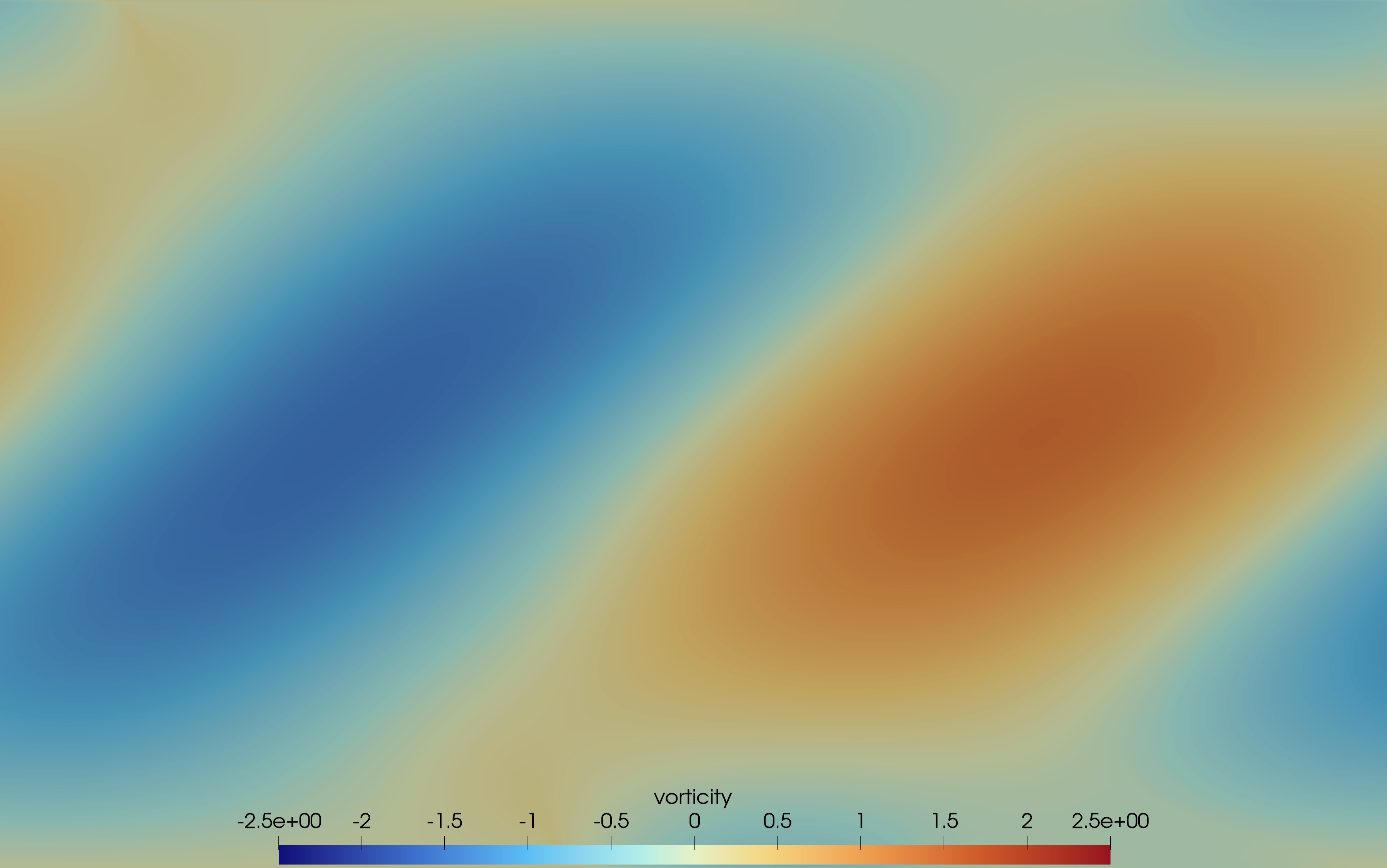}\,%
\includegraphics[width=0.49\textwidth]{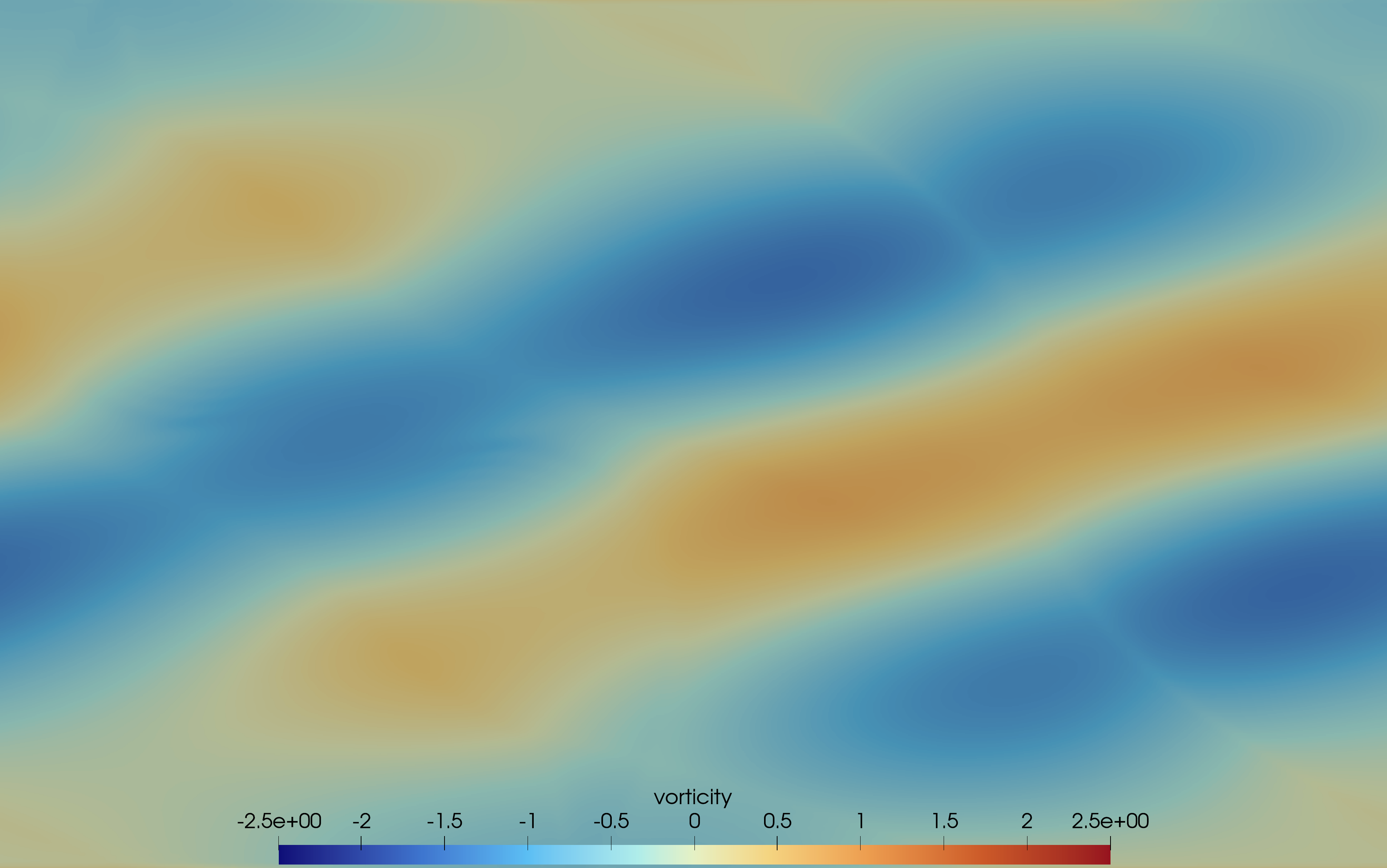}\\%
\vspace{1.5pt}%
\includegraphics[width=0.49\textwidth]{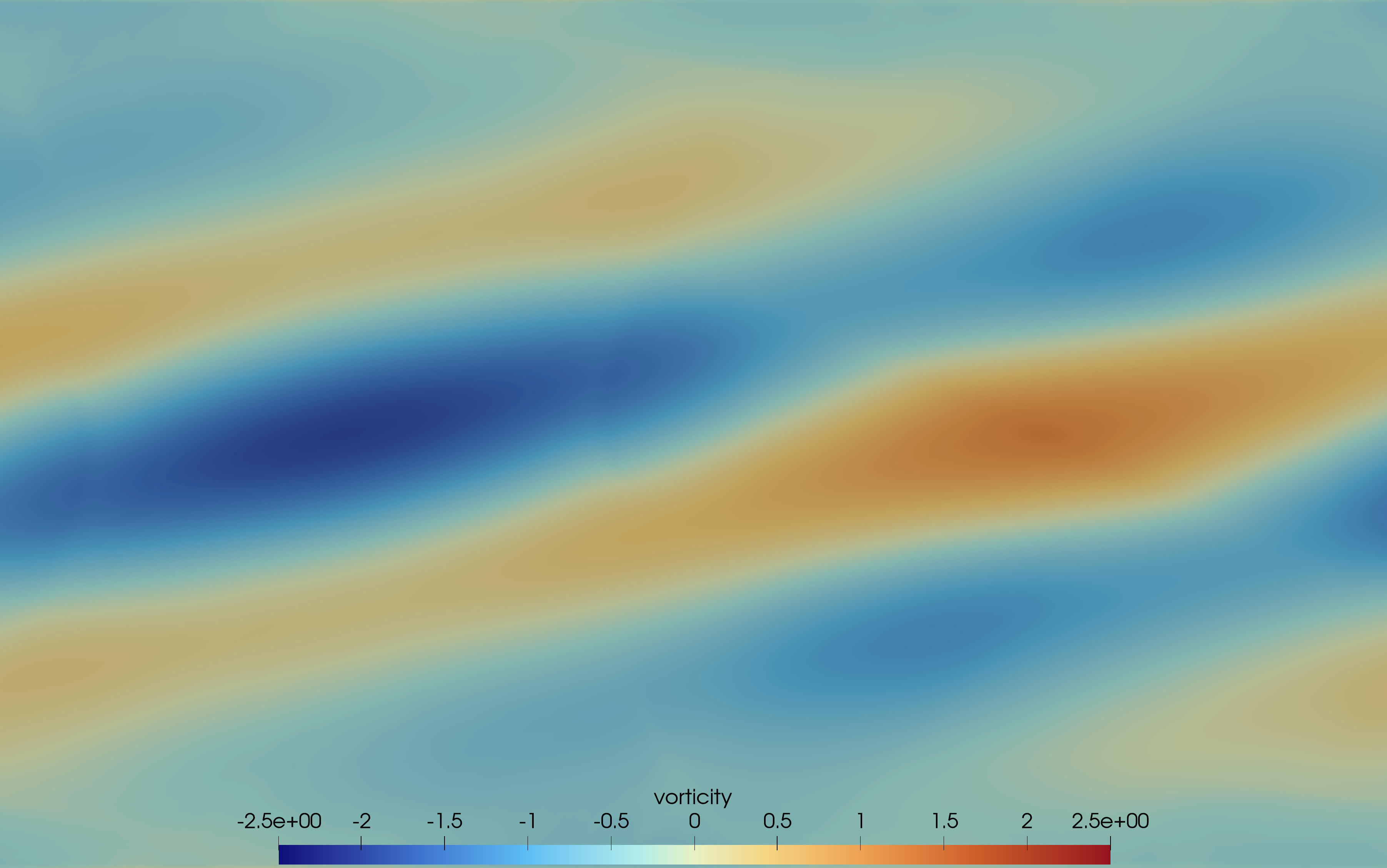}\,%
\includegraphics[width=0.49\textwidth]{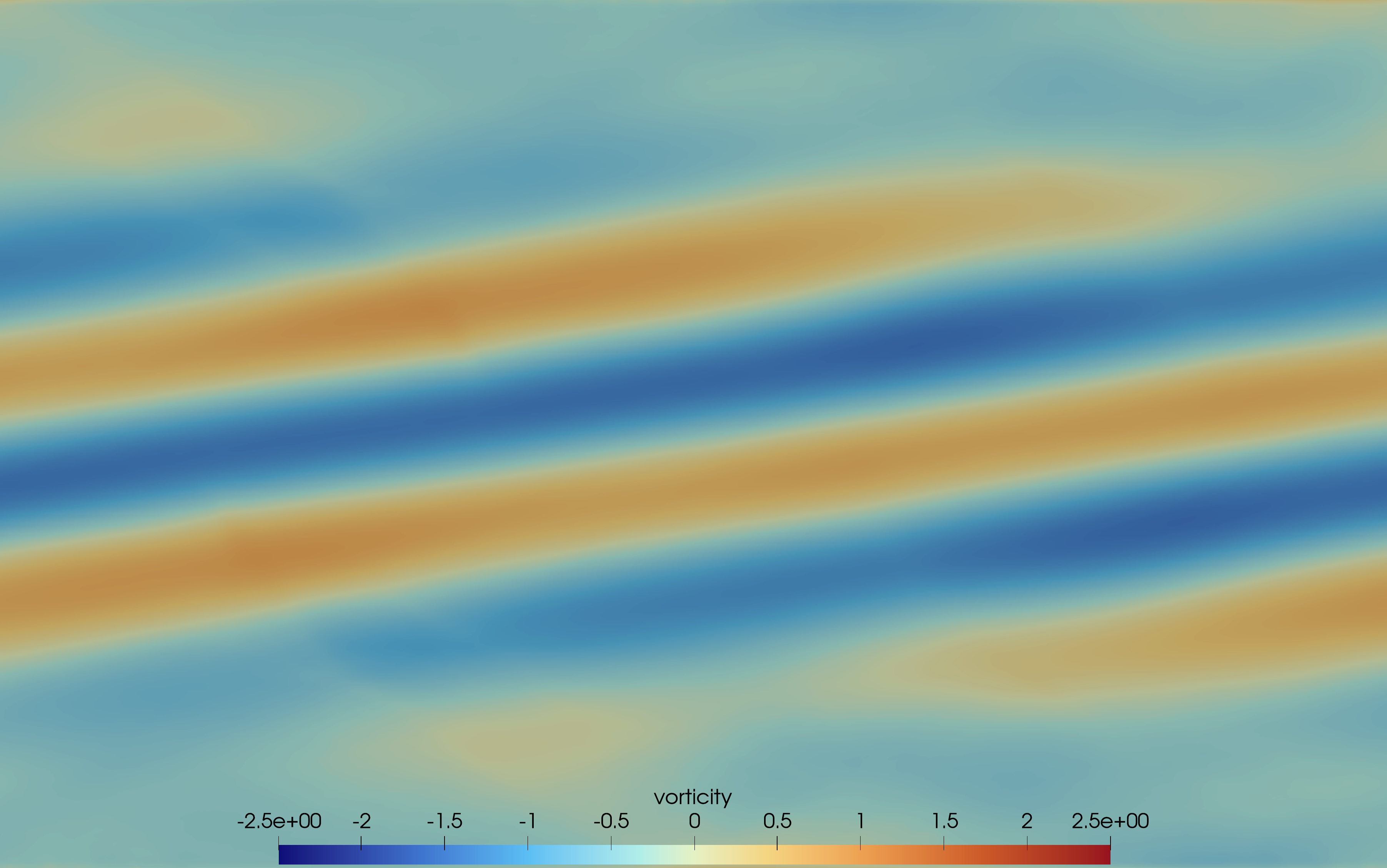}%
\caption{Vorticity fluctuations in the Couette flow in the presence of
  the van der Waals effect. Upper-left -- at 0.01 seconds, upper-right
  -- at 0.02 seconds, lower-left -- at 0.03 seconds, lower-right -- at
  0.05 seconds.}
\label{fig:vorticity_Couette}
\end{figure}

Similarly to the Poiseuille flow above in
Section~\ref{sec:Poiseuille}, next we simulate the flow with the same
initial density deviation as in
Figure~\ref{fig:density_Couette_novdW}, however this time the van der
Waals effect is included in \eqref{eq:div_vort}.  The resulting
snapshots of the deviations in $\rho$, $\chi$ and $\omega$ from their
background states are shown in Figures~\ref{fig:density_Couette},
\ref{fig:divergence_Couette} and~\ref{fig:vorticity_Couette},
respectively, for the elapsed times $t=0.01$, $0.02$, $0.03$ and
$0.05$ seconds. Similarly to the Poiseuille flow in
Figures~\ref{fig:density_Poiseuille}--\ref{fig:vorticity_Poiseuille},
here we can see the development of wave-like nontrivial dynamics,
where large-scale fluctuations eventually become small-scale
fluctuations as a result of a direct cascade, as predicted in
\cite{Abr27}.

Similarly to the Poiseuille flow scenario in
Section~\ref{sec:Poiseuille}, here the fluctuations in $\rho$ and
$\omega$ remain small relative to their background states ($\sim$ 1\%
of the total magnitude), yet the developed patterns look chaotic,
especially in the velocity divergence. Although we do not show it
here, the flow remains macroscopically laminar in the same manner as
it does in the Poiseuille scenario in
Figure~\ref{fig:tracer_Poiseuille}; namely, while the fluctuations
exhibit a complex behavior, they at the same time remain small enough
so that the tracer streaks remain intact and do not break, although a
minor distortion and smudging are visible upon closer inspection.

\subsection{Power decay of the Fourier spectra}

\begin{figure}[t]%
\includegraphics[width=0.5\textwidth]{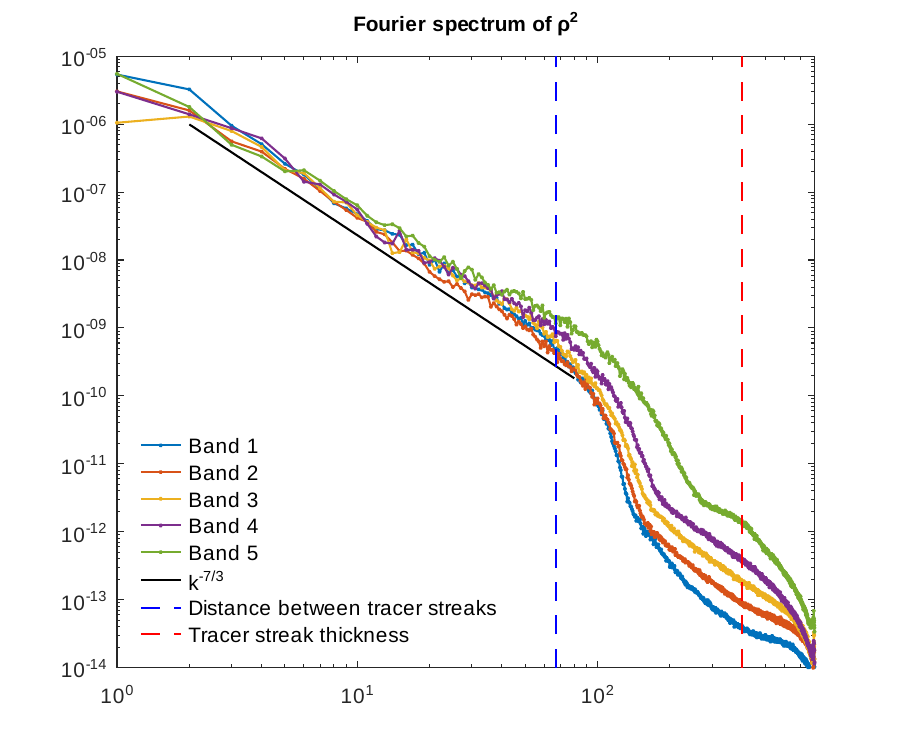}%
\includegraphics[width=0.5\textwidth]{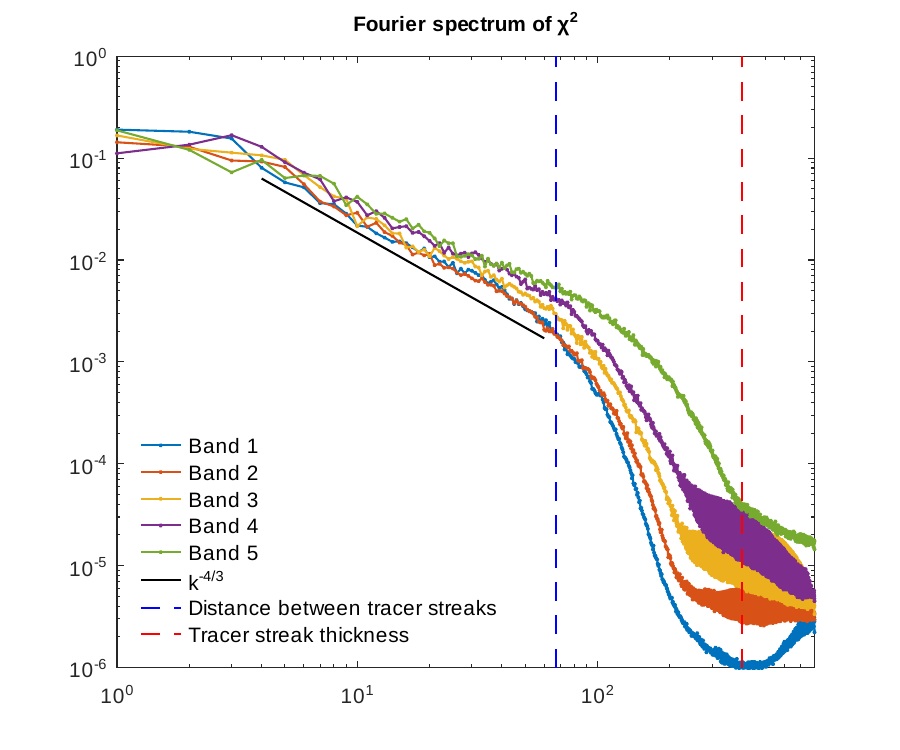}\\%
\includegraphics[width=0.5\textwidth]{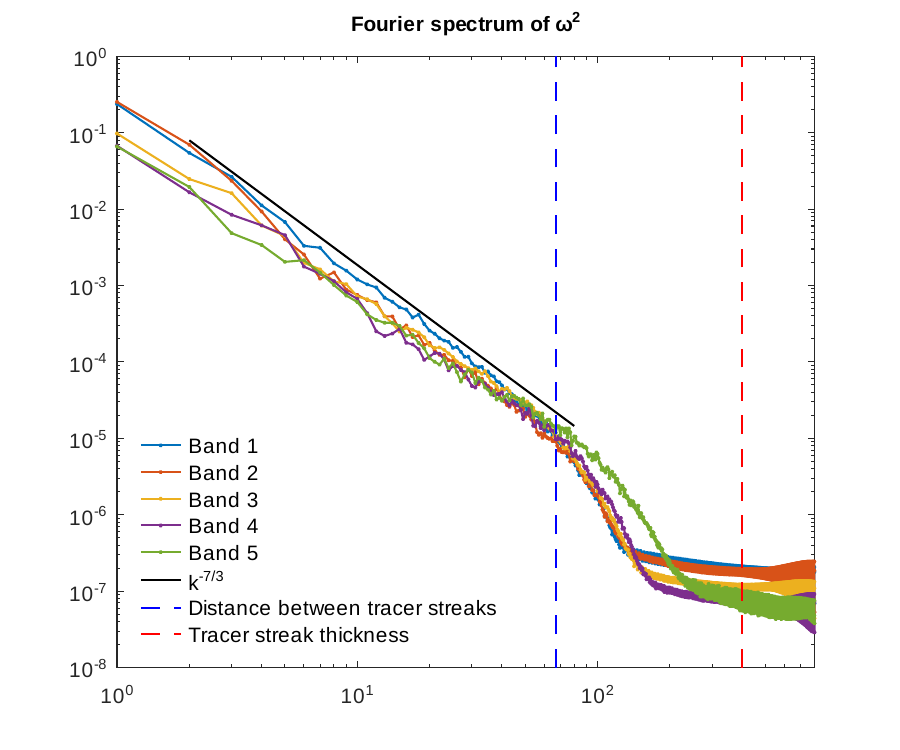}%
\vspace{-1EM}
\caption{The Fourier spectrum of $\rho^2$ (upper-left), $\chi^2$
  (upper-right), and $\omega^2$ (lower-center), Couette flow.}
\label{fig:rho_div_omega_Couette_spectrum}
\end{figure}
\begin{figure}[t]%
\includegraphics[width=0.5\textwidth]{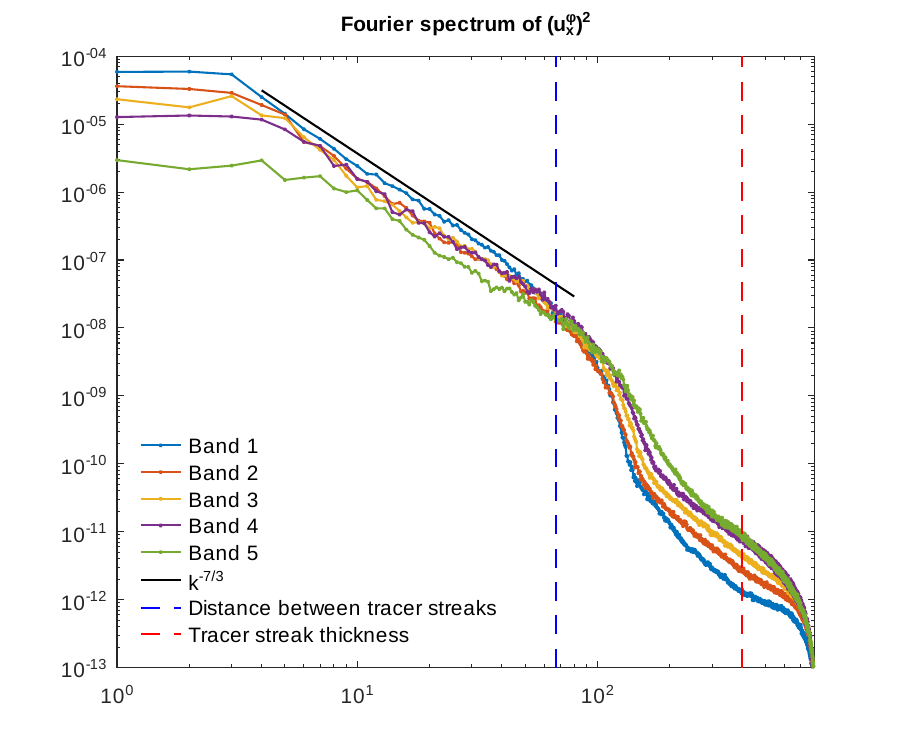}%
\includegraphics[width=0.5\textwidth]{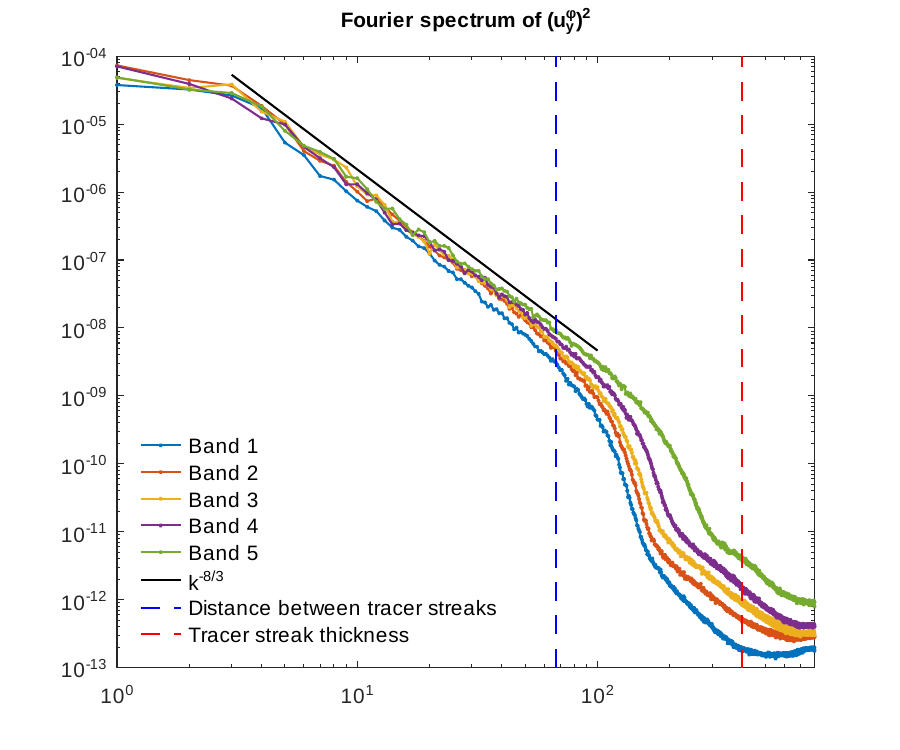}\\%
\includegraphics[width=0.5\textwidth]{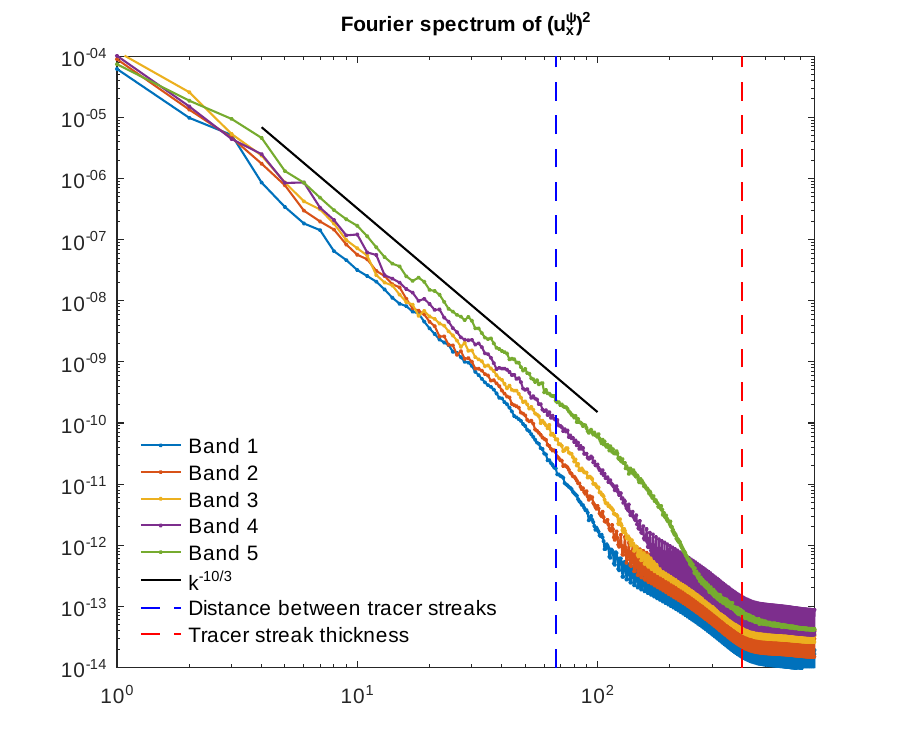}%
\includegraphics[width=0.5\textwidth]{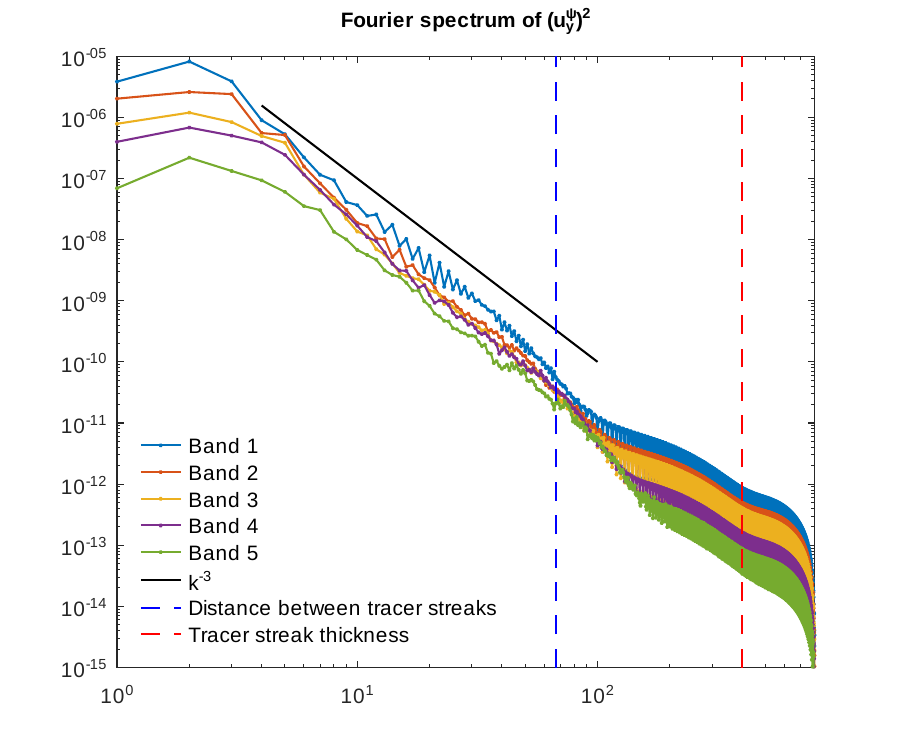}\\%
\vspace{-1EM}
\caption{The Fourier spectrum of $(u^\varphi_x)^2$ (upper-left),
  $(u^\varphi_y)^2$ (upper-right), $(u^\psi_x)^2$ (lower-left), and
  $(u^\psi_y)^2$ (lower-right), Couette flow.}
\label{fig:U_Couette_spectrum}
\end{figure}
\begin{table}[t]%
\begin{tabular}{|c||c|c|c|c|c|c|c|}%
\hline
Powers & $\rho$ & $\chi$ & $\omega$ & $u_x^\varphi$ & $u_y^\varphi$ &
$u_x^\psi$ & $u_y^\psi$ \\
\hline\hline
Poiseuille & $-7/3$ & $-5/3$ & $-8/3$ & $-8/3$ & $-3$ & $-11/3$ & $-10/3$ \\
Couette    & $-7/3$ & $-4/3$ & $-7/3$ & $-7/3$ & $-8/3$ & $-10/3$ & $-3$ \\
\hline
\end{tabular}%
\vspace{0.5EM}
\caption{A summary of powers of the Fourier spectra of $\rho$, $\chi$,
  $\omega$, $\BV u^\varphi$ and $\BV u^\psi$ for both the Poiseuille
  and Couette flows.}
\label{tab:powers}
\end{table}%

For the Couette flow, in
Figures~\ref{fig:rho_div_omega_Couette_spectrum}
and~\ref{fig:U_Couette_spectrum} we compute the time averages of the
power spectra in the same manner, and of the same quantities, as we
did for the Poiseuille flow above in Section~\ref{sec:Poiseuille}.
Just as in the Poiseuille flow scenario, here the power decay of all
displayed Fourier spectra occurs on the spatial scales larger than the
distance between the tracer streaks, that is, where the flow is
technically laminar.

For the squares of the fluctuations of $\rho$, $\chi$ and $\omega$,
the computed time-averages are shown in
Figure~\ref{fig:rho_div_omega_Couette_spectrum}. The density spectrum,
shown in the upper-left pane of
Figure~\ref{fig:rho_div_omega_Couette_spectrum}, exhibits the $\sim
k^{-7/3}$ power decay for all five channel bands, up until the
wavenumber $k\sim 80$. The spectrum of the velocity divergence, shown
in the upper-right pane of
Figure~\ref{fig:rho_div_omega_Couette_spectrum}, also exhibits power
decay, however the rate of decay is different at $\sim k^{-4/3}$. The
vorticity spectrum has the most rapid decay (out of the three
aforementioned quantities) at $\sim k^{-7/3}$ rate, shown in the lower
pane of Figure~\ref{fig:rho_div_omega_Couette_spectrum}.  It is
interesting that, while the density decay rate for the Couette flow is
identical to that of the Poiseuille flow in
Figure~\ref{fig:rho_div_omega_Poiseuille_spectrum}, the decay rates
for both the velocity divergence and vorticity are slower by
$\sqrt[3]k$.

\begin{figure}[t]%
\includegraphics[width=0.49\textwidth]{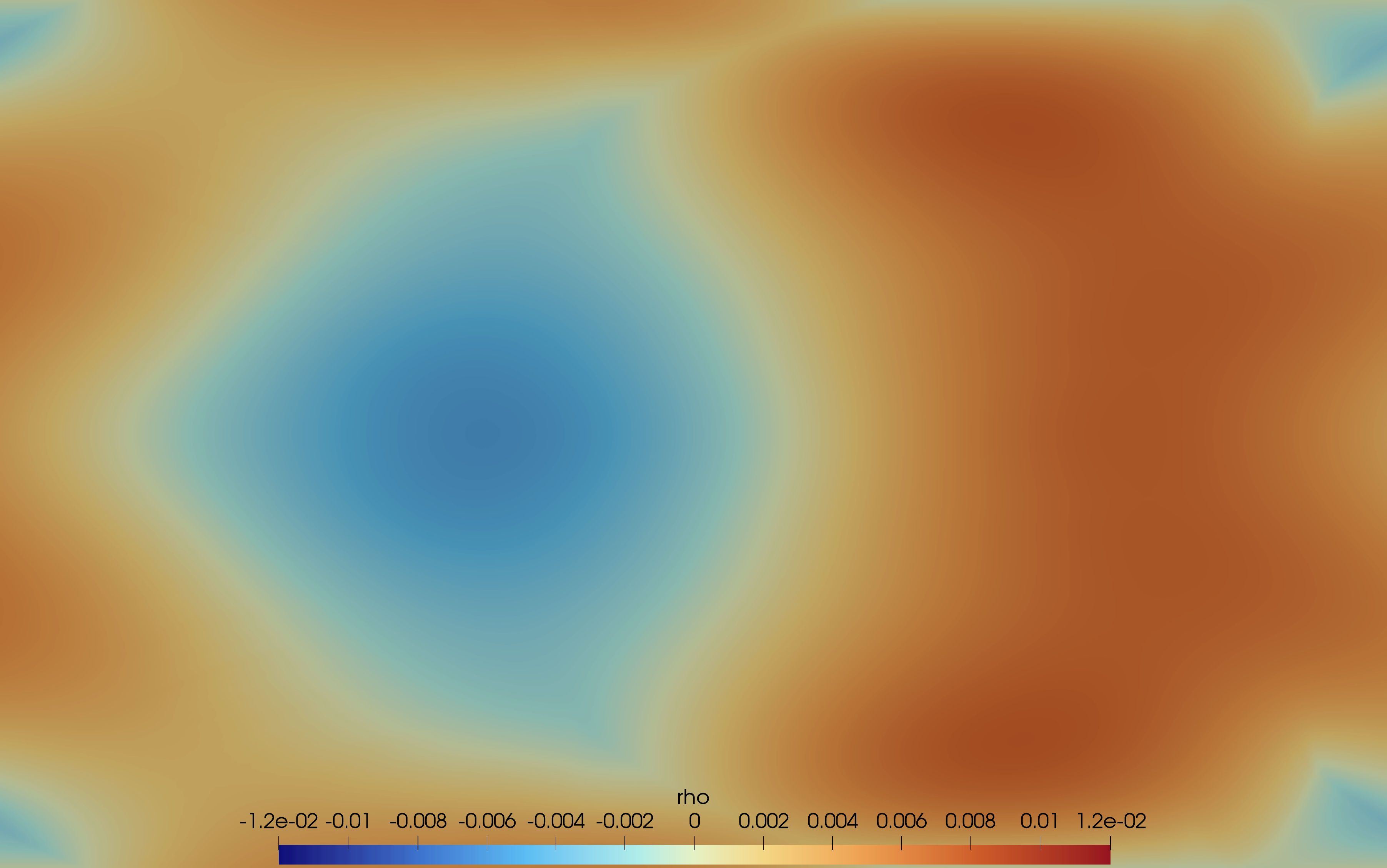}\,%
\includegraphics[width=0.49\textwidth]{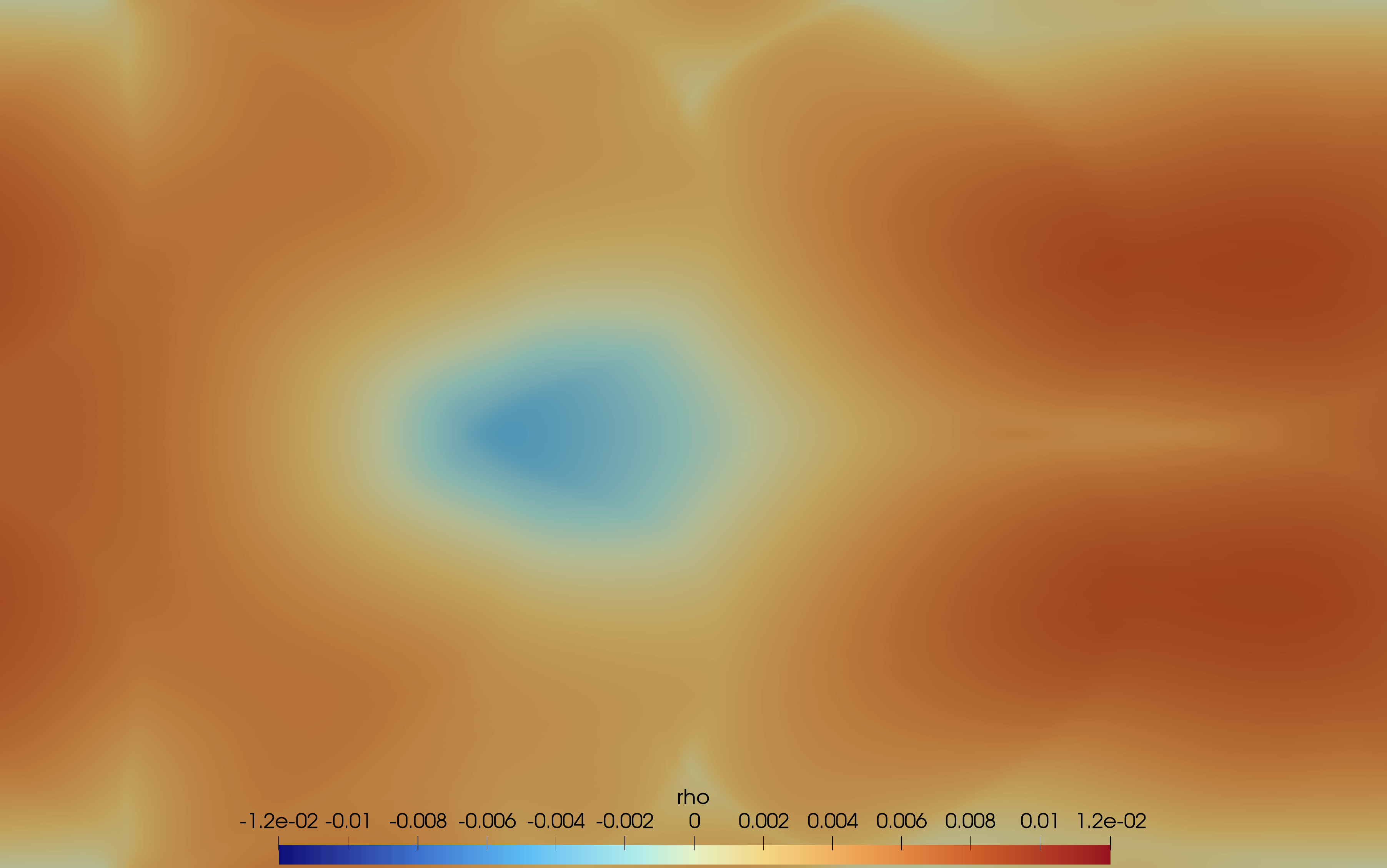}\\%
\vspace{1.5pt}%
\includegraphics[width=0.49\textwidth]{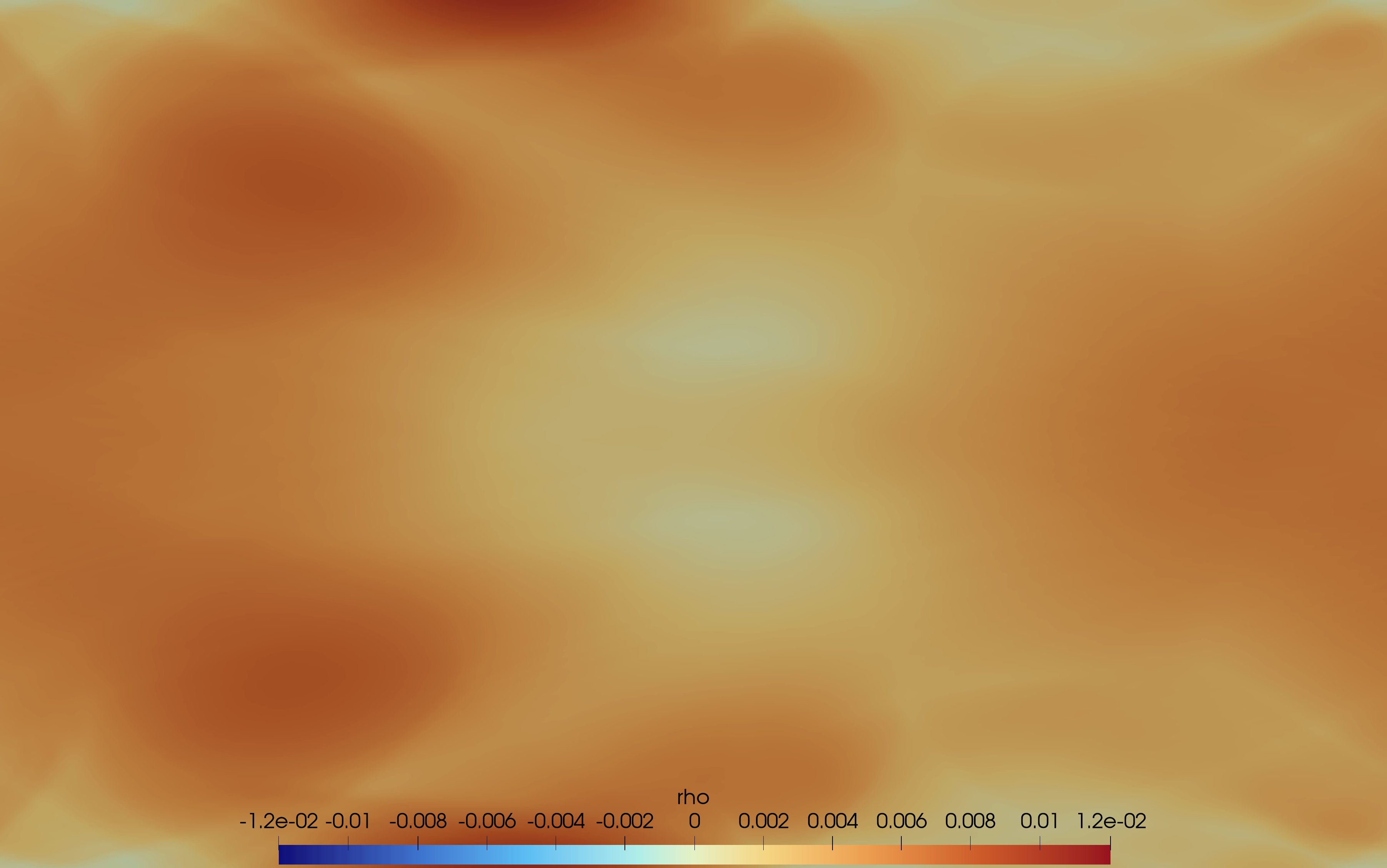}\,%
\includegraphics[width=0.49\textwidth]{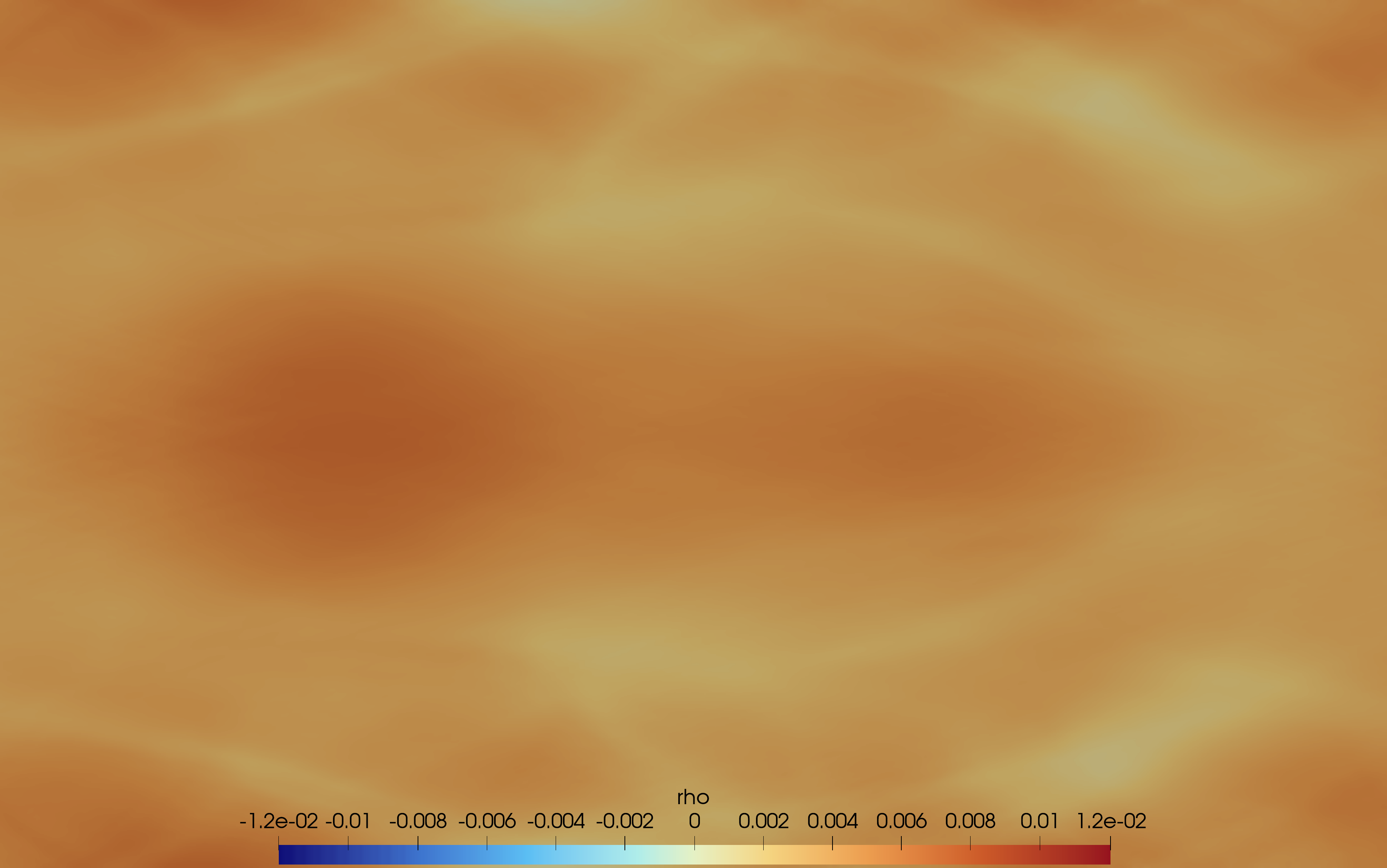}%
\caption{Density fluctuations in the Poiseuille flow with stationary
  vorticity in the presence of the van der Waals effect. Upper-left --
  at 0.01 seconds, upper-right -- at 0.02 seconds, lower-left -- at
  0.03 seconds, lower-right -- at 0.05 seconds.}
\label{fig:density_Poiseuille_noVorticity}
\end{figure}

Just as we did above for the Poiseuille flow, here in
Figure~\ref{fig:U_Couette_spectrum} we show the time averages for the
squares of the fluctuations of the streamwise and transversal
components of velocities $\BV u^\varphi$ and $\BV u^\psi$, defined in
\eqref{eq:Helmholtz}, which represent different parts of the total
kinetic energy of the flow. The spectrum of the streamwise component
$u^\varphi_x$, which corresponds to the potential part of the
streamwise kinetic energy of the flow, is shown in the upper-left pane
of Figure~\ref{fig:U_Couette_spectrum}, and decays at the rate of
$\sim k^{-7/3}$ in all channel bands. The spectrum of the transversal
component $u^\varphi_y$, which corresponds to the potential part of
the transversal kinetic energy of the flow, is shown in the
upper-right pane of Figure~\ref{fig:U_Couette_spectrum}, and decays at
the rate of $\sim k^{-8/3}$ in all channel bands. The spectrum of the
streamwise component $u^\psi_x$, which corresponds to the stream
function part of the streamwise kinetic energy of the flow, is shown
in the lower-left pane of Figure~\ref{fig:U_Couette_spectrum}, and
decays at the rate of $\sim k^{-10/3}$ in all channel bands. Finally,
the spectrum of the transversal component $u^\psi_y$, which
corresponds to the stream function part of the transversal kinetic
energy of the flow, is shown in the lower-right pane of
Figure~\ref{fig:U_Couette_spectrum}, and decays at the rate of $\sim
k^{-3}$ in all channel bands. It is remarkable that the rates of decay
of different parts of the kinetic energy of the flow in
Figure~\ref{fig:U_Couette_spectrum} are all distinctly different, just
as it was for the Poiseuille flow. Also, here the decay rate for all
components of the kinetic energy is slower than that for the
Poiseuille flow by a cubic root of the wavenumber. We summarize all
observed decay powers for the Fourier transforms of all variables of
both the Poiseuille and Couette flows in Table~\ref{tab:powers}. As we
can see, only the density powers are identical between the Poiseuille
and Couette flows, while all others differ by $\sqrt[3]k$.

\begin{figure}%
\includegraphics[width=0.49\textwidth]{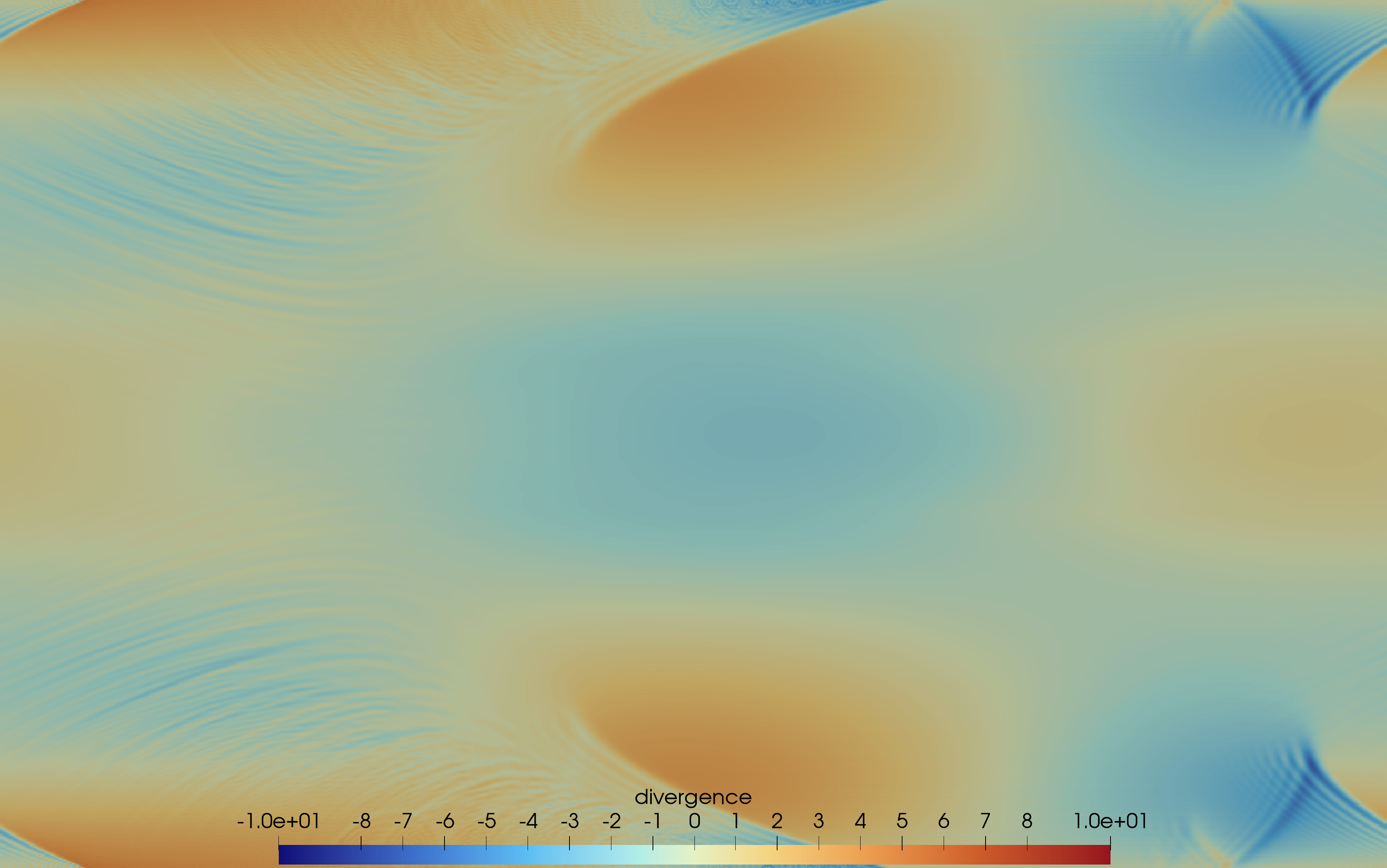}\,%
\includegraphics[width=0.49\textwidth]{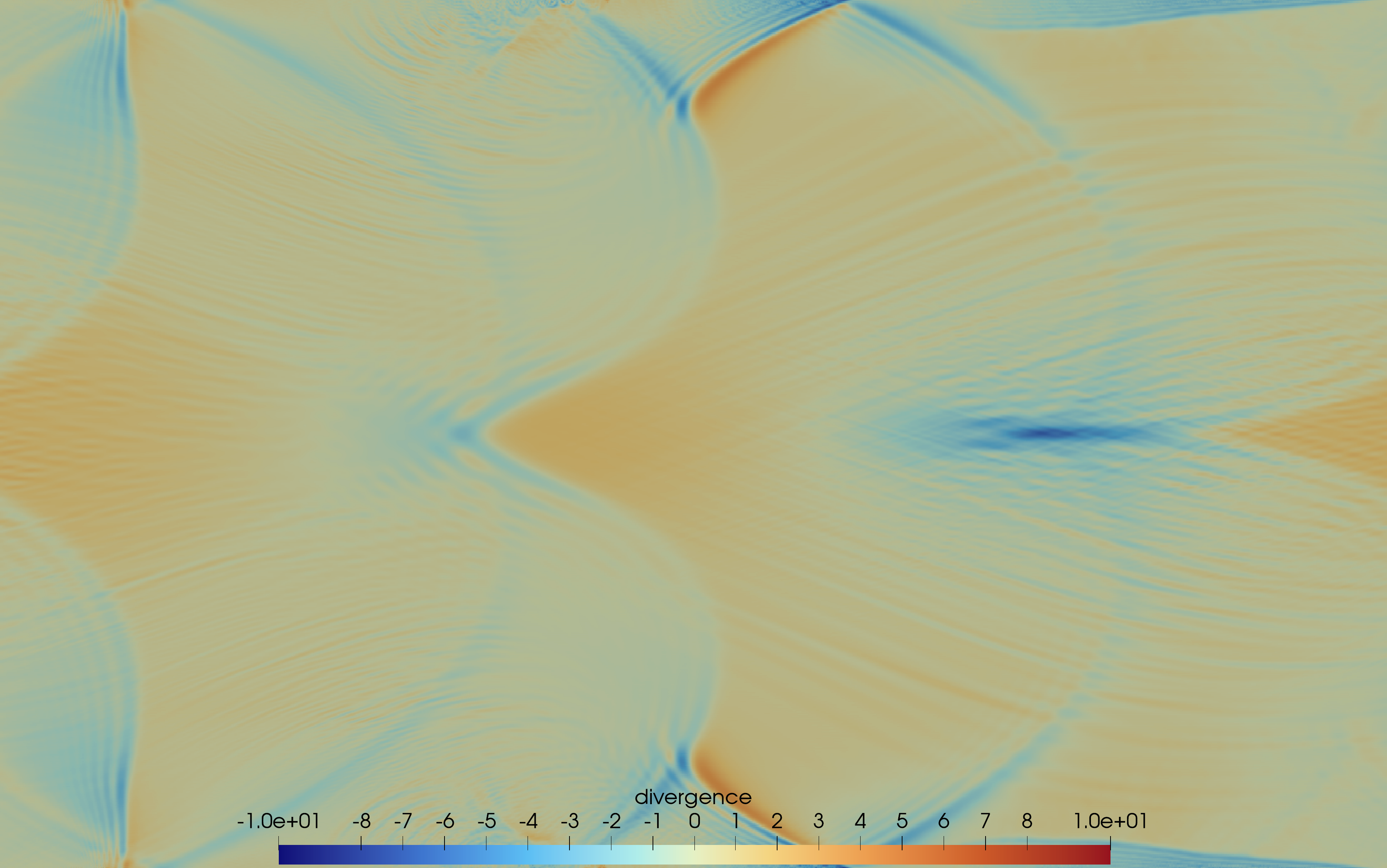}\\%
\vspace{1.5pt}%
\includegraphics[width=0.49\textwidth]{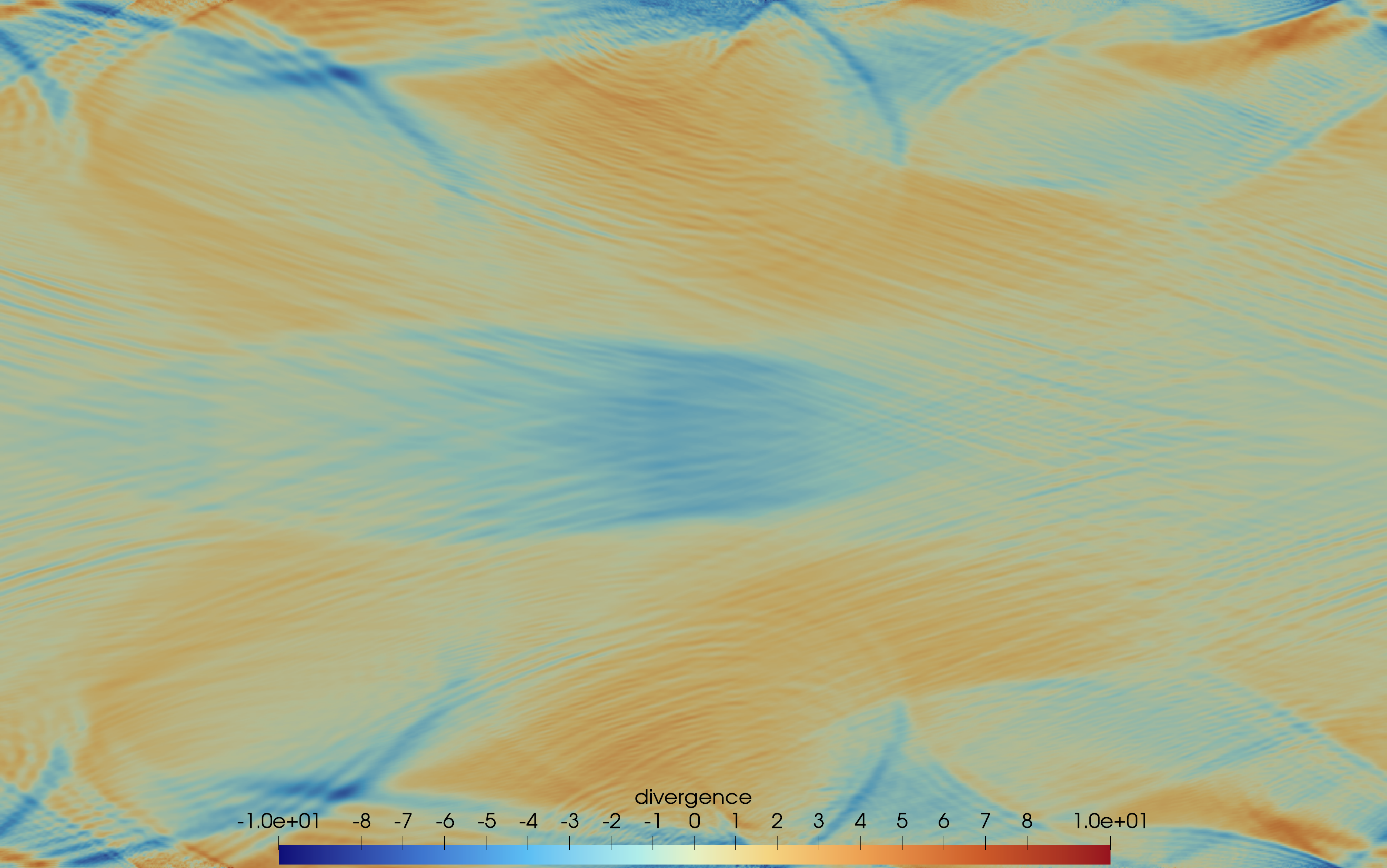}\,%
\includegraphics[width=0.49\textwidth]{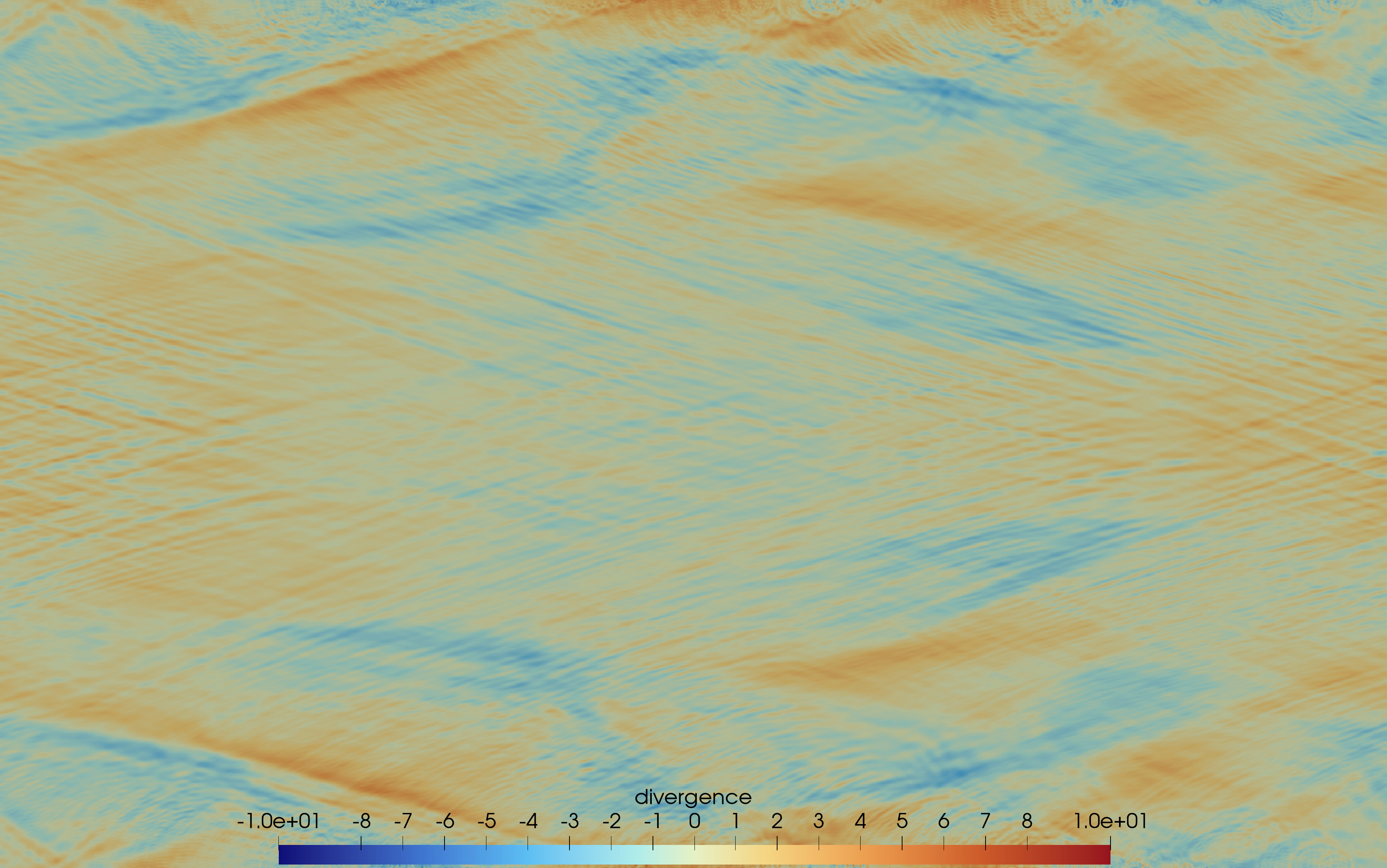}%
\caption{Divergence fluctuations in the Poiseuille flow with
  stationary vorticity in the presence of the van der Waals
  effect. Upper-left -- at 0.01 seconds, upper-right -- at 0.02
  seconds, lower-left -- at 0.03 seconds, lower-right -- at 0.05
  seconds.}
\label{fig:divergence_Poiseuille_noVorticity}
\end{figure}

\section{Numerical simulation of the Poiseuille flow with stationary
vorticity}

Observe that, in the numerical simulations above for both the
Poiseuille and Couette flows, the fluctuations the vorticity variable
$\omega$ develops around its respective Poiseuille or Couette
background state are very small, about $\sim 1$\% of its overall
magnitude.  Also, according to our theory in \cite{Abr27}, the
vorticity $\omega$ is associated with a stable eigenvector, while the
turbulent instability is spanned by the density $\rho$ and velocity
divergence $\chi$.  Therefore, an interesting question is: what if we
remove the vorticity fluctuations altogether, and replace $\omega$ in
\eqref{eq:div_vort} with its own stationary state (that is, Poiseuille
or Couette)?

\begin{figure}[t]%
\includegraphics[width=0.5\textwidth]{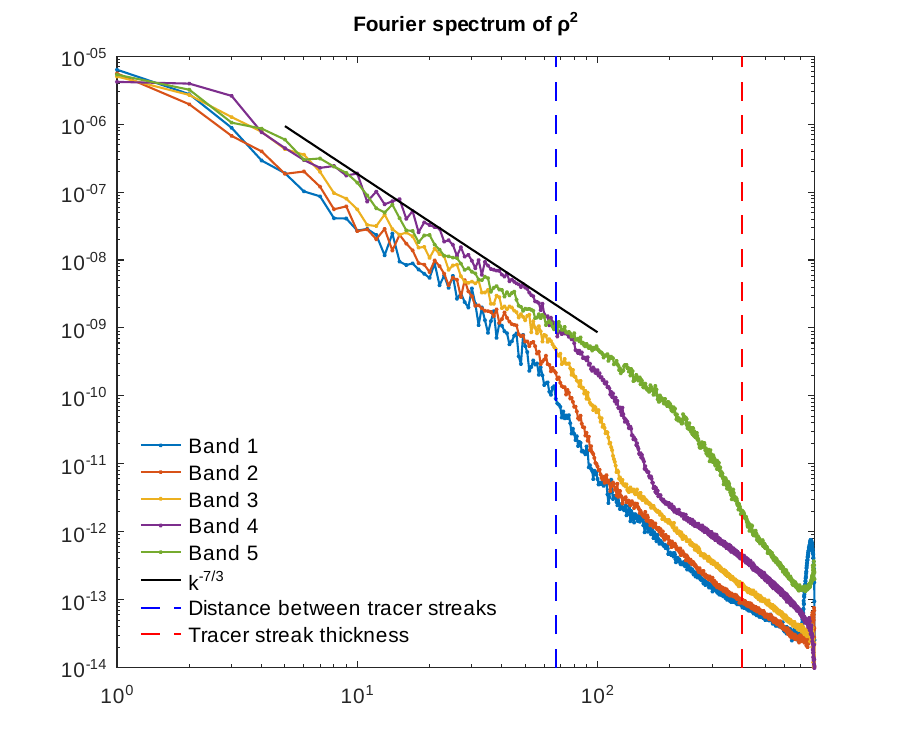}%
\includegraphics[width=0.5\textwidth]{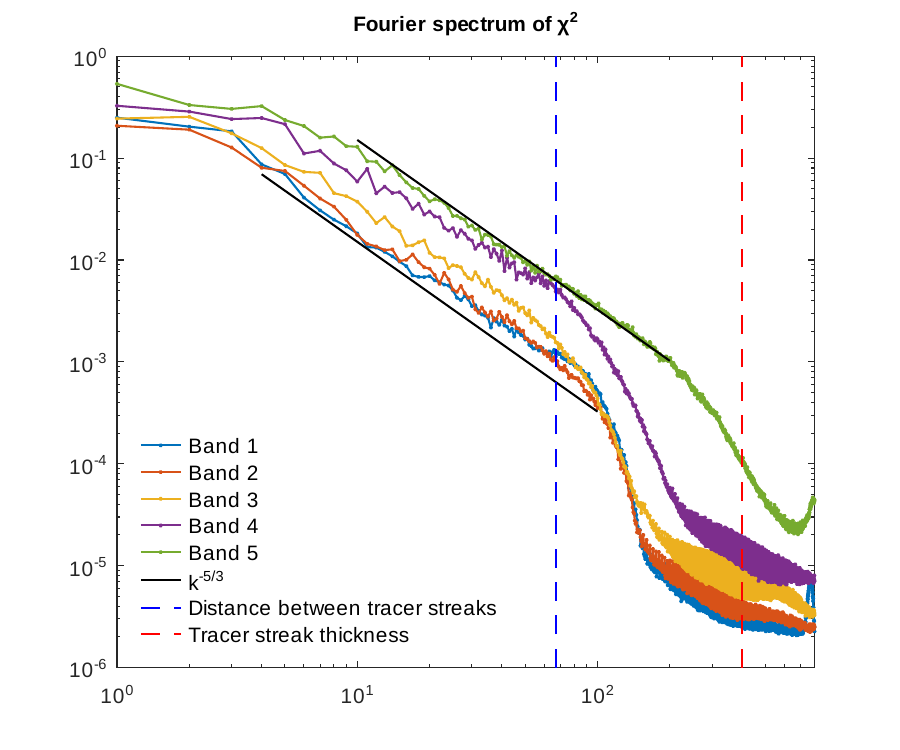}%
\vspace{-1EM}
\caption{The Fourier spectrum of $\rho^2$ (left), and $\chi^2$
  (right), Poiseuille flow with stationary vorticity.}
\label{fig:rho_div_Poiseuille_noVort_spectrum}
\end{figure}
\begin{figure}[t]%
\includegraphics[width=0.5\textwidth]{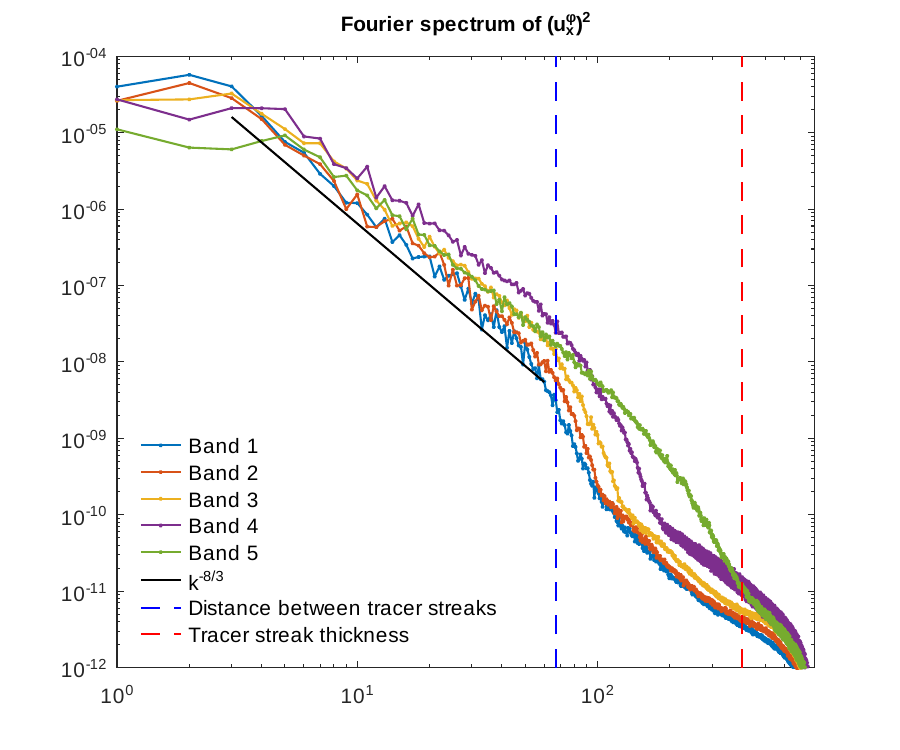}%
\includegraphics[width=0.5\textwidth]{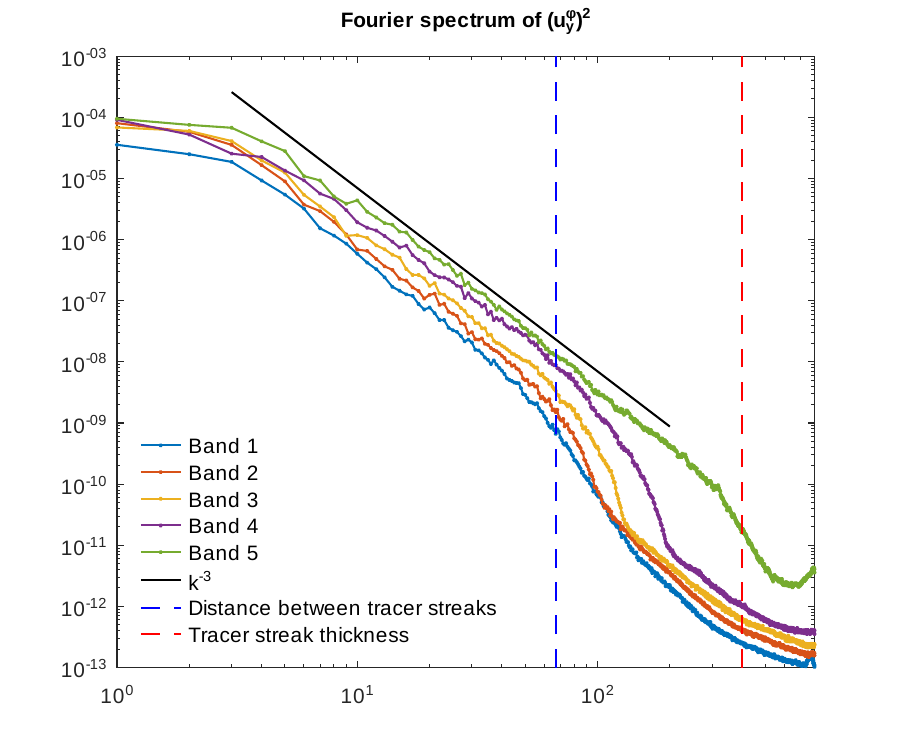}%
\vspace{-1EM}
\caption{The Fourier spectrum of $(u^\varphi_x)^2$ (left), and
  $(u^\varphi_y)^2$ (right), Poiseuille flow with stationary
  vorticity.}
\label{fig:U_Poiseuille_noVort_spectrum}
\end{figure}

The resulting reduced system for the density $\rho$ and velocity
divergence $\chi$ is given via
\begin{subequations}
\label{eq:div}
\begin{equation}
\parderiv\rho t+\nabla\cdot(\rho\BV u)=0,\qquad
\parderiv{\chi}t+\nabla\cdot(\chi\BV u)-2\det(\nabla\BV u)+\frac{4p_0
}{\rho_\HS}\nabla\cdot\left(\frac{\nabla\rho}\rho\right)=\frac 43\nu
\Delta\chi,
\end{equation}
\begin{equation}
\Delta\varphi=\chi,\qquad\BV u=\nabla\varphi+\BV u_0.
\end{equation}
\end{subequations}
Above, $\BV u_0$ is the corresponding stationary background velocity
state, either Poiseuille in \eqref{eq:u_Poiseuille}, or Couette in
\eqref{eq:u_Couette}, depending on the simulation. It is easy to
verify that the turbulent instability of \eqref{eq:div_vort},
described in our work \cite{Abr27}, persists in the reduced
system~\eqref{eq:div}, because the latter retains both the term
$-2\det(\nabla\BV u)$ and the van der Waals effect
$4p_0\nabla\cdot(\rho^{-1}\nabla\rho) /\rho_\HS$ in the divergence
equation.

Another reason why the density-divergence system in \eqref{eq:div} is
interesting, is because it is somewhat opposite to the incompressible
Navier--Stokes equations; in 2D, the incompressible dynamics are
purely rotational and consist of a sole vorticity transport equation,
whereas the density is constant and the velocity divergence is zero.
Conversely, in \eqref{eq:div} it is the vorticity which is a
stationary quantity, while the density and velocity divergence are
variables which remain fully coupled via the van der Waals effect. The
flow of \eqref{eq:div} is, therefore, comprised of a stationary
rotational (or shear) component, with small fluctuations around it
which consist solely of the compressibility effects.

\begin{figure}[t]%
\includegraphics[width=0.49\textwidth]{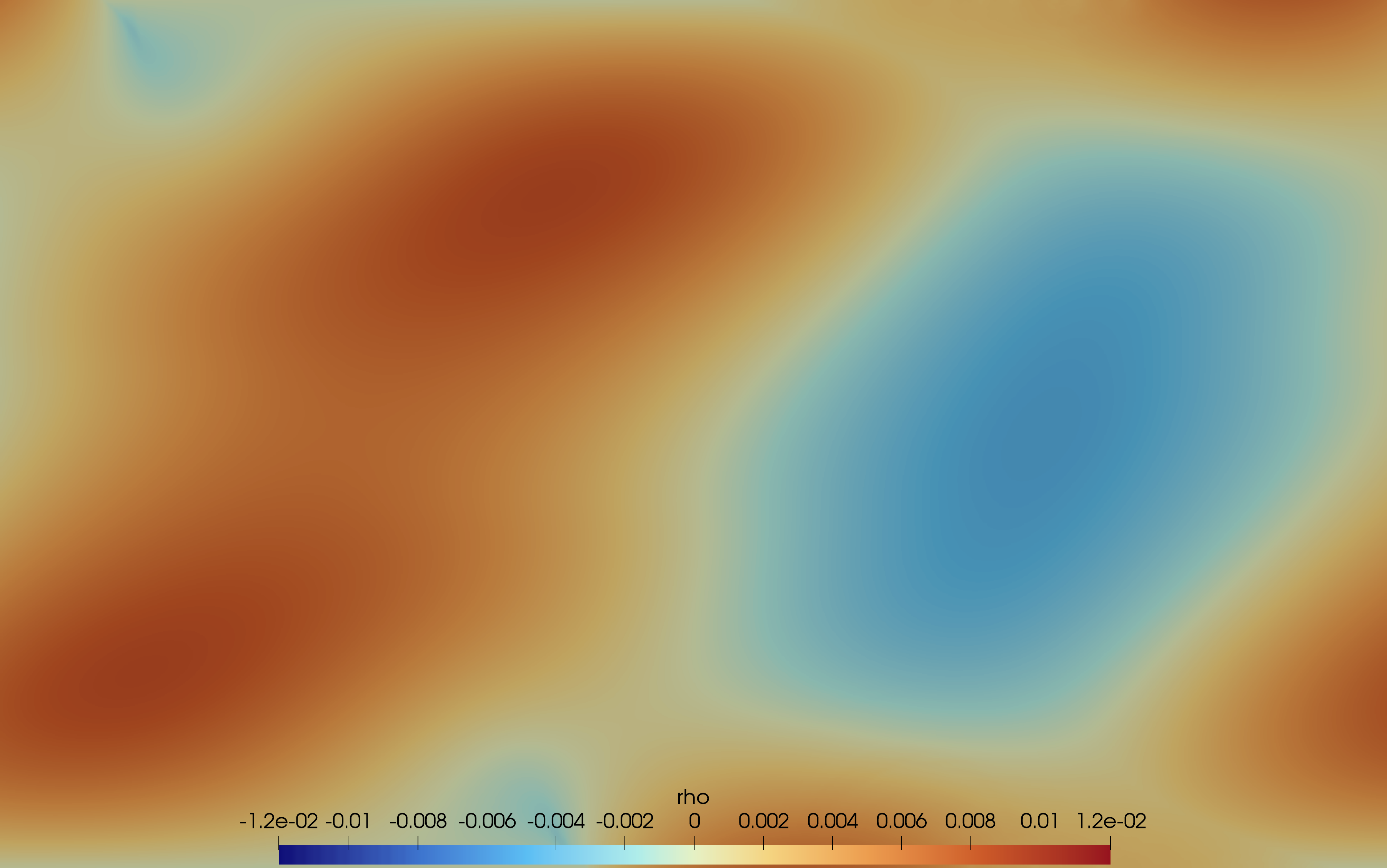}\,%
\includegraphics[width=0.49\textwidth]{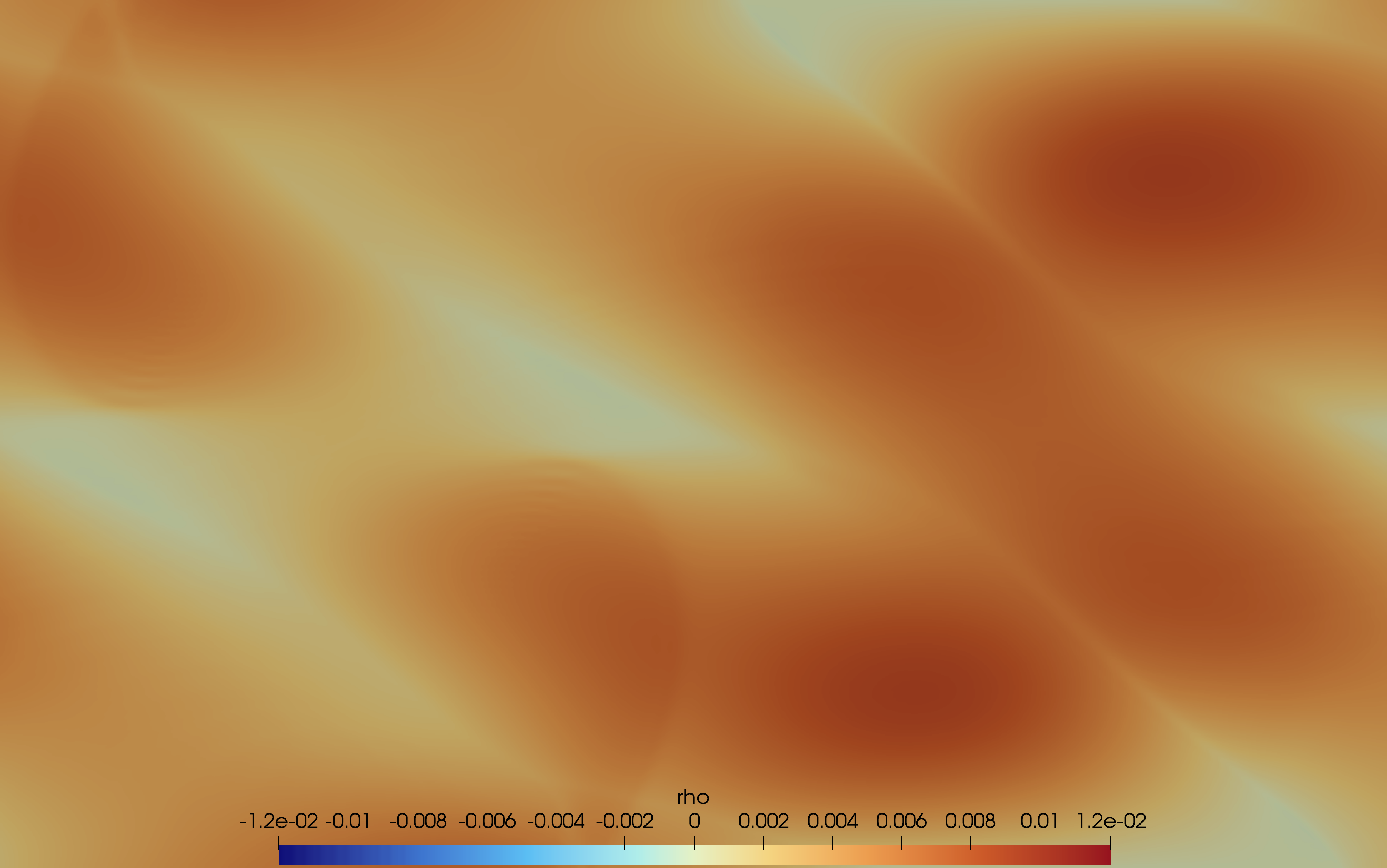}\\%
\vspace{1.5pt}%
\includegraphics[width=0.49\textwidth]{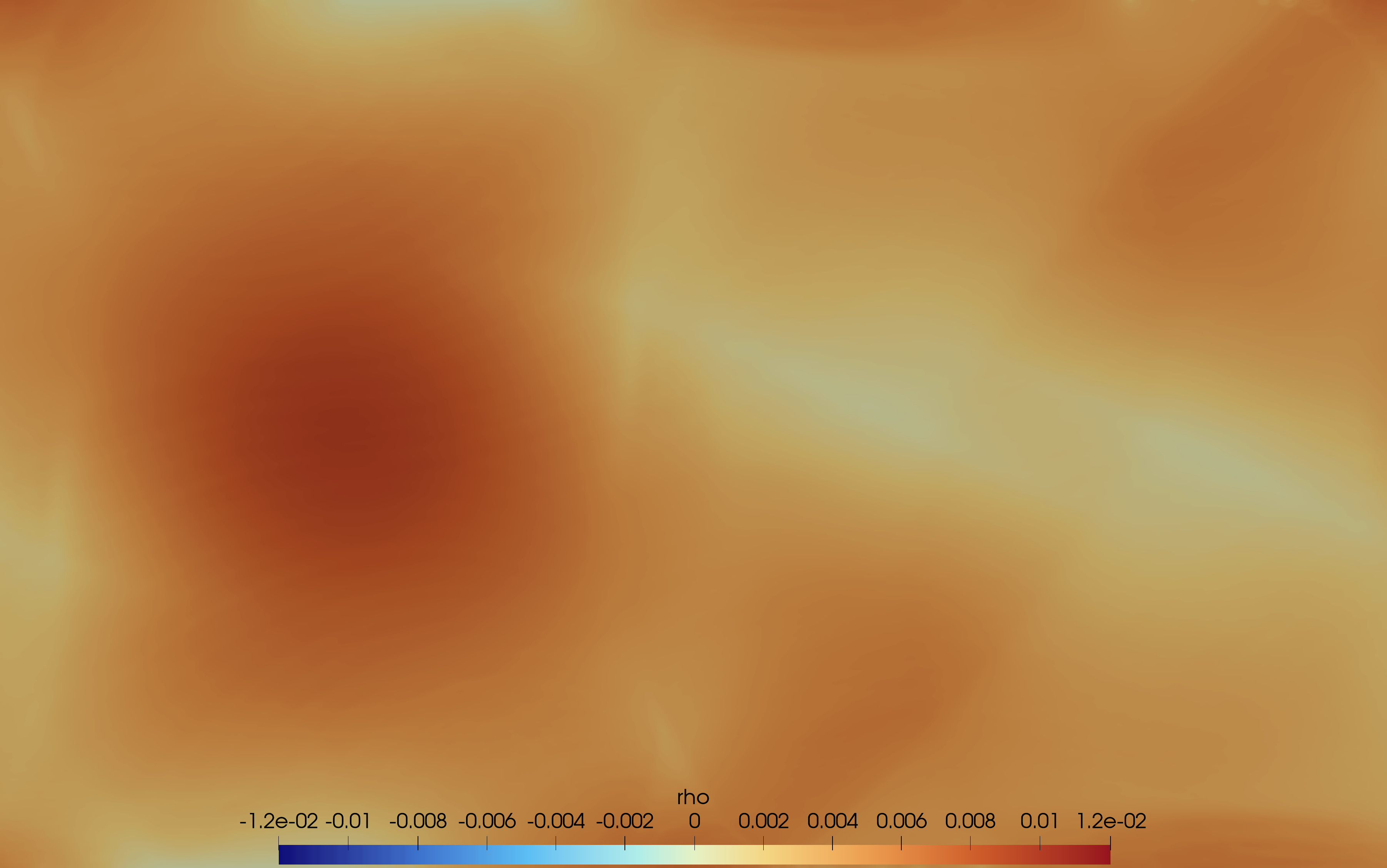}\,%
\includegraphics[width=0.49\textwidth]{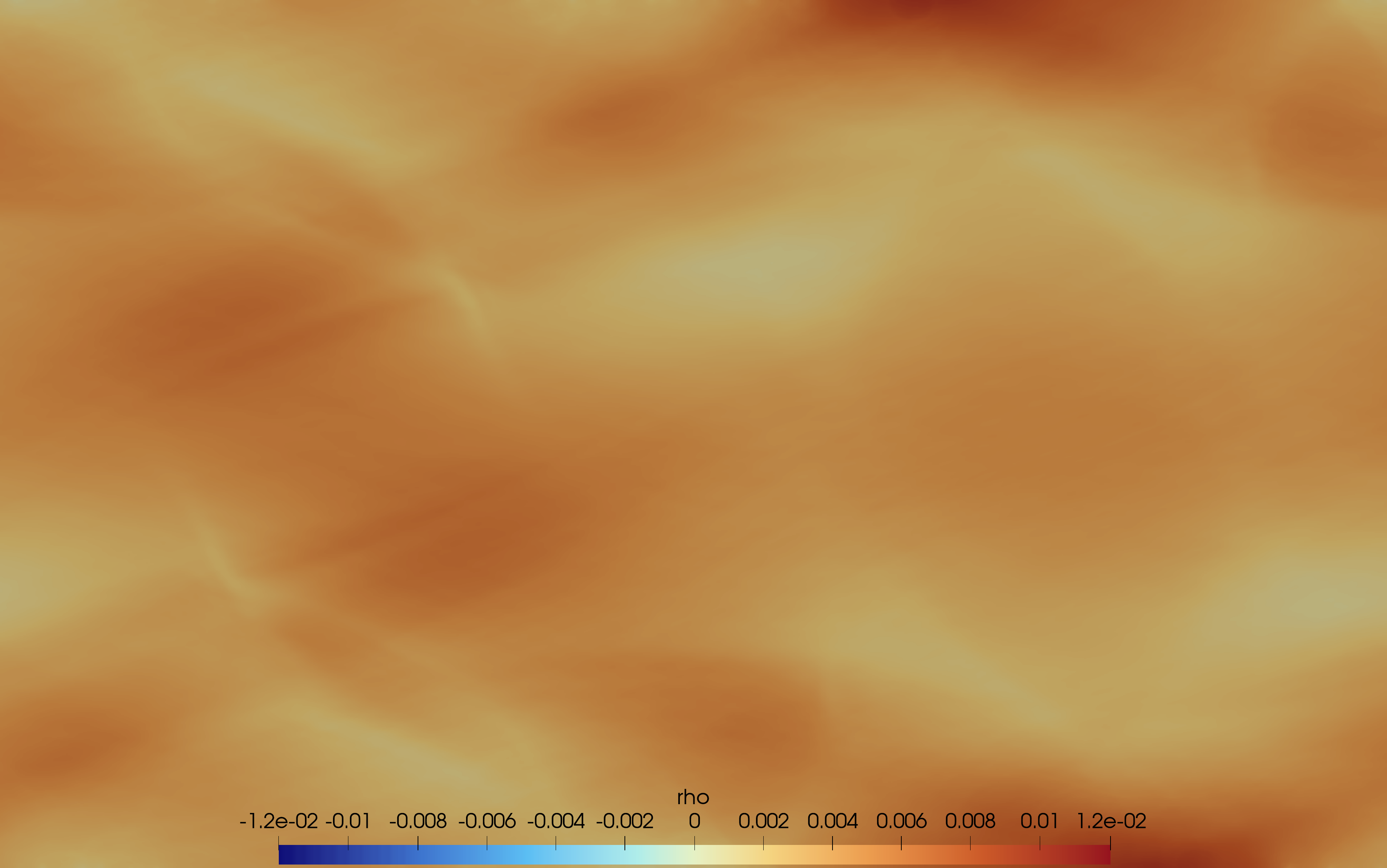}%
\caption{Density fluctuations in the Couette flow with stationary
  vorticity in the presence of the van der Waals effect. Upper-left --
  at 0.01 seconds, upper-right -- at 0.02 seconds, lower-left -- at
  0.03 seconds, lower-right -- at 0.05 seconds.}
\label{fig:density_Couette_noVorticity}
\end{figure}

We numerically simulate the density-divergence system in
\eqref{eq:div} for the same Poiseuille flow profile and initial
conditions in \eqref{eq:Poiseuille_rho_chi_omega} as we did with the
full system in \eqref{eq:div_vort} back in
Section~\ref{sec:Poiseuille}. The resulting snapshots of the
deviations in $\rho$ and $\chi$ from their background states are shown
in Figures~\ref{fig:density_Poiseuille_noVorticity}
and~\ref{fig:divergence_Poiseuille_noVorticity}, respectively, for the
same elapsed times $t=0.01$, $0.02$, $0.03$ and $0.05$ seconds. Direct
comparison with Figures~\ref{fig:density_Poiseuille}
and~\ref{fig:divergence_Poiseuille} shows profound similarities
between the numerical solutions of \eqref{eq:div_vort} and
\eqref{eq:div}; in fact, the final snapshots of the velocity
divergence at $t=0.05$ seconds (lower-right panes in
Figures~\ref{fig:divergence_Poiseuille}
and~\ref{fig:divergence_Poiseuille_noVorticity}), while not strictly
identical, are quite similar despite complex wave structures appearing
in both plots. The simulation confirms the result of \cite{Abr27},
where it was found that the vorticity $\omega$ does not play a key
role in the development of the direct cascade in the density and
velocity divergence fluctuations.

\begin{figure}[t]%
\includegraphics[width=0.49\textwidth]{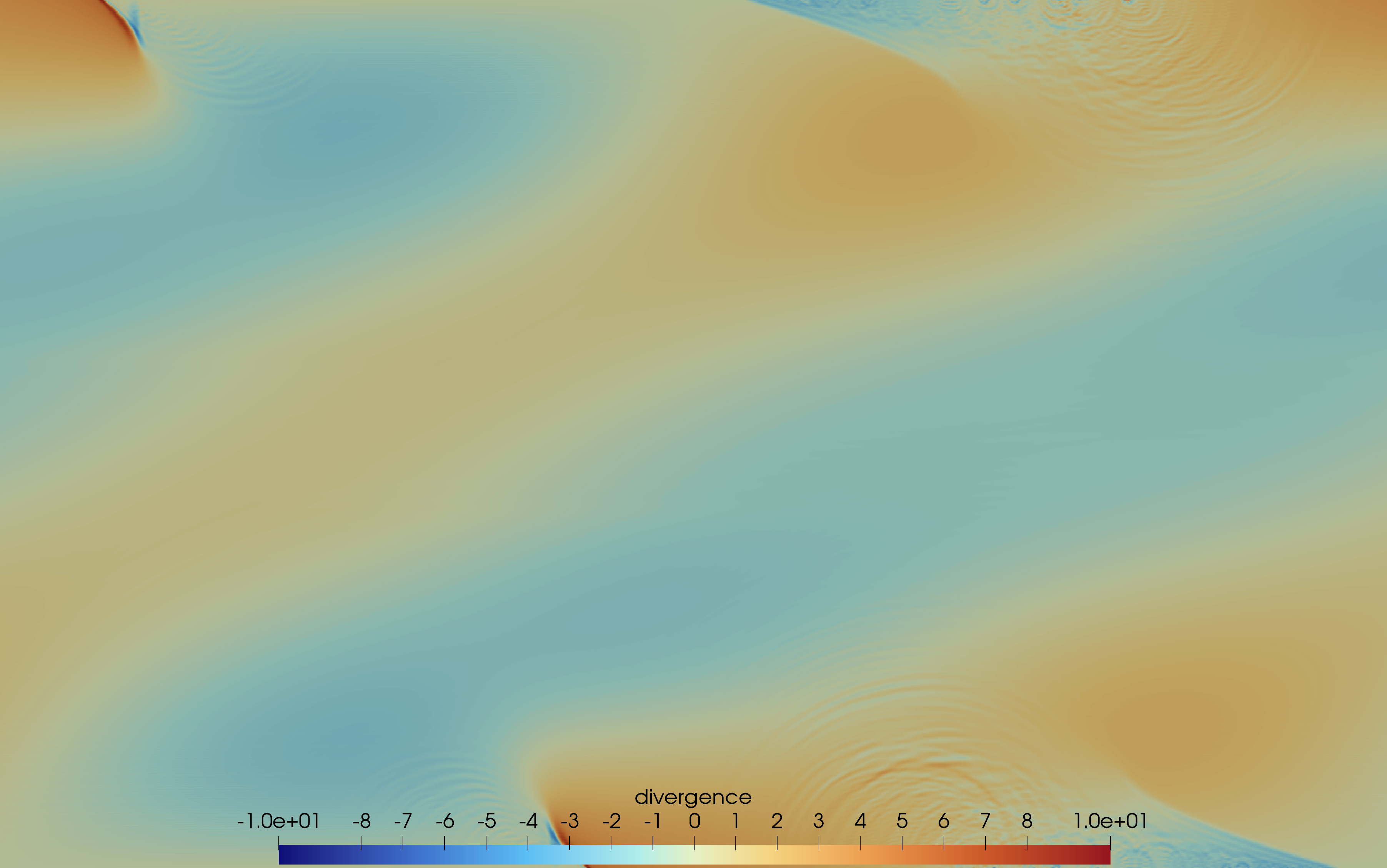}\,%
\includegraphics[width=0.49\textwidth]{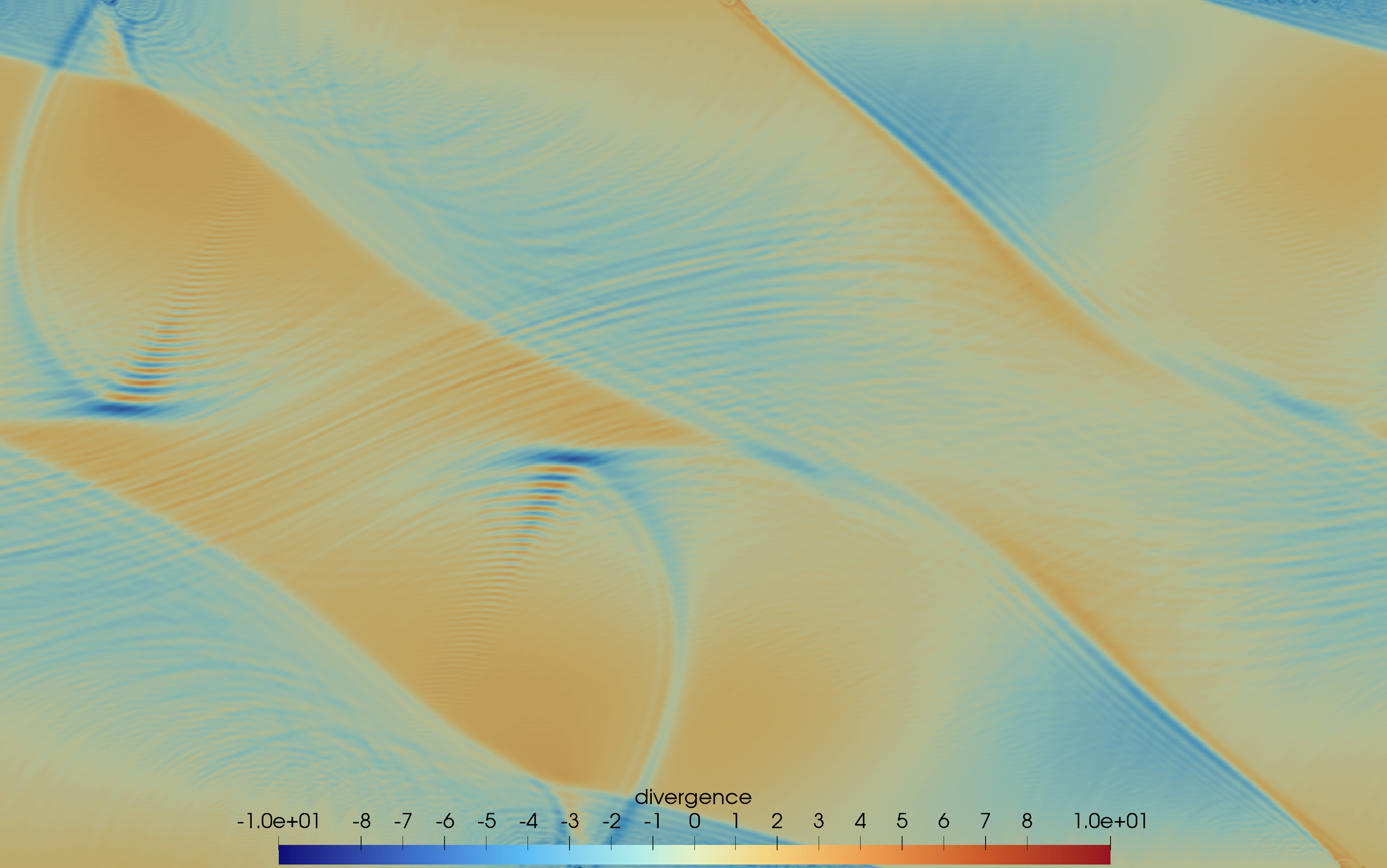}\\%
\vspace{1.5pt}%
\includegraphics[width=0.49\textwidth]{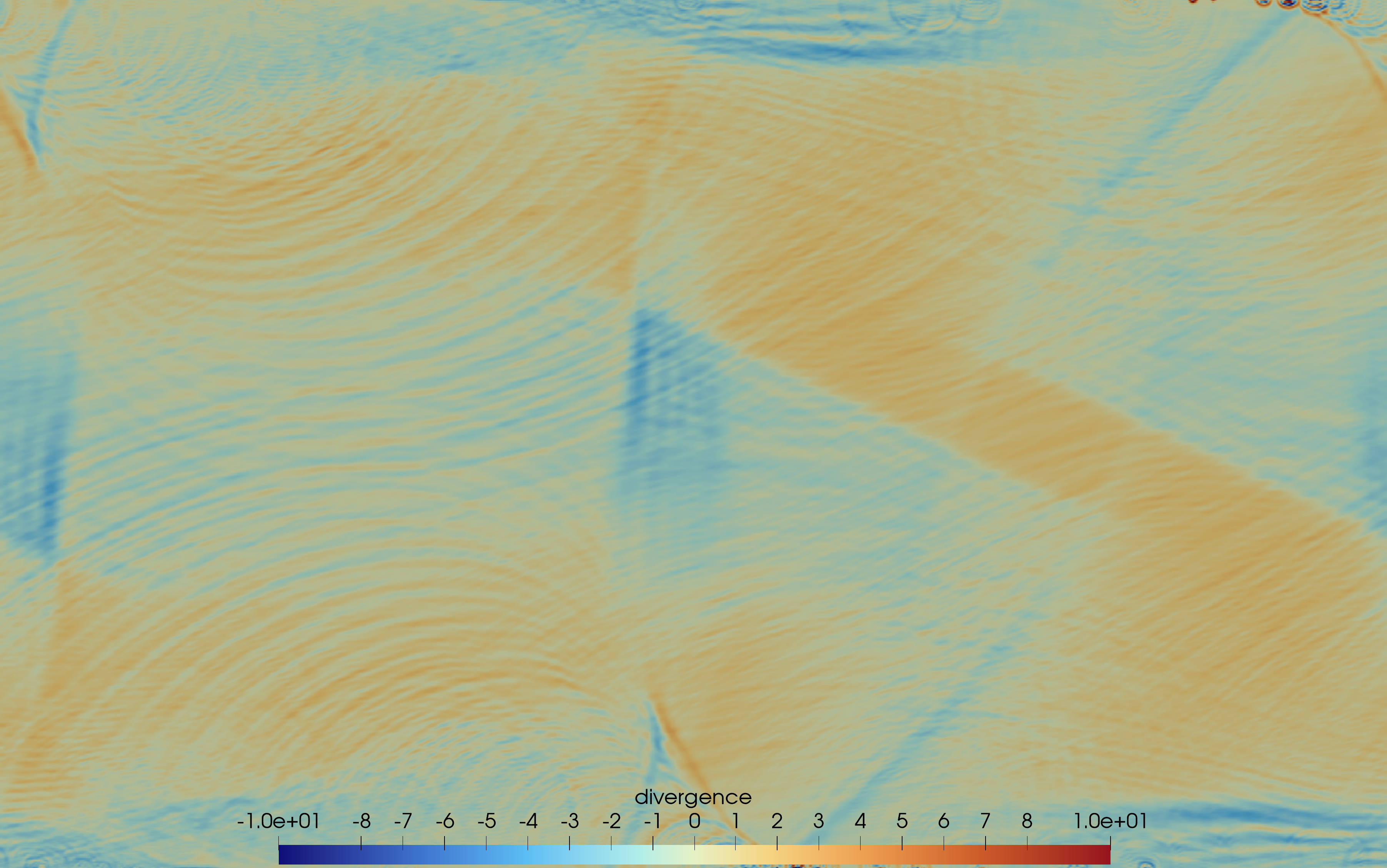}\,%
\includegraphics[width=0.49\textwidth]{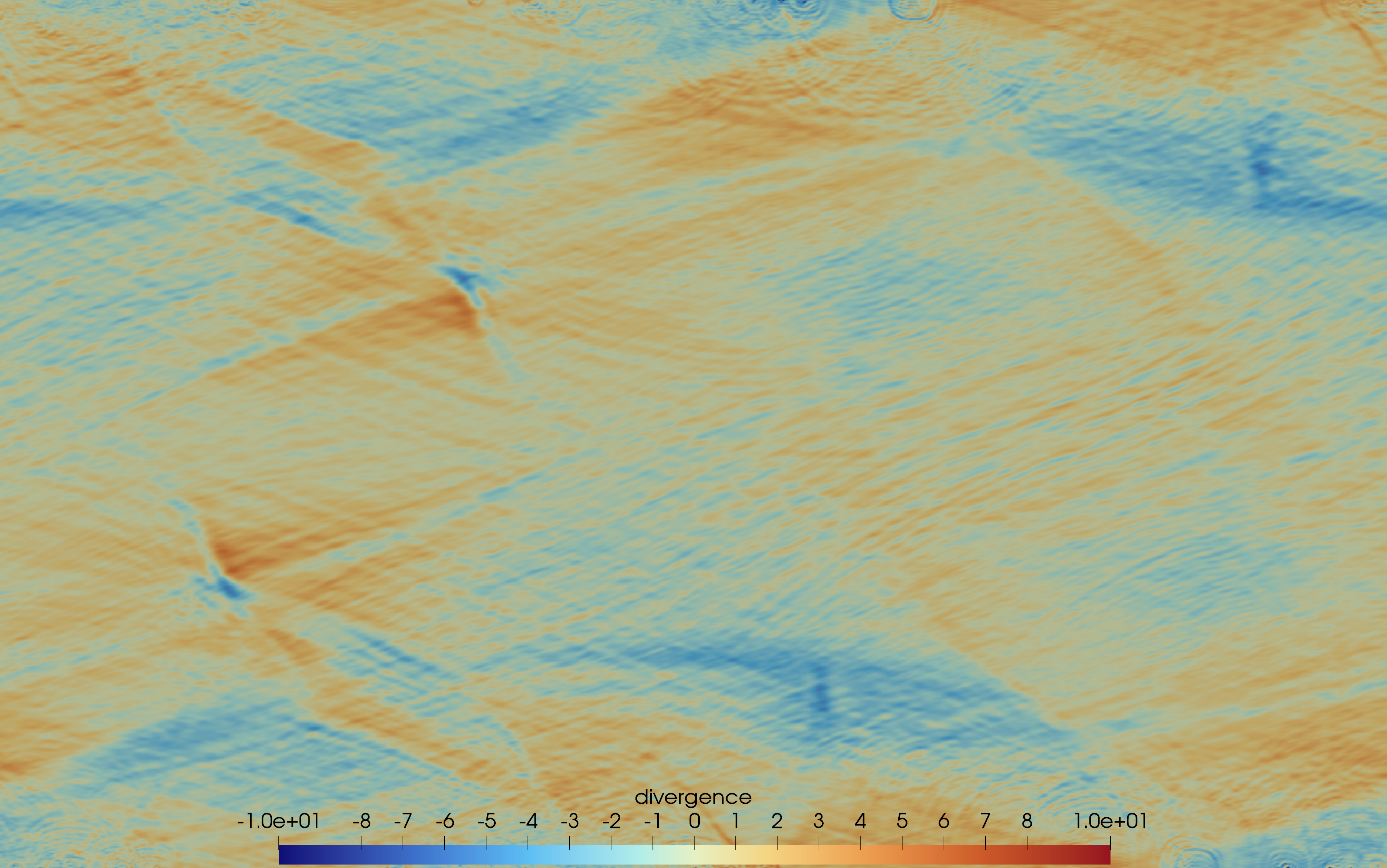}%
\caption{Divergence fluctuations in the Couette flow with stationary
  vorticity in the presence of the van der Waals effect. Upper-left --
  at 0.01 seconds, upper-right -- at 0.02 seconds, lower-left -- at
  0.03 seconds, lower-right -- at 0.05 seconds.}
\label{fig:divergence_Couette_noVorticity}
\end{figure}

\subsection{Power decay of the Fourier spectra}

Here, we compute the time averages of the power spectra in the same
manner as we did for the Poiseuille flow of the full
system~\eqref{eq:div_vort} above in Section~\ref{sec:Poiseuille}. The
computed time-averages of the squares of the fluctuations of $\rho$
and $\chi$ are shown in
Figure~\ref{fig:rho_div_Poiseuille_noVort_spectrum}, and those of
squares of components of $\BV u^\varphi$ in
Figure~\ref{fig:U_Poiseuille_noVort_spectrum}.  Remarkably, all
computed spectra have the same power slopes as their respective
counterparts of the full system \eqref{eq:div_vort} in
Figures~\ref{fig:rho_div_omega_Poiseuille_spectrum}
and~\ref{fig:U_Poiseuille_spectrum}. Namely, the density spectrum,
shown in the left-hand pane of
Figure~\ref{fig:rho_div_Poiseuille_noVort_spectrum}, shows the $\sim
k^{-7/3}$ power decay for all five channel bands. The spectrum of the
velocity divergence, shown in the right-hand pane of
Figure~\ref{fig:rho_div_Poiseuille_noVort_spectrum}, also shows power
decay at the rate $\sim k^{-5/3}$, same as in
Figure~\ref{fig:rho_div_omega_Poiseuille_spectrum}. The spectrum of
the streamwise component $u^\varphi_x$, which corresponds to the
potential part of the streamwise kinetic energy of the flow, is shown
in the left-hand pane of
Figure~\ref{fig:U_Poiseuille_noVort_spectrum}, and decays at the rate
of $\sim k^{-8/3}$ in all channel bands. The spectrum of the
transversal component $u^\varphi_y$, which corresponds to the
potential part of the transversal kinetic energy of the flow, is shown
in the right-hand pane of
Figure~\ref{fig:U_Poiseuille_noVort_spectrum}, and decays at the rate
of $\sim k^{-3}$ in all channel bands.

\begin{figure}[t]%
\includegraphics[width=0.5\textwidth]{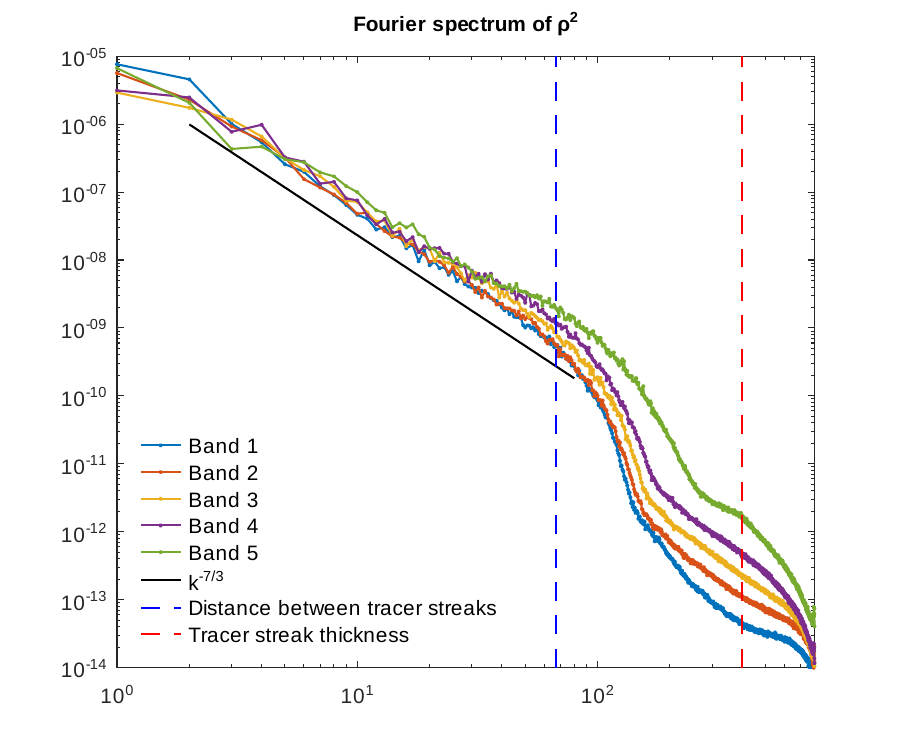}%
\includegraphics[width=0.5\textwidth]{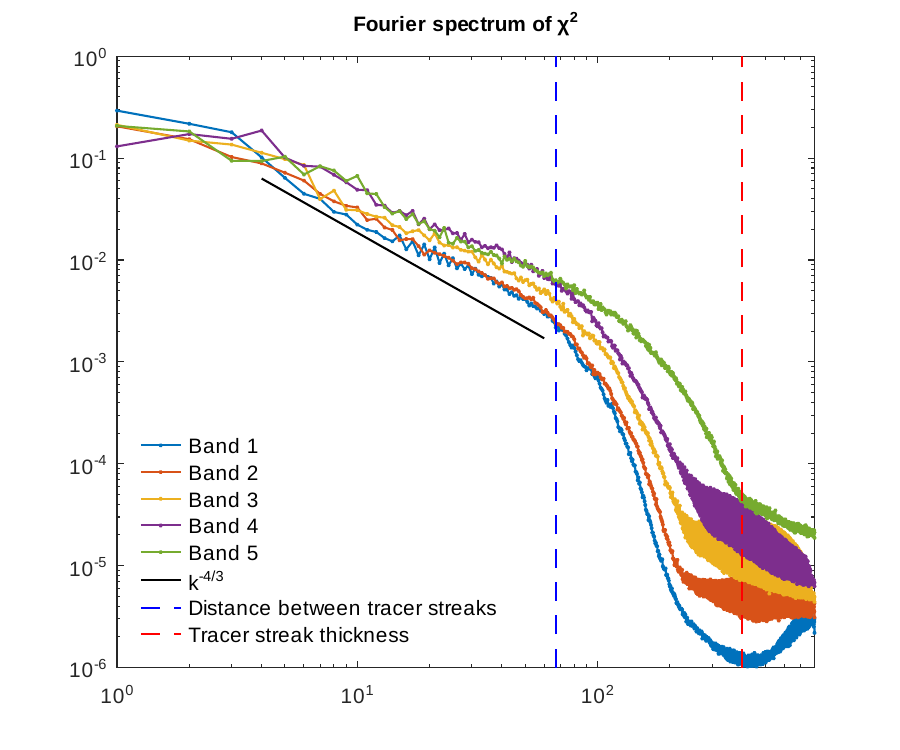}%
\vspace{-1EM}
\caption{The Fourier spectrum of $\rho^2$ (left), and $\chi^2$
  (right), Couette flow with stationary vorticity.}
\label{fig:rho_div_Couette_noVort_spectrum}
\end{figure}
\begin{figure}[t]%
\includegraphics[width=0.5\textwidth]{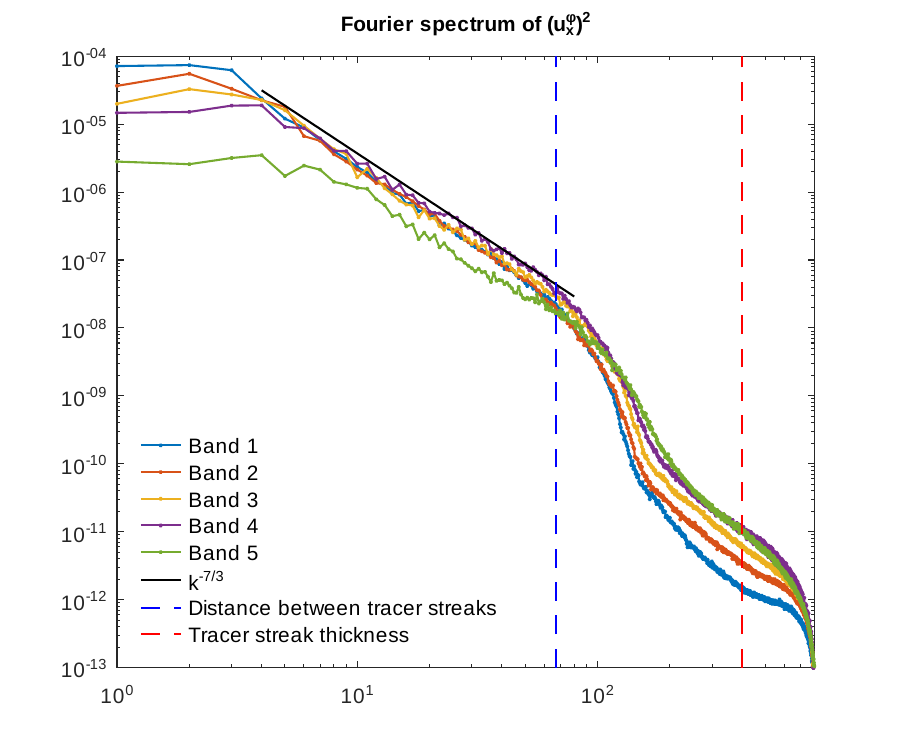}%
\includegraphics[width=0.5\textwidth]{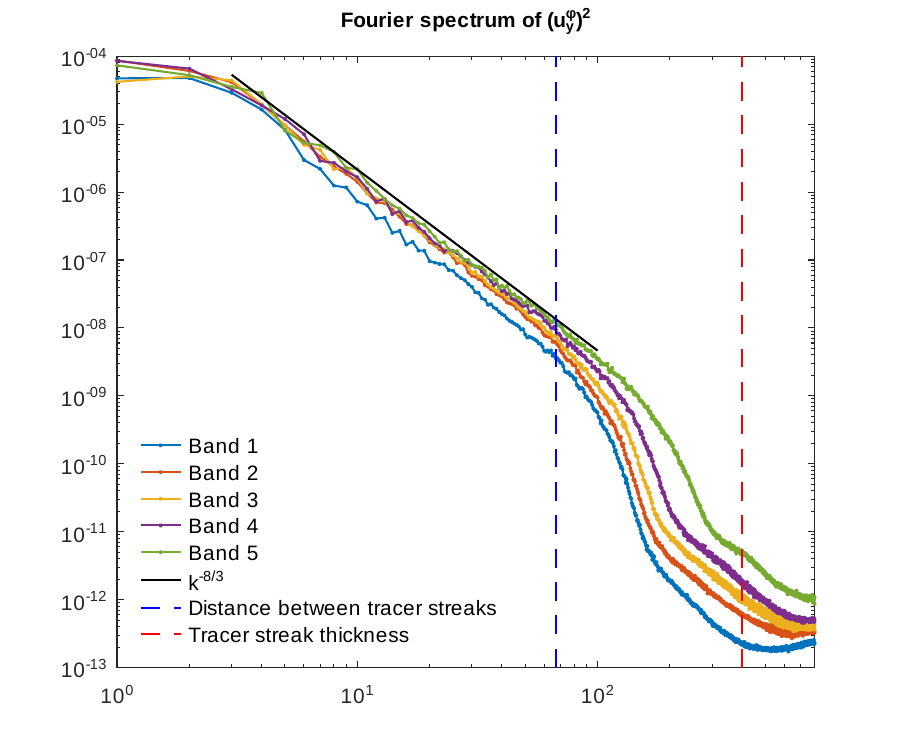}%
\vspace{-1EM}
\caption{The Fourier spectrum of $(u^\varphi_x)^2$ (left), and
  $(u^\varphi_y)^2$ (right), Couette flow with stationary vorticity.}
\label{fig:U_Couette_noVort_spectrum}
\end{figure}

\section{Numerical simulation of the Couette flow with stationary
  vorticity}

Here, we numerically simulate the reduced density-divergence system in
\eqref{eq:div} for the same Couette flow profile and initial
conditions in \eqref{eq:Couette_rho_chi_omega} as we did with the full
system in \eqref{eq:div_vort} back in Section~\ref{sec:Couette}. The
resulting snapshots of the deviations in $\rho$ and $\chi$ from their
background states are shown in
Figures~\ref{fig:density_Couette_noVorticity}
and~\ref{fig:divergence_Couette_noVorticity}, respectively, for the
same elapsed times $t=0.01$, $0.02$, $0.03$ and $0.05$ seconds. Just
as in the Poiseuille flow scenario, here the direct comparison with
Figures~\ref{fig:density_Couette} and~\ref{fig:divergence_Couette}
reveals profound similarities between the numerical solutions of
\eqref{eq:div_vort} and \eqref{eq:div}; in fact, the final snapshots
of the velocity divergence at $t=0.05$ seconds (lower-right panes in
Figures~\ref{fig:divergence_Couette}
and~\ref{fig:divergence_Couette_noVorticity}), while not strictly
identical, are quite similar despite complex wave structures appearing
in both plots. Again, the simulation confirms that vorticity does not
play a key role in the development of the direct cascade in density
and velocity divergence, which was predicted in \cite{Abr27}.

\subsection{Power decay of the Fourier spectra}

Here, we compute the time averages of the power spectra in the same
manner as we did for the Couette flow of the full
system~\eqref{eq:div_vort} above in Section~\ref{sec:Couette}. The
computed time-averages of the squares of the fluctuations of $\rho$
and $\chi$ are shown in
Figure~\ref{fig:rho_div_Couette_noVort_spectrum}, and those of squares
of components of $\BV u^\varphi$ in
Figure~\ref{fig:U_Couette_noVort_spectrum}. As was the case with the
Poiseuille flow, here all computed spectra have the same power slopes
as their respective counterparts of the full system
\eqref{eq:div_vort} in
Figures~\ref{fig:rho_div_omega_Couette_spectrum}
and~\ref{fig:U_Couette_spectrum}. Namely, the density spectrum, shown
in the left-hand pane of
Figure~\ref{fig:rho_div_Couette_noVort_spectrum}, shows the $\sim
k^{-7/3}$ power decay for all five channel bands. The spectrum of the
velocity divergence, shown in the right-hand pane of
Figure~\ref{fig:rho_div_Couette_noVort_spectrum}, also shows power
decay at the rate $\sim k^{-4/3}$, same as in
Figure~\ref{fig:rho_div_omega_Couette_spectrum}. The spectrum of the
streamwise component $u^\varphi_x$, which corresponds to the potential
part of the streamwise kinetic energy of the flow, is shown in the
left-hand pane of Figure~\ref{fig:U_Couette_noVort_spectrum}, and
decays at the rate of $\sim k^{-7/3}$ in all channel bands. The
spectrum of the transversal component $u^\varphi_y$, which corresponds
to the potential part of the transversal kinetic energy of the flow,
is shown in the right-hand pane of
Figure~\ref{fig:U_Couette_noVort_spectrum}, and decays at the rate of
$\sim k^{-8/3}$ in all channel bands.

\section{Discussion}

In the current work, we use the equations for inertial flow with the
van der Waals effect \cite{Abr22,Abr23,Abr24,Abr26,Abr27} to simulate
the dynamics of small perturbations around the Poiseuille and Couette
flows in a straight two-dimensional channel. We re-cast the
two-dimensional inertial flow equations into the divergence--vorticity
formulation to separate the effects of compressibility and
rotation. The main results are as follows:
\begin{enumerate}[label=\arabic*)]
\item Unlike what we observed in \cite{Abr23}, in the current scenario
  the numerically simulated flow does not break down into fully
  chaotic turbulent motions. The fluctuations around both the
  Poiseuille and Couette background flows remain small enough on the
  time scale of the simulation, so that the flow remains
  macroscopically ``pseudo-laminar'' (in particular, the tracer
  streaks, seeded in the initial condition, do not mix and remain
  separate, although minor distortion and smudging can be observed).
\item Yet, small fluctuations around both the Poiseuille and Couette
  flows become chaotic. Our theory in \cite{Abr27} predicts the
  existence of the direct cascade in the inertial flow with the van
  der Waals effect, and the numerical simulations here seem to confirm
  that; namely, the initially large-scale density fluctuations are
  eventually converted into small scale chaotic dynamics.
\item Despite the absence of the turbulent breakdown of the flow, we
  consistently observe the manifestation of the power spectra in the
  time-averages of the Fourier transforms of all variables of the
  system, that is, the density, velocity divergence, vorticity, and
  both the potential and stream function components of the kinetic
  energy of the flow. It is interesting that, first, different
  variables have different powers of their Fourier spectra, and,
  second, the slope also depends on the background flow -- namely, the
  power slopes of the Poiseuille flow are steeper by a cubic root of
  the wavenumber than those of the Couette flow for all variables
  except the density (see Table~\ref{tab:powers} for a summary). It
  seems that the Kolmogorov power of the energy spectrum ($\sim
  k^{-5/3}$) tends to manifest universally in a fully broken-down,
  turbulent flow \cite{Abr22,Abr23,Abr24,Abr26}, whereas a
  slightly perturbed ``pseudo-laminar'' regime can have different
  powers depending on the background profile.
\item Remarkably, setting the vorticity to its background state, and
  leaving only the density and the velocity divergence as variables
  does not qualitatively change the dynamics of the flow. Namely, the
  coupled density and velocity divergence by themselves develop
  chaotic dynamics with the same power spectra of the Fourier
  transforms as does the full system. This is an indication that the
  dynamical mechanism of the power spectra generation resides in the
  density and velocity divergence, which are coupled via the van der
  Waals effect in the momentum equation, while the vorticity seems to
  be irrelevant. This is supported by our theory in \cite{Abr27}.
\end{enumerate}
Based on these results, the following observations can be made:
\begin{enumerate}[label=\alph*)]
\item Typically, it is presumed that there is a clear dichotomy
  between the ``laminar'' and ``turbulent'' flow regimes. This stems
  from the seminal work of Reynolds \cite{Rey83}, where an initially
  laminar flow suddenly became fully turbulent with little or no
  transitional stage in between. However, in the current work the
  simulated flow is technically laminar (in the sense of the tracer
  streaks in Figure~\ref{fig:tracer_Poiseuille}), since its small
  fluctuations have insufficient strength to break and mix the tracer
  streaks.  Yet, these small fluctuations develop chaotic dynamics and
  power spectra, which are normally associated with turbulence. This
  suggests that turbulence, as a phenomenon, may consist of more than
  one distinct ``parts'', and in certain flow regimes some parts may
  be present, and some absent.
\item It is sufficient to have a 2D flow to generate power spectra.
  This was suggested by our theory in \cite{Abr27}, where we found
  that the instability which causes the direct cascade can be
  explained entirely in two dimensions, and is now confirmed directly
  by a numerical simulation. Moreover, as a variable, the small-scale
  vorticity plays no discernible role in the dynamics of the power
  spectra, and the latter are generated largely by the instability in
  the density and velocity divergence variables, albeit in the
  presence of a stationary large scale vorticity, which serves as an
  ``external forcing'' (see \cite{Abr27} for more details).
\item In the same 2D divergence-vorticity setting as
  \eqref{eq:div_vort}, the incompressible Navier--Stokes equations
  \cite{MajBer} are given by
  \begin{equation}
    \label{eq:Navier-Stokes}
    \parderiv\omega t+\nabla^\perp\psi\cdot\nabla\omega=\nu\Delta
    \omega,\qquad\Delta\psi=\omega.
  \end{equation}
  Remarkably, \eqref{eq:Navier-Stokes} completely lacks the
  compressibility mechanism of \eqref{eq:div_vort}; in particular, the
  density $\rho$ and velocity divergence $\chi$ are no longer present,
  and the 2D incompressible flow consists solely of the vorticity
  $\omega$ and the associated rotation. This naturally raises the
  following question: are the incompressible Navier--Stokes equations
  \eqref{eq:Navier-Stokes} suitable for modeling turbulence?  The
  complete absence of the dynamical mechanism of power spectra casts
  doubt on the overall ability of \eqref{eq:Navier-Stokes} to
  accurately predict the observed turbulent behavior of real-world
  flows.
\item In classical turbulence, power slopes of the Fourier spectra of
  various quantities are usually explained via Kolmogorov's
  dimensional hypothesis \cite{Kol41a,Kol41b,Kol62}, which, in turn,
  relies mainly on the Buckingham $\pi$ theorem \cite{Buck} to produce
  an estimate, and avoids taking into account the actual physical
  mechanism behind the observed dynamics (e.g.~the van der Waals
  effect in our case). While we do not dispute the validity of the
  $\pi$ theorem itself, here we cannot help but question the practical
  limitations of this rather popular approach. Recall that the various
  components of the kinetic energy, whose spectra are shown throughout
  our work in Figures~\ref{fig:U_Poiseuille_spectrum},
  \ref{fig:U_Couette_spectrum},
  \ref{fig:U_Poiseuille_noVort_spectrum},
  and~\ref{fig:U_Couette_noVort_spectrum}, all have, first, the same
  physical units, and, second, the same physical origin. This suggests
  that they all must be treated identically in the context of
  Kolmogorov's hypothesis, which would result in the same power slope
  estimate. Yet, not only all components of the kinetic energy have
  different power slopes for a given background flow configuration,
  but even changing the background flow profile (i.e.~Poiseuille to
  Couette) affects the slopes. We, of course, do not doubt the
  observational prevalence of the Kolmogorov $k^{-5/3}$-slope of the
  kinetic energy (which indeed suggests that there must be a certain
  universality to it), but attempting to ``explain'' it without even
  identifying its underpinning physical mechanism does seem to be a
  bit of a stretch.
\item One of the reasons why Kolmogorov's dimensional hypothesis is
  used to explain the power decay of the Fourier spectra, is that the
  latter are normally attributed to a fully developed turbulent flow,
  and that makes is quite difficult (if not impossible) to produce
  better estimates. Conversely, here we established that the power
  spectra exist in small perturbations around relatively simple
  laminar stationary states, which seems rather promising. Since the
  fluctuations of numerical solutions of both \eqref{eq:div_vort} and
  \eqref{eq:div} remain small relative to the background states, it
  seems plausible that the dynamics captured by our numerical
  simulations can be described via a linearization near the
  corresponding background state (Poiseuille or Couette). We already
  studied the linearization of \eqref{eq:div_vort} near the Couette
  flow in \cite{Abr27}, and found that it leads to a $3\times 3$
  system of linear non-autonomous ODE along the characteristics.  In
  that system, we only managed to examine the initial development of
  the instability and its asymptotic limit, but not the intermediate,
  ``inertial'' stage of the solution. However, the corresponding
  linearization of the reduced system \eqref{eq:div} should lead to a
  smaller, $2\times 2$ system of linear ODE, which is likely easier to
  study. With enough luck, perhaps we could even find some explicit
  solutions which exhibit the power decay in their Fourier spectra.
\end{enumerate}
The natural next step is to examine the linearization of the reduced
system \eqref{eq:div} around the Couette profile, as was done in
\cite{Abr27} for the full system \eqref{eq:div_vort}. Such a
linearization will lead to a $2\times 2$ system of non-autonomous
linear ODE. The matrix entries of this system will be at most rational
functions of the time variable, with at most a quadratic power in both
the numerator and denominator. Structurally, the dynamics will consist
of the self forcing-damping term in the velocity divergence variable,
coupled to the density variable via the van der Waals effect to form
an oscillator. While it is unclear at this time whether the solution
can be obtained explicitly, we hope to at least understand
qualitatively why its solutions in the inertial stage tend to decay as
a power law of the wavenumber.

\ack The work was supported by the Simons Foundation grant \#636144.

\end{document}